%% file: ms.tex
\documentclass[journal=ancac3,manuscript=article,layout=twocolumn]{achemso}

\makeatletter
\let\l@addto@macro\relax
\makeatother
\usepackage[fontsize=10pt]{scrextend}

\usepackage{multirow}
\usepackage{graphicx}
\usepackage{wrapfig}
\usepackage{subcaption}
\usepackage{amsmath} 
\usepackage{amssymb}
\usepackage{siunitx}
\usepackage{booktabs}
\usepackage[export]{adjustbox}
\usepackage{cleveref}
\usepackage{booktabs}
\usepackage{gensymb}
\usepackage{float}
\usepackage{diagbox}
\usepackage{array}

\SectionNumbersOn

\usepackage[utf8]{inputenc}
\DeclareUnicodeCharacter{03C9}{$\omega$}

\newcolumntype{P}[1]{>{\centering\arraybackslash}p{#1}}
\newcolumntype{M}[1]{>{\centering\arraybackslash}m{#1}}
\newcommand{\PreserveBackslash}[1]{\let\temp=\\#1\let\\=\temp}
\newcolumntype{C}[1]{>{\PreserveBackslash\centering}p{#1}}
\newcolumntype{R}[1]{>{\PreserveBackslash\raggedleft}p{#1}}
\newcolumntype{L}[1]{>{\PreserveBackslash\raggedright}p{#1}}

\title{Capturing Subdiffusive Solute Dynamics and Predicting Selectivity in Nanoscale Pores with Time Series Modeling}
\author{Benjamin J. Coscia}
\affiliation{Department of Chemical and Biological Engineering, University of Colorado Boulder, Boulder, CO 80309, USA}
\author{Michael R. Shirts}
\email{michael.shirts@colorado.edu}
\affiliation{Department of Chemical and Biological Engineering, University of Colorado Boulder, Boulder, CO 80309, USA}

\begin{document}
  
  \begin{abstract}
  
  Mathematically modeling complex transport phenomena can be a powerful tool
  for extracting important physical information from molecular simulations. 
  In this study, we present two new approaches that use stochastic time series
  modeling to predict long time-scale behavior and macroscopic properties 
  from molecular simulation which can be generalized to other molecular systems
  where complex diffusion occurs. Specifically, we parameterize our models using
  long molecular dynamics (MD) simulation trajectories of a cross-linked 
  H\textsubscript{II} phase lyotropic liquid crystal (LLC) membrane in order to
  predict solute mean squared displacements (MSDs), solute flux, and
  solute selectivity in macroscopic length pores.

  First, using anomalous diffusion theory, we show how solute dynamics can be
  modeled as a fractional diffusion process subordinate to a continuous time 
  random walk. From the MD simulations, we parameterize the distribution of 
  dwell times, hop lengths between dwells and correlation between hops. We 
  explore two variations of the anomalous diffusion modeling approach. The first 
  variation applies a single set of parameters to the solute displacements and 
  the second applies two sets of parameters based on the solute's radial distance
  from the closest pore center. 

  Next, we present an approach that generalizes Markov state models, treating 
  the configurational states of the system as a Markov process where each 
  state has distinct transport properties. For each state and transition between
  states, we parameterize the distribution and temporal correlation structure of
  positional fluctuations as a means of characterization and to allow us to
  predict solute MSDs. 
 
  We show that both stochastic models reasonably reproduce the MSDs calculated 
  from MD simulations. However, qualitative differences between MD and Markov 
  state dependent model-generated trajectories may in some cases limit their 
  usefulness. With these parameterized stochastic models, we demonstrate how 
  one can estimate flux of a solute across a macroscopic-length pore and, based
  on those quantities, the membrane's selectivity towards each solute. This
  work therefore helps to connect microscopic, chemically-dependent solute 
  motions that do not follow simple diffusive behavior with long time-scale 
  behavior, in an approach generalizable to many types of molecular systems 
  with complex dynamics.

  \end{abstract}

  \maketitle
  
  \graphicspath{{./figures/arxiv_figures/}{./supporting_figures/arxiv_figures/}}

  \section{Introduction}

  There is a large disparity between the time scales accessible to atomistic 
  molecular dynamics (MD) simulations and experiment. The dynamics of molecules 
  over nanoseconds of simulation may tell a very different story than that 
  inferred by bulk measurements on the time-scale of seconds. Understanding 
  complex dynamical behavior on short time-scales and connecting them with much
  longer time-scales can provide fundamental insight which can be used for 
  atomic level design of materials. In this work, we pursue this connection 
  applied to highly selective aqueous filtration membranes and demonstrate how
  one can gain chemical intuition which serves both theoretical and experimental
  researchers.

  Highly selective separation membranes are desirable in numerous
  applications. The ability to efficiently separate ions from saline water
  sources using membranes has been actively pursued for years in an effort to
  create potable water for people in water-scarce
  regions.~\cite{werber_materials_2016} Even in relatively safe municipal water
  supplies, there is a need for membranes that can specifically separate
  potentially harmful organic micropollutants such as fertilizers, pesticides,
  pharmaceuticals and personal care products.~\cite{barbosa_occurrence_2016} 
  Lyotropic liquid crystals (LLCs) are a class of amphiphilic molecules that
  can be cross-linked into mechanically strong and highly selective nanoporous
  membranes.~\cite{gin_polymerized_2008} They may provide a promising
  alternative to conventional membrane separation techniques by being selective
  based not only on solute size and charge, but on solute chemical
  functionality as well.~\cite{dischinger_application_2017} One can tune the
  functionality of LLC monomers in order to enhance or weaken specific
  solute-membrane interactions.~\cite{dischinger_effect_2017} 

  In this work, we study the inverted hexagonal, H\textsubscript{II}, LLC phase. 
  The H\textsubscript{II} phase has densely packed, uniform-sized pores on the 
  order of 1 nm in size. A perfectly aligned H\textsubscript{II} phase has 
  an ideal geometry for high throughput separations.~\cite{feng_scalable_2014,feng_thin_2016}

  Molecular modeling may make it possible to efficiently evaluate solute-specific
  separation membranes using the available chemical space of LLC monomers by 
  allowing researchers to understand their macroscopic behavior based on 
  microscopic, chemically-dependent solute motions. To date, only a limited 
  subset of H\textsubscript{II} phase-forming LLC monomers have been studied
  experimentally in membrane applications.~\cite{carter_glycerol-based_2012,hatakeyama_nanoporous_2010,smith_ordered_1997,zhou_assembly_2003,resel_structural_2000}
  Even within this small subset, they have demonstrated selectivities that
  could not be explained beyond speculation and vague empirical
  correlations.~\cite{dischinger_application_2017} Atomistic molecular modeling
  can provide the resolution necessary to identify molecular interactions that
  are key to separation mechanisms, allowing us to move beyond Edisonian design
  approaches.

  In our previous work,~\cite{coscia_chemically_2019} we used molecular
  dynamics (MD) simulations to study the transport of 20 small polar molecules
  in an H\textsubscript{II} phase LLC membrane. 
  In general,
  we observed subdiffusive transport behavior characterized by intermittent
  hops separated by periods of entrapment. We identified three mechanisms
  responsible for the solute trapping behavior: entanglement among monomer
  tails, hydrogen bonding with monomer head groups, and association with the
  monomer's sodium counter ions.
  
  Up through our previous studies, our molecular models have provided valuable
  qualitative mechanistic insight with some quantitative support. This insight
  already allows us to speculate about new LLC monomer designs. However, they
  would be of greater value to a larger set of researchers if we could provide
  quantitative predictions of macroscopic observables such as solute flux and
  selectivity. Due to the size of the types of systems we are studying (at
  least 62,000 atoms) it is prohibitively expensive and time-consuming to run
  simulations longer than those performed for this work ($\sim$5 $\mu$s). Even
  with the relatively large size of our system, we only simulate 24 solute
  trajectories per unit cell in order to minimize solute-solute interactions.
  We are in need of an efficient method which can use our limited atomistic 
  simulation data sets in order to predict long timescale solute behavior.  

  Mathematical descriptions of transport in complex separations membranes are a
  powerful way to understand mechanisms and formulate design principles.
  \cite{vinh-thang_predictive_2013,geens_transport_2006,darvishmanesh_mass_2016}
  In dense homogeneous membranes, the solution-diffusion model
  can extract diffusion and partition coefficients and has successfully
  predicted solute transport rates.~\cite{wijmans_solution-diffusion_1995}
  Analogously, pore-flow models yield predictions of diffusion coefficients and
  solute transport rates in nanoporous membranes.~\cite{paul_diffusive_1974}
  Modern single particle tracking approaches have taken researchers beyond
  continuum modeling allowing them to characterize complex diffusive
  behavior.~\cite{manzo_review_2015} At the molecular level, one can use
  molecular dynamics (MD) simulations to study both single particle dynamics
  and bulk transport properties with atomic-level
  insight.~\cite{coscia_chemically_2019,maginn_best_2018} All of these
  approaches facilitate generation of hypotheses about the molecular origins of
  separations by attempting to give a more intuitive understanding of how
  solutes move as a function of their environment, in turn suggesting
  experiments that could be performed.

  Using a bottom-up modeling approach, it may be possible to parameterize single
  particle behavior by extracting particle trajectories from MD simulations and
  studying the properties that have the greatest influence on solute dynamics.
  With this information, one can learn how to construct an ensemble of characteristic
  single solute trajectories with these properties which would be useful for
  making long timescale predictions with computational ease and lower
  uncertainties. There is an abundance of information encoded in
  the complex solute trajectories. One can incorporate fluctuations and 
  time-dependent correlations in solute position into transport models, 
  as well as the length of time solutes are trapped. It is possible to add further
  detail by integrating this time series analysis with existing knowledge of the
  primary trapping mechanisms as well as the solute's changing chemical environment
  within the heterogeneous membrane. 

  In this work, we use the output of our MD simulations to construct two
  classes of mathematical models which aim to predict membrane performance
  while providing quantitative mechanistic insights. The functional forms of
  these models are driven by mechanistic observations from our previous work and
  their inputs are parameterized using a substantial amount of data generated
  by long ($\geq 5~\mu$s) MD simulations. Where sufficiently close to observed data,
  these fitted models provide a way to propagate information about solute 
  trajectories gathered at the microsecond time scale to realistic experimental 
  time scales. 

  We constructed our first model by applying the existing rigorous theoretical
  foundation which describes the motion of particles that exhibit non-Brownian,
  or anomalous, transport behavior.~\cite{metzler_random_2000,bouchaud_anomalous_1990} 
  The tools introduced by fractional calculus are instrumental to this 
  theory.~\cite{gorenflo_fractional_1997} They allow us to generalize the normally
  linear diffusion equation to fractional derivative orders, providing descriptions
  of a much more diverse set of behavior, including subdiffusion, a type of 
  anomalous diffusion exhibited by solutes on the simulation time scales of this
  study.~\cite{klages_anomalous_2008}

  We treat our system in terms of two well-known classes of anomalous
  subdiffusion: fractional Brownian motion (FBM) subordinate to a 
  continuous time random walk (CTRW), or subordinated FBM (sFBM) for short. 
  FBM is common in crowded, viscoelastic environments where each jump comes 
  from a Gaussian distribution but is anti-correlated to its previous steps.
  ~\cite{mandelbrot_fractional_1968,jeon_fractional_2010,banks_anomalous_2005}
  A pure CTRW is characterized by a distribution of hop lengths and dwell times, 
  where trajectories consist of sequential independent random draws from each
  distribution.~\cite{montroll_random_1965} If the draws are not independent,
  then the hop sequence generation process can be described as subordinate to
  the CTRW. In our case, we observe anti-correlated draws from the hop distribution
  suggesting an FBM process is subordinate to the CTRW.
  
  
  If one has knowledge of the primary mechanisms leading to anomalous diffusion
  behavior, one may gain additional insight by formulating a state-based model,
  such as Markov state models (MSMs). MSMs are a popular class of models used
  to project long timescale system properties based on molecular simulation
  trajectories by identifying different dynamical modes and quantifying the
  rates of transitions between them. MSMs are frequently used to study systems
  with slow dynamics, such as protein
  folding.~\cite{snow_how_2005,chodera_automatic_2007} Researchers typically
  aim to come up with a low dimensional representation of the system based on
  features which preserve the process kinetics. This still often results in
  hundreds to thousands of distinct states.~\cite{chodera_markov_2014}

  Our second modeling approach applies an extended MSM framework to a
  relatively small set of known states based on the three previously observed
  solute trapping mechanisms. Standard MSMs are typically applied to determine
  equilibrium populations of states and the kinetics of transitions between
  those states.~\cite{bowman_using_2009} We extend the framework to include
  state-dependent fluctuations and correlations in solute position. The model
  determines the magnitude of a solute's fluctuations from its average trapped
  position and the degree of correlation with previous fluctuations by its current
  state. To distinguish our approach from standard MSMs, we have named it the
  Markov state-dependent dynamical model (MSDDM).
  
  We determine the degree of success of our modeling approaches in two ways.
  First, if we can closely reproduce solute MSDs measured from MD simulations
  with realizations of our models, then it is likely that the model
  sufficiently captures solute dynamics and can be used in a predictive
  capacity. Even if a model fails to reproduce the MD MSDs, there is value in
  uncovering the cause of the deviation. The second measure of success is based
  on the qualitative comparison between individual realizations of solute 
  trajectories generated by our models and those observed from MD simulations. 
  Even if we can reproduce the solute MSDs based on realizations of our models,
  the absence or inappropriate reproduction of hopping and trapping behavior may
  indicate underlying model issues. 
  
  The goal of this work is not to definitively determine which modeling
  approach is better but to evaluate their performance independently because
  they both have potential value dependent on a given research study's goals.
  The AD approach provides a systematic way to compare the dynamical behavior
  of different solutes based solely on the time series of their center of mass
  positions. The process of choosing the correct AD approach provides its own
  mechanistic insight since it requires a thorough analysis of solute time
  series behavior. One can simulate different solutes, compare the fit model
  parameters and relate them back to differences in solute size and chemical
  composition. The MSDDM characterizes explicitly defined trapping mechanisms,
  providing a clear picture of solute behavior while in each trapped state as
  well as the equilibrium occupation of each trapping state.  It is possible to
  use the two modeling approaches in tandem, the AD approach to identify
  mechanisms, and the MSDDM to characterize mechanisms. 
  
  We evaluate the two modeling approaches by using them to characterize the
  dynamical behavior of the four fastest moving solutes studied in our previous
  work: methanol, urea, ethylene glycol and acetic acid. In addition to moving
  more quickly than other solutes studied previously, allowing them to extensively
  explore membrane structural space, these solutes have a range of chemical 
  functionality and experience each of the three trapping mechanisms to different extents.
  
  Finally, we demonstrate how one can use stochastically generated realizations
  of our models in order to predict solute flux and selectivity in pores of 
  macroscopic length, thus achieving a better understanding of macroscopic 
  properties on the basis of microscopic dynamics. By improving our 
  understanding of macroscopic behavior in this way, we can begin to think more
  critically about how to design membranes in order to selectively pass or 
  reject specific solutes.
    
  \section{Methods}
    
  We ran all MD simulations and energy minimizations using GROMACS
  2018~\cite{bekker_gromacs:_1993,berendsen_gromacs:_1995,van_der_spoel_gromacs:_2005,hess_gromacs_2008}.
  We performed all post-simulation trajectory analysis using Python scripts
  which are available online at
  \texttt{https://github.com/shirtsgroup/LLC\_Membranes}. The appropriate
  scripts to use for subsequent calculations are summarized in
  Table~\ref{S-table:python_scripts} of the Supporting Information.
  
  \subsection{Molecular Dynamics Simulations}

\begin{figure*}
  \centering
  \begin{subfigure}{0.4\textwidth}
  \centering
  \vspace{1.1cm}
  \includegraphics[width=\textwidth]{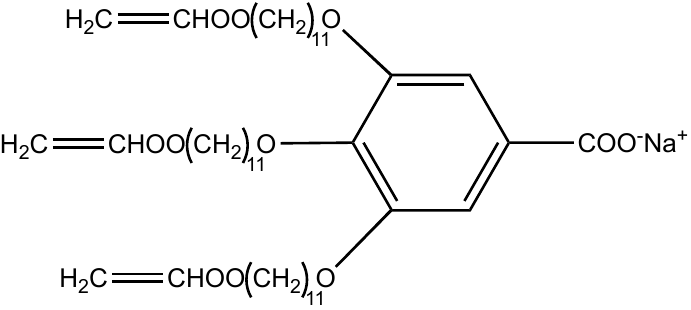}
  \vspace{0.6cm}
  \caption{}\label{fig:monomer_structure}
  \end{subfigure}
  \begin{subfigure}{0.5\textwidth}
  \centering
  \includegraphics[width=\textwidth]{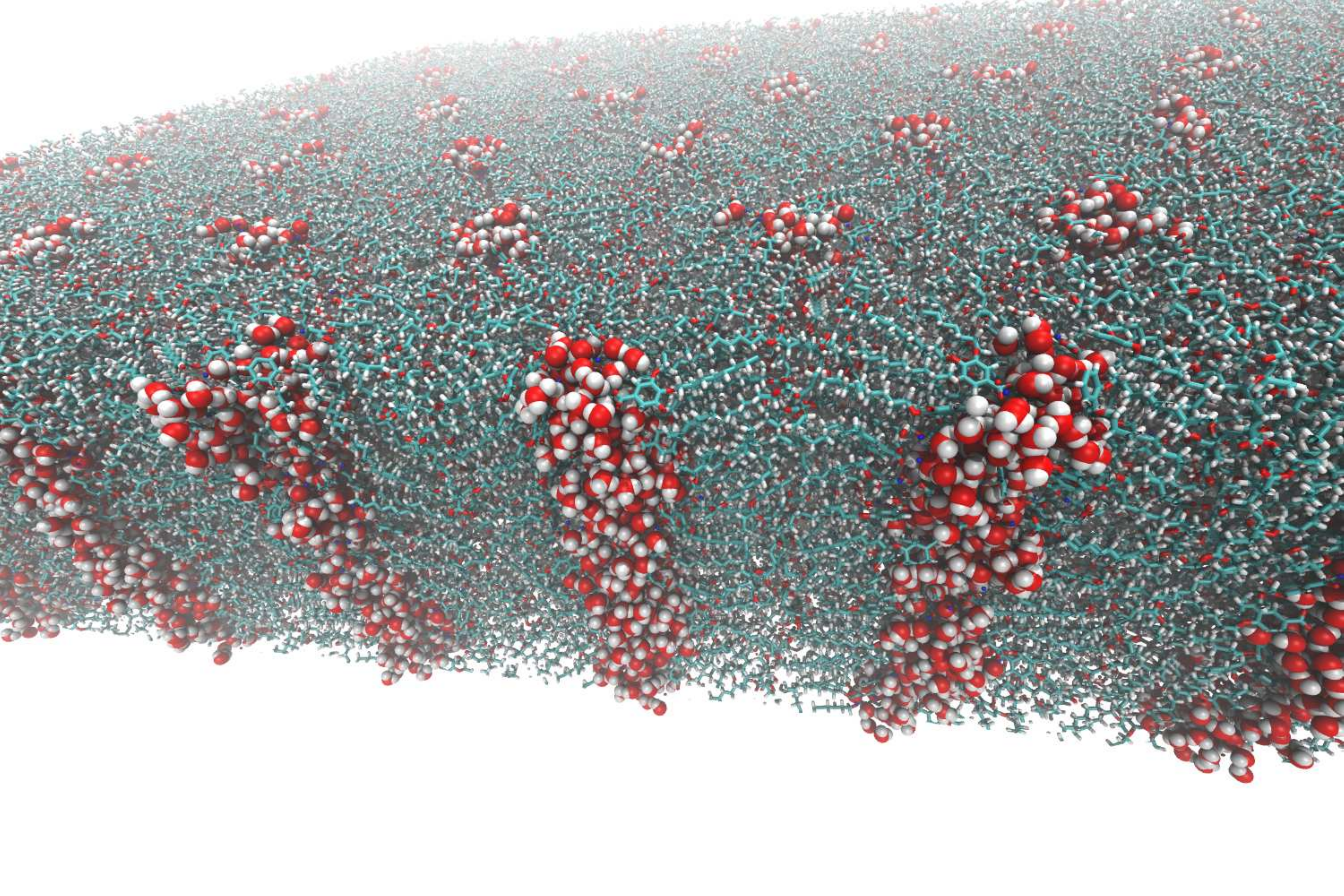}
  \caption{}\label{fig:membrane_profile}
  \end{subfigure}
  \caption{(a) The wedge-shaped liquid crystal monomer Na-GA3C11 will form the inverted
  hexagonal phase in the presence of water where the carboxylate head groups occupy the
  pore centers. (b) A cross-section of a periodically replicated atomistic unit cell 
  used for simulations in this study reveals the membrane's aqueous, hexagonally packed,
  straight and uniform sized pores. Water molecules (red and white spheres) present in 
  the tail region are omitted for clarity.}\label{fig:membrane_structure}
  \end{figure*}

  We studied transport of solutes in the H\textsubscript{II} phase using an
  atomistic molecular model of four pores in a monoclinic unit cell with 10\%
  water by weight (see Figure~\ref{fig:membrane_structure}). Approximately one
  third of the water molecules occupy the tail region with the rest near the
  pore center.
  
  We chose to study a subset of the 4 fastest moving solutes from our previous
  work: methanol, acetic acid, urea and ethylene glycol. For each solute we 
  created a separate system and to each system we added 6 solutes per pore 
  for a total of 24 solutes. On the time scales which we simulate, this number
  of solutes per pore provides a sufficient amount of data from which to 
  generate statistics. It also maintains a low degree of interaction between
  solutes since, at present, we are primarily interested in solute-membrane 
  interactions. Further details on the setup and equilibration of these systems
  are described in our previous work.\cite{coscia_chemically_2019}
  
  We extended the 1 $\mu$s simulations of our previous work to 5 $\mu$s in order
  to collect ample data. We ran MD simulations using the leapfrog integrator with 
  hydrogen bonds constrained by the LINCS algorithm. We simulated a time step of 2 fs at
  a pressure of 1 bar and 300K controlled by the Parrinello-Rahman barostat and
  the v-rescale thermostat respectively. We recorded frames every 0.5 ns.
  
  We considered each system to be equilibrated when the solute partitioning between the 
  pore and tail region reached apparent equilibrium and use only data after that point
  in our analysis of solute kinetics. We plot the solute partition versus time in
  Figure~\ref{S-fig:equilibration} of the Supporting Information in order to justify
  our choice of equilibration times for each solute.

  \subsection{The Anomalous Diffusion Model}\label{method:model_sFBM}

  On the time scales simulated in this study, solutes in this system very 
  clearly exhibit subdiffusive behavior, a type of anomalous diffusion. 
  During an anomalous diffusion process, the mean squared displacement (MSD)
  does not grow linearly with time, but instead it follows a power law of 
  the form: 
  \begin{equation} 
  \langle z^2(t) \rangle = K_{\alpha}t^{\alpha}
  \label{eqn:msd_form}
  \end{equation} 
  where $\alpha$ is the anomalous exponent and $K_{\alpha}$ is the generalized 
  diffusion coefficient. In this work, we only consider the MSD with respect to
  the solutes' center of mass $z$-coordinate, which is oriented along the pore 
  axis. A value of $\alpha < 1$ indicates a subdiffusive process, while values
  of $\alpha = 1$ and $\alpha > 1$ are characteristic of Brownian and 
  superdiffusive motion respectively.
 
  While it is theoretically possible to extract the value of $\alpha$ by fitting
  Equation~\ref{eqn:msd_form} to an MSD generated directly from MD simulations, 
  we do not simulate enough independent solute trajectories to obtain a reliable
  estimate. It is also possible that solute dynamics will become diffusive on
  long simulation timescales. Therefore, the transition from subdiffusion to regular
  diffusion may interfere with estimates of $\alpha$ obtained by fitting to the MSD
  curves. We can obtain higher precision estimates of $\alpha$ using the dwell
  time distributions as described in subsequent sections.
  
  In this study, we primarily use the MSD as a tool for characterizing the average
  dynamic behavior of solute trajectories. Therefore, it is only important that we use a
  consistent definition for calculating the MSD between modeled trajectories and
  directly observed MD trajectories. We chose to calculate the time-averaged MSD from 
  our MD simulations and stochastic models which measures all observed displacements
  over time lag $\tau$: 
  \begin{equation}
  \overline{z^2(\tau)} = \dfrac{1}{T - \tau}\int_{0}^{T - \tau} (z(t + \tau) - z(t))^2 dt
  \label{eqn:tamsd}
  \end{equation}
  where T is the length of the trajectory. In some situations, it may make
  more sense to use the ensemble MSD. We present further discussion of this
  choice in Section~\ref{S-section:msd} of the Supporting Information.

  \subsubsection{Subordinated Fractional Brownian Motion}\label{method:sfbm}

  One can characterize the CTRW component of an sFBM process by the parameters
  which describe its dwell time and hop length distributions. We used the
  \texttt{ruptures} Python package in order to automatically identify mean
  shifts in solute trajectories, indicating hops.\cite{truong_ruptures:_2018}
  We used the corresponding hop lengths and dwell times between hops to
  construct empirical distributions.
  
  \textit{Dwell time distributions:} For subdiffusive transport, the distribution 
  of dwell times is expected to fit a power law distribution 
  proportional to $t^{-1-\alpha}$.~\cite{meroz_toolbox_2015}
  Because we are limited to taking measurements at discrete intervals dictated by the output 
  frequency of our simulation trajectories, we fit the empirical dwell times
  to a discrete power law distribution whose maximum likelihood $\alpha$ 
  parameter we calculated by maximizing the log likelihood function: 
  \begin{equation}
	\mathcal{L}(\beta) = -n\ln \zeta(\beta, t_{min}) -
	\beta\sum_{i=1}^{n} \ln t_i 
  \label{eqn:powerlaw_likelihood}
  \end{equation}
  where $\beta = 1 + \alpha$, $t_i$ are collected dwell time data points,
  $n$ the total number of data points, and $\zeta$ is the Hurwitz zeta function
  where $t_{min}$ is the smallest measured value of $t$.~\cite{clauset_power-law_2009}
  
  In practical applications, the heavy tail of power law distributions can result in 
  arbitrarily long dwell times that are never observed in MD simulations. 
  In order to directly compare our anomalous diffusion model to finite-length MD 
  trajectories we need to bound the dwell time distribution. A standard way of 
  doing this, with easily estimated parameters, is by adding an exponential 
  cut-off to the power law so the dwell time distribution is now proportional 
  to $t^{-1 - \alpha}e^{-\lambda t}$.~\cite{newman_power_2005,clauset_power-law_2009}  
  We determine MLEs of $\alpha$ and $\lambda$ by maximizing the log 
  likelihood function:~\cite{clauset_power-law_2009}
  \begin{equation}
  	\begin{split}
    \mathcal{L}(\alpha, \lambda) = n(1 - \alpha)\ln\lambda - n\ln\Gamma(1 - \alpha, t_{min}\lambda) \\
    -\alpha\sum_{i=1}^{n}\ln t_i - \lambda\sum_{i=1}^n t_i
    \end{split}
  \label{eqn:powerlaw_cutoff_likelihood}
  \end{equation}
  
  \textit{Correlated hop length distributions:} The distribution of hop 
  lengths by solutes undergoing an sFBM process is Gaussian, therefore we 
  parameterize it by its standard deviation, $\sigma$.
  ~\cite{metzler_random_2000, metzler_anomalous_2014,neusius_subdiffusion_2009}
  The measured mean of the hop length distribution is always very close to zero so
  we assume that it is exactly zero in our time series simulations since we have 
  no reason to expect drift in either direction based on pore symmetry.

  sFBM implies that hops are correlated which we describe using the Hurst
  parameter, $H$. The autocovariance function of hop lengths has the
  analytical form:~\cite{mandelbrot_fractional_1968}
  \begin{equation}
    \gamma(k) = \dfrac{\sigma^2}{2}\bigg[|k-1|^{2H} - 2|k|^{2H} + |k+1|^{2H}\bigg]
  \label{eqn:fbm_autocorrelation}
  \end{equation}
  where $\sigma^2$ is the variance of the underlying Gaussian distribution from
  which hops are drawn and $k$ is the time lag, or number of increments between
  hops. The hop autocorrelation function is simply
  Equation~\ref{eqn:fbm_autocorrelation} normalized by the variance.  When $H <
  0.5$, hops are negatively correlated, when $H = 0.5$ we recover Brownian
  motion and when $H > 0.5$, one observes positive correlation between hops. 
  We obtained $H$ by performing a least squares fit of 
  Equation~\ref{eqn:fbm_autocorrelation} to the first ten 0.5 ns time lags of 
  the empirically measured autocorrelation function (see section~\ref{S-section:H_estimate}
  of the Supporting Information for a more in depth justification of this method).
 

  \subsubsection{Subordinated Fractional L\'evy Motion}\label{method:sflm}

  Because we also want to account for the possibility that the distribution of
  hops is not Gaussian, we can model them with the more general class of L\'evy
  stable distributions. For independent and identically distributed random
  variables, the generalized central limit theorem assures convergence of the
  associated probability distribution function (PDF) to a L\'evy stable PDF.
  \cite{klages_anomalous_2008} The characteristic equation which describes the
  Fourier transform of a L\'evy stable PDF is: 
  \begin{equation}
  \begin{split}
    p_{\alpha_h, \beta}(k;\mu,\sigma) =~~~~~~~~~~~~~~~~~~~~~~~~~~~~~~~~~~~~~~~~~~~~~ \\
    \exp\left[i\mu k - \sigma^{\alpha_h}|k|^{\alpha_h}\left(1 - i\beta\frac{k}{|k|}\omega(k, \alpha_h)\right)\right]
  \end{split}
  \end{equation}
  where \\
  \[\omega(k, \alpha_h) = \begin{cases}
  	\tan{\frac{\pi \alpha_h}{2}} & \text{if}~\alpha_h \neq 1, 0 < \alpha_h < 2, \\
  	-\frac{2}{\pi}\ln |k| & \text{if}~\alpha_h = 1
  	 \end{cases}
  \]
  $\alpha_h$ is the index of stability or L\'evy index, $\beta$ is the skewness 
  parameter, $\mu$ is the shift parameter and $\sigma$ is a scale parameter. The most
  familiar case, and one of three that can be expressed in terms of elementary functions,
  is the Gaussian PDF ($\alpha_h$ = 2, $\beta$ = 0). We assume symmetric distributions
  centered about 0 implying that $\beta$ and $\mu$ are both 0.
  
  For correlated hops, solute behavior may be described by subordinated
  fractional L\'evy motion (sFLM). The Hurst parameter can again be used to
  describe hop correlations because they share the same autocorrelation
  structure.~\cite{tikanmaki_fractional_2010} The autocovariance function for
  FLM is:
  \begin{equation}
  \begin{gathered}
    \gamma(k) = \dfrac{C}{2}\bigg[|k-1|^{2H} - 2|k|^{2H} + |k+1|^{2H}\bigg], \\
    ~C = \frac{E\big[L(1)^2\big]}{\Gamma(2H + 1)\sin(\pi H)}
  \end{gathered}
  \label{eqn:flm_autocovariance}
  \end{equation}
  where $E\big[L(1)^2\big]$ is the expected value of squared draws from the 
  underlying L\'evy distribution, effectively the distribution's 
  variance.~\cite{bishwal_maximum_2011} In general, most L\'evy stable distributions
  have an undefined variance due to their heavy tails. However, normalizing
  Equation~\ref{eqn:flm_autocovariance} by the variance of a finite number of draws
  from a L\'evy stable distribution results in the same autocorrelation structure as FBM.
  See Section~\ref{S-section:H_estimate} of the Supporting Information for numerical
  simulations illustrating this point.
  
  Analogous to power law dwell times, the heavy tails of L\'evy stable hop
  length distributions result in rare but arbitrarily long hops. These long and
  unrealistic hops result in over-estimated simulated MSDs (see
  Figure~\ref{fig:ad_realizations_1mode} for example). We observe that the
  distribution of hops observed in our MD simulations are well approximated by
  L\'evy stable distributions close to the mean, but they significantly
  under-sample the tails. We chose to truncate the L\'evy stable distributions
  based on where the theoretical PDF starts
  to deviate from the empirically measured PDF (see
  Section~\ref{S-section:truncation} of the Supporting Information).~\cite{mantegna_stochastic_1994}

  \subsubsection*{Multiple Anomalous Diffusion Regimes}
  
  We observe different dynamical behavior when solutes move while inside the
  pore versus while in the tail region. This suggests two anomalous diffusion
  models of varying complexity. We first create a simple, single mode model
  with a single set of parameters fit to solute trajectories. Our second, two
  mode model assigns a set of parameters to each of 2 modes based on the
  solute's radial location. We define the first mode as the pore region,
  defined as less than 0.75 nm from any pore center. Solutes outside the pore
  region are in the second mode, the tail region. We determined this cut-off by
  maximizing the difference in dynamical behavior as described in our previous
  work.~\cite{coscia_chemically_2019} Unfortunately, there were not enough
  sufficiently long sequences of hops in each mode to reliably calculate a
  Hurst parameter for each mode so we used the single, average Hurst parameter
  from the single mode model for both modes of the two mode model.
 
  For the two mode model, we defined a transition matrix describing the rate at
  which solutes moved between the tail and pore region. We assumed Markovian
  transitions between modes, meaning each transition had no memory of
  previously visited modes. We populated a 2$\times$2 count matrix by
  incrementing the appropriate entry by 1 each time step and then generated a
  transition probability matrix by normalizing the entries in each row of the
  count matrix so that they summed to unity.
  
  \subsubsection*{Simulating Anomalous Diffusion}

  We simulated models with all combinations of the types of dwell and hop length
  distributions described above, summarized in Table~\ref{table:anomalous_models}.
  All models include correlation between hops.

  \begin{table*}[!htb]
	  \centering
	  \begin{tabular}{ccc}
            \hline
	  \hline
	  Dwell Distribution                & ~~~~~~Hop Length Distribution~~~~~~ & Abbreviation \\
	  \hline
      Power Law                         & Gaussian                & sFBM         \\
      Power Law w/ Exponential Cut-off  & Gaussian                & sFBMcut      \\
      Power Law                         & L\'evy Stable           & sFLM         \\
      Power Law w/ Exponential Cut-off  & L\'evy Stable           & sFLMcut      \\
	  \hline
          \hline
	  \end{tabular}
	  \caption{We tested four anomalous diffusion models with various modifications 
	  to the dwell and hop length distributions. We incorporate hop correlation 
	  into all models.}\label{table:anomalous_models}
          
 \end{table*}

  For each solute, we simulated 1,000 anomalous diffusion trajectories of
  length $T_{sim}$ in order to directly compare our model's predictions to MD
  simulations. $T_{sim}$ varied between solutes due to differing solute
  equilibration times. We constructed trajectories by simulating sequences of
  dwell times and correlated hop lengths generated based on parameters randomly
  chosen from our bootstrapped parameter distributions. We propagated each
  trajectory until the total time equaled or exceeded $T_{sim}~ \mu$s, then
  truncated the last data point so that the total time exactly equaled
  $T_{sim}~ \mu$s since valid comparisons are only possible between fixed
  length sFBM simulations. 
  
  We used Equation~\ref{eqn:tamsd} to calculate the time-averaged MSD of the MD
  and AD model trajectories then estimated their uncertainty using statistical
  bootstrapping. For each bootstrap trial, we randomly chose $n$ solute
  trajectories, where $n$ is the number of independent trajectories, with
  replacement, from the ensemble of trajectories and then calculated the MSD of
  the subset. We reported the time-averaged MSD up to a 1000 ns time lag with
  corresponding 1$\sigma$ confidence intervals. 

  When simulating 2 mode models, we determined the state sequence based on
  random draws weighted by the appropriate row of the probability transition
  matrix. We then drew hops and dwells based on the current state of the
  system. Since we calculated the transition probabilities from a finite data
  set, they have an associated uncertainty which we incorporated by re-sampling
  each row from a two dimensional Dirichlet distribution (which is also a beta
  distribution for the 2D case) with concentration parameters defined by the
  count matrix.~\cite{bacallado_bayesian_2009}
  
  We used the Python package \texttt{fbm}~\cite{flynn_exact_2019} to generate exact simulations of FBM
  and our own Python implementation (see Table~\ref{S-table:python_scripts} of
  the Supporting Information) of the algorithm by Stoev and Taqqu to
  simulate FLM.~\cite{stoev_simulation_2004} Note that, to our knowledge, there
  are no known exact simulation algorithms for generating FLM trajectories.
  However, the algorithm we used sufficiently approximates draws from the
  marginal L\'evy stable distribution and reasonably approximates the
  correlation structure on MD simulation timescales. We added an empirical
  correction to enhance the accuracy of the correlation structure (see
  Section~\ref{S-section:flm_correlation} of the Supporting Information for
  validation of the approach).

  \subsection{The Markov State-Dependent Dynamical Model}\label{method:MSMs}  

  A Markov state model (MSM) decomposes a time series into a set of discrete
  states with transitions between states defined by a transition probability
  matrix, $T$. $T$ describes the conditional probability of moving to a
  specific state given the previously observed
  state.~\cite{pande_everything_2010,wehmeyer_introduction_2018}

  In this work, we define a total of 8 discrete states based on the 3 trapping
  mechanisms observed in our previous work. Therefore, there is no need to
  apply any algorithmic approaches to identify and decompose our system into
  discrete states. The states we have chosen include all combinations of
  trapping mechanisms in the pore and out of the pore (see
  Table~\ref{table:states}). They assume that there are no significant kinetic
  effects resulting from solute conformational changes or pore size
  fluctuations. We use the same radial cut-off (0.75 nm) as in the AD approach to
  differentiate the pore and tail region. We define a hydrogen bond to exist
  if the distance between donor, D, and acceptor, A, atoms is less than 3.5
  \AA~and the angle formed by $D-H \cdots A$ is less than
  30$\degree$.~\cite{luzar_effect_1996} We define a sodium ion to be associated
  with an atom if they are within 2.5 \AA~of each other, as determined in our
  previous work.~\cite{coscia_chemically_2019}
  
  \begin{table*}[!htb]
	  \centering
	  \begin{tabular}{ll}
            \hline
            \hline
	  \multicolumn{2}{c}{Markov State-Dependent Dynamical Model State Definitions} \\
	  \hline
	  1. In tails, not trapped                      & 5. In pores, not trapped                     \\
	  2. In tails and hydrogen bonding              & 6. In pores and hydrogen bonding             \\
	  3. In tails and associated with sodium        & 7. In pores and associated with sodium       \\
	  4. In tails, hydrogen bonding and associated  ~~~~& 8. In pores, hydrogen bonding and associated \\
	  \hline
          \hline
	  \end{tabular}
	  \caption{We defined 8 discrete states based on all combinations of previously observed solute
	  trapping mechanisms.}\label{table:states}  
  \end{table*}
  
  We constructed the state transition probability matrix, $T$, based on
  observed solute trajectories. Using methods described in our previous work,
  we determined each solute's radial location and which, if any, trapping
  mechanisms affected it at each time step, then assigned the observation to a
  specific state according to the definitions in
  Table~\ref{table:states}.~\cite{coscia_chemically_2019} Analogous to the mode
  transition matrix in Section~\ref{method:model_sFBM}, and based on the
  current and previous state observation, we incremented the appropriate entry
  of an $n~\times~n$ count matrix by 1, where $n$ is the number of states. We
  verified the Markovianity of state transitions as described in
  Section~\ref{S-section:markov_validation} of the Supporting Information.
  
  Adding to the standard MSM framework, we incorporated the dynamics of the
  solutes within each state as well as the dynamics of state transitions, which
  includes the overall configurational state of the solute and it's
  environments. While MSMs are often used to estimate equilibrium populations
  of various states, adding state-dependent dynamics allows us to simulate
  solute trajectories. Hence why we refer to them as Markov state-dependent
  dynamical models (MSDDMs). 
  
  We recorded the $z$-direction displacement at each time step in order to
  construct individual emission distributions for each state and transition
  between states. This results in 64 distinct emission distributions with some
  far more populated than others. We modeled all of the emission distributions
  as symmetric L\'evy stable distributions in order to maintain flexibility in
  parameterizing the distributions.
  
  We use the Hurst parameter to describe negative time series correlation.
  However, there is not sufficient data to accurately measure a Hurst parameter
  for each type of transition. We avoided this problem by combining all
  distributions associated with state transitions and treating all transitions
  as correlated emissions from a single L\'evy stable distribution. This
  reduces the number of emission distributions from 64 to 9 (1 for each of the
  8 states and 1 for transitions between states).  
  
  We simulated realizations of the MSDDM using the probability transition
  matrix and emission distributions. For each trajectory simulated, we chose an
  initial state randomly from a uniform distribution. We generated a full state
  sequence by randomly drawing subsequent states weighted by the rows of the
  probability transition matrix corresponding to the particle's current state.
  Again, because we are working with a finite data set, we incorporated
  transition probability uncertainties into the rows of the transition matrix
  by resampling them from a Dirichlet distribution. For each same-state
  subsequence of the full state sequence, we simulated an FLM process using the
  Hurst parameter of that state and the parameters of the corresponding
  emission distribution. Independently, we simulated the transition between
  each same-state sequence with an FLM process based on the Hurst parameter of
  transition sequences and the parameters of the single transition emission
  distribution. We used the same FLM simulation procedure described in
  Section~\ref{method:model_sFBM}.

  \subsection{Estimating Solute Flux}\label{method:mfpt}
  
  We determine the rate at which solutes cross macroscopic-length pores based
  on the Hill relation:~\cite{hill_free_1989}
  \begin{equation}
  J = \frac{1}{MFPT}
  \label{eqn:hill_relation}
  \end{equation}
  where $J$ is the single particle solute flux and MFPT refers to the mean
  first passage time. To account for input concentration dependence of the
  flux, assuming that particles are independent, one can multiply
  Equation~\ref{eqn:hill_relation} by the total number of particles to get the
  total flux. In the context of our work, the MFPT describes the average length
  of time it takes a particle to move from the pore entrance to the pore exit. 
  
  We generated particle trajectories, parameterized with the above models, in
  order to construct a distribution of first passage times across a membrane
  pore of length $L$. For each pore length, we simulated 10,000 realizations
  of an AD approach model
  all released at the pore entrance ($z=0$).  In the case of uncorrelated hops,
  one can continuously draw from the hop length distribution until $z \geq L$
  (or $-L$ for the sake of computational efficiency). The length of time
  between the last time the particle crossed $z=0$ and the end of the
  trajectory gives a single passage time. When particle hops are correlated, as
  they are in all cases of this work, we cannot continuously construct the
  particle trajectories. Rather, we must generate trajectories of length $n$
  and measure the length of the sub-trajectory which traverses from $0$ to $L$
  without becoming negative.
  
  We calculated the expected value of analytical fits to the passage time
  distributions in order to determine the MFPT for a given solute and pore
  length. One should not use the mean of the empirical passage time
  distribution because it is highly likely that the true MFPT will be
  underestimated unless 100\% of a very large number of trajectories reach $L$.
  If a trajectory does not reach $L$ within $n$ steps, it is possible that a
  very long passage time has been excluded from the distribution.
  
  To derive an analytical equation describing the passage time distributions,
  one can frame the problem as a pulse of particles instantaneously released at
  the pore inlet ($z=0$) which moves at a constant velocity, $v$, and spreads
  out as it approaches $L$. This spreading is parameterized by an effective
  diffusivity parameter, $D$. This approach gives results equivalent to if we
  had released each particle individually and then analyzed the positions of
  the ensemble of trajectories as a function of time. The analytical expression
  describing the distribution of first passage times
  is:~\cite{cussler_diffusion:_2009}
  \begin{equation}
  \begin{split}
  P(t) =~~~~~~~~~~~~~~~~~~~~~~~~~~~~~~~~~~~~~~~~~~~~~~~~~~~~~~~\\
  -\frac{1}{\sqrt{\pi}}e^{-(L - vt)^2 / (4Dt)}\bigg(-\frac{D(L - vt)}{4(Dt)^{3/2}} - \frac{v}{2\sqrt{Dt}}\bigg)
  \end{split}
  \label{eqn:passage_times}
  \end{equation} 
  where the only free parameters for fitting are $v$ and $D$. A derivation of
  Equation~\ref{eqn:passage_times} is given in
  Section~\ref{S-section:fpt_derivation} of the Supporting Information. We
  calculated the expected value of Equation~\ref{eqn:passage_times} in order to
  get the MFPT.

  We used the ratio of solute fluxes in order to determine membrane
  selectivity, $S_{ij}$, towards solutes. Selectivity of solute $i$ versus $j$
  is defined in terms of the ratio of solute permeabilities,
  $P$:~\cite{guo_pervaporation_2004}
  \begin{equation}
  S_{ij} = \frac{P_i}{P_j}
  \end{equation}
  We can relate this to solute flux using Kedem and Katchalsky's equations for
  solvent volumetric flux, $J_v$, and solute flux,
  $J_s$:~\cite{kedem_permeability_1963,al-zoubi_rejection_2007}
  \begin{equation}
  J_v = L_p(\Delta P - \sigma\Delta \pi)
  \end{equation} 
  \begin{equation}
  J_s = P_s \Delta C + (1 - \sigma)CJ_v
  \label{eqn:solute_flux}
  \end{equation}
  where $L_p$ is the pure water permeability, $\Delta P$ and $\Delta \pi$ are
  the trans-membrane hydraulic and osmotic pressure differences, $\sigma$ is
  the reflection coefficient, $P_s$ is the solute permeability, $\Delta C$ is
  the trans-membrane solute concentration difference and $C$ is the mean solute
  concentration. Since our simulations do not include convective solute flux,
  we eliminate the second term of Equation~\ref{eqn:solute_flux} which allows
  us to derive a simple expression for selectivity in terms of solute flux:
  \begin{equation}
  S_{ij} = \frac{J_i / \Delta C_i}{J_j / \Delta C_j}
  \label{eqn:selectivity}
  \end{equation}
  
  \section{Results and Discussion}
  
  \subsection{Anomalous Diffusion Modeling}\label{section:sFBM}
  
  \subsubsection{Parameterizing Subordinated Fractional Brownian Motion}\label{section:AD_parameterization}

\begin{figure*}
  \centering
  \begin{subfigure}{0.45\textwidth}
  \includegraphics[width=\textwidth]{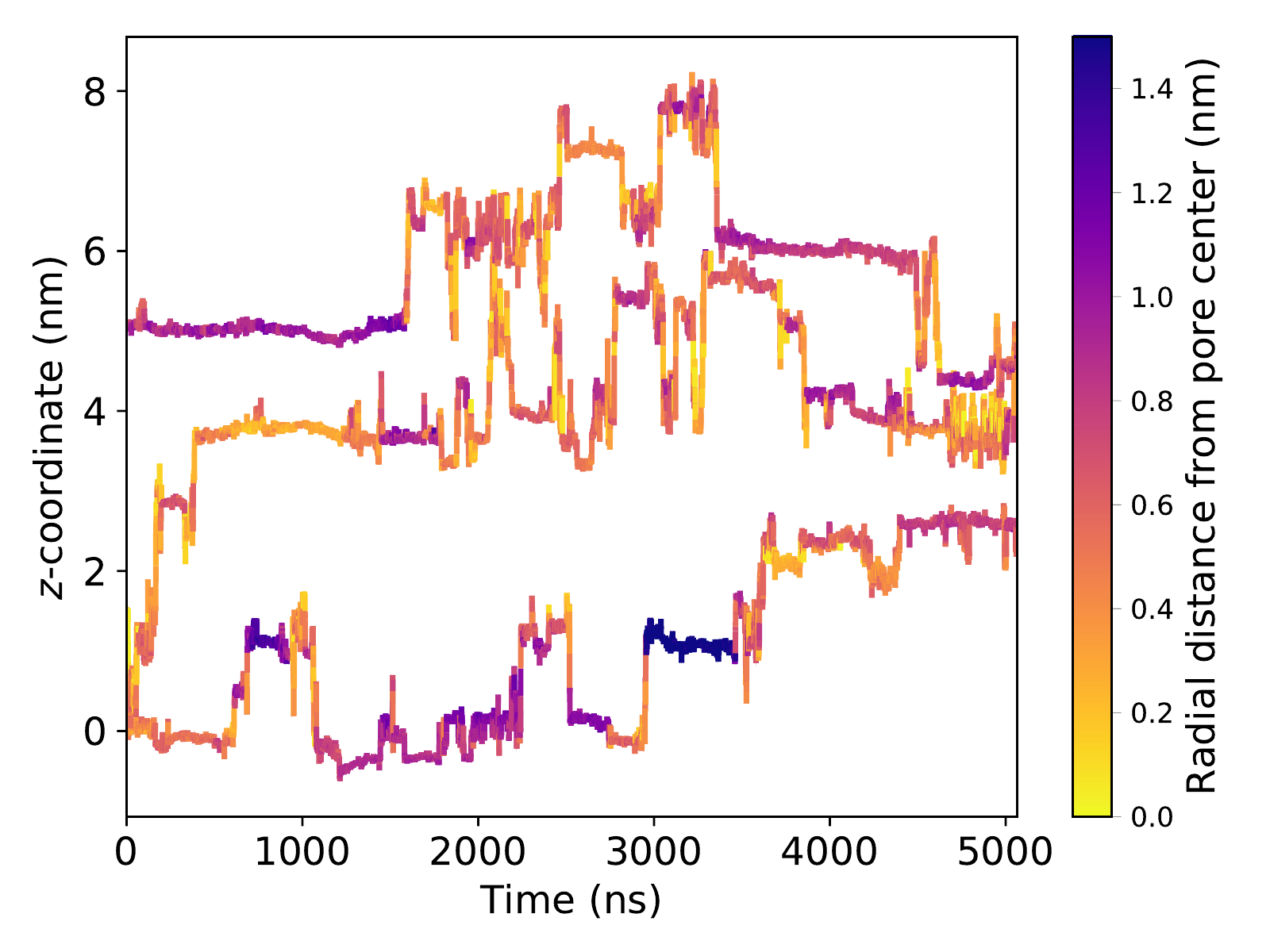}
  \caption{urea}\label{fig:URE_trajectories}
  \end{subfigure}
  \begin{subfigure}{0.45\textwidth}
  \includegraphics[width=\textwidth]{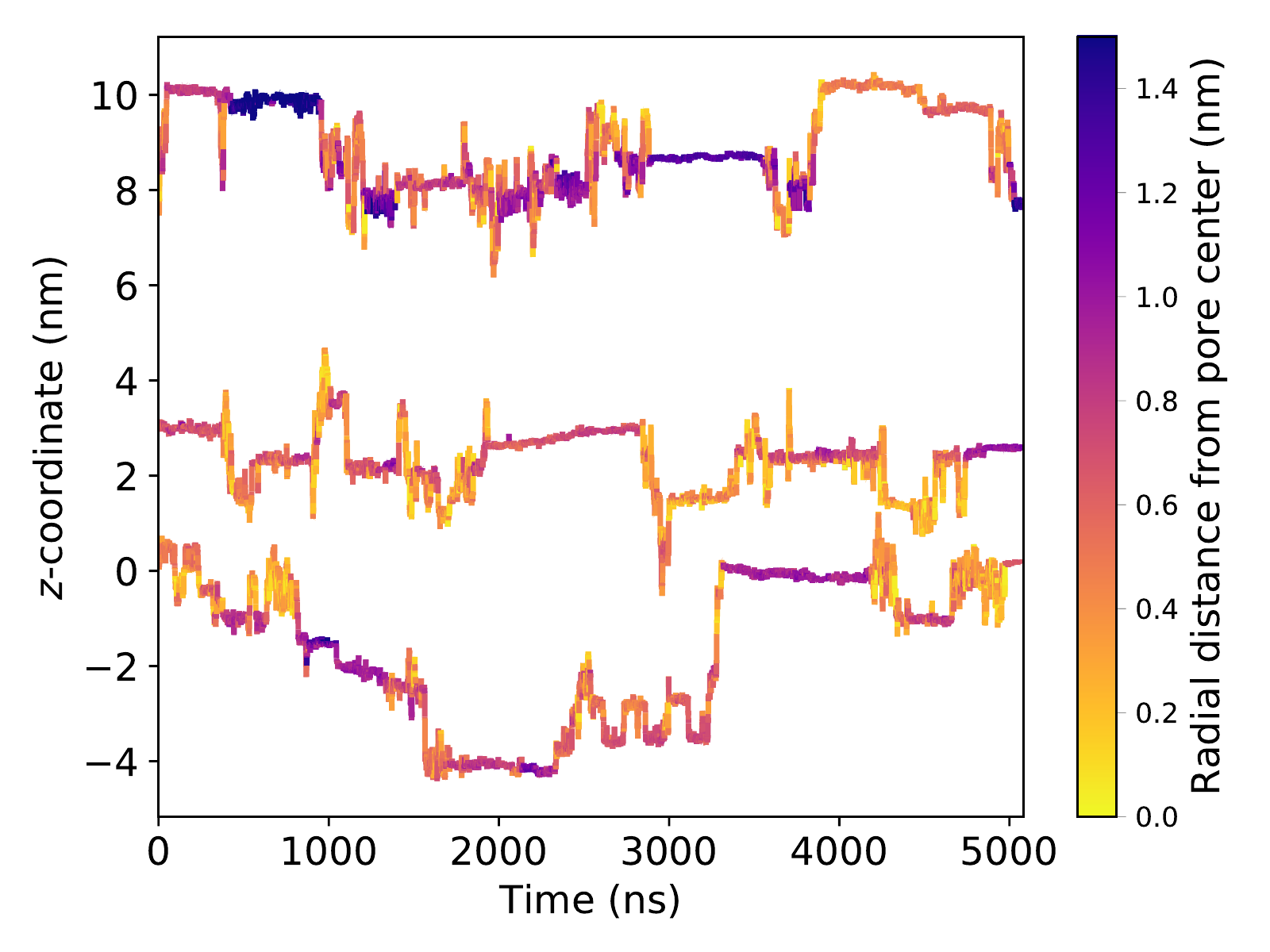}
  \caption{ethylene glycol}\label{fig:GCL_trajectories}
  \end{subfigure}
  \begin{subfigure}{0.45\textwidth}
  \includegraphics[width=\textwidth]{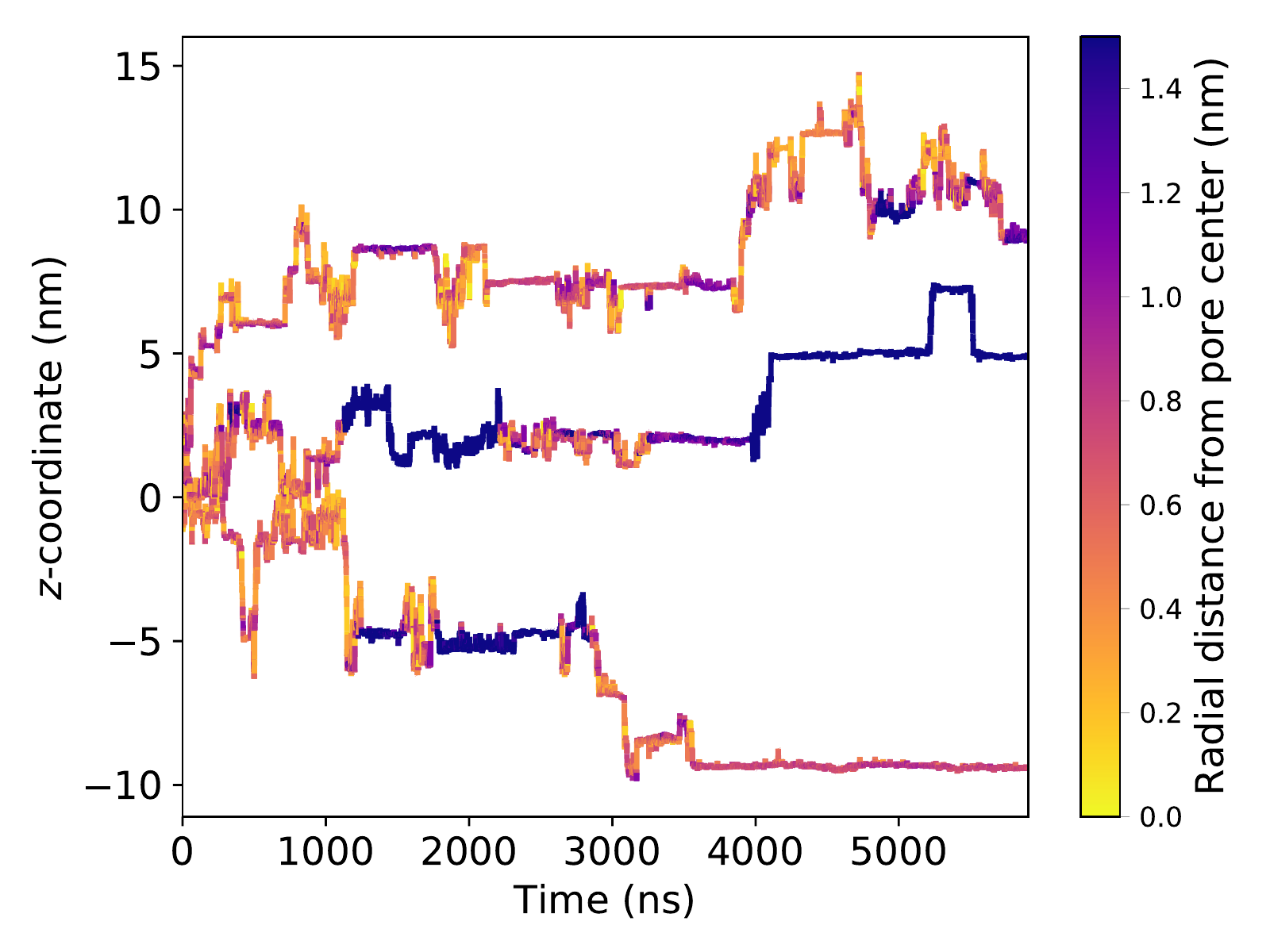}
  \caption{methanol}\label{fig:MET_trajectories}
  \end{subfigure}
  \begin{subfigure}{0.45\textwidth}
  \includegraphics[width=\textwidth]{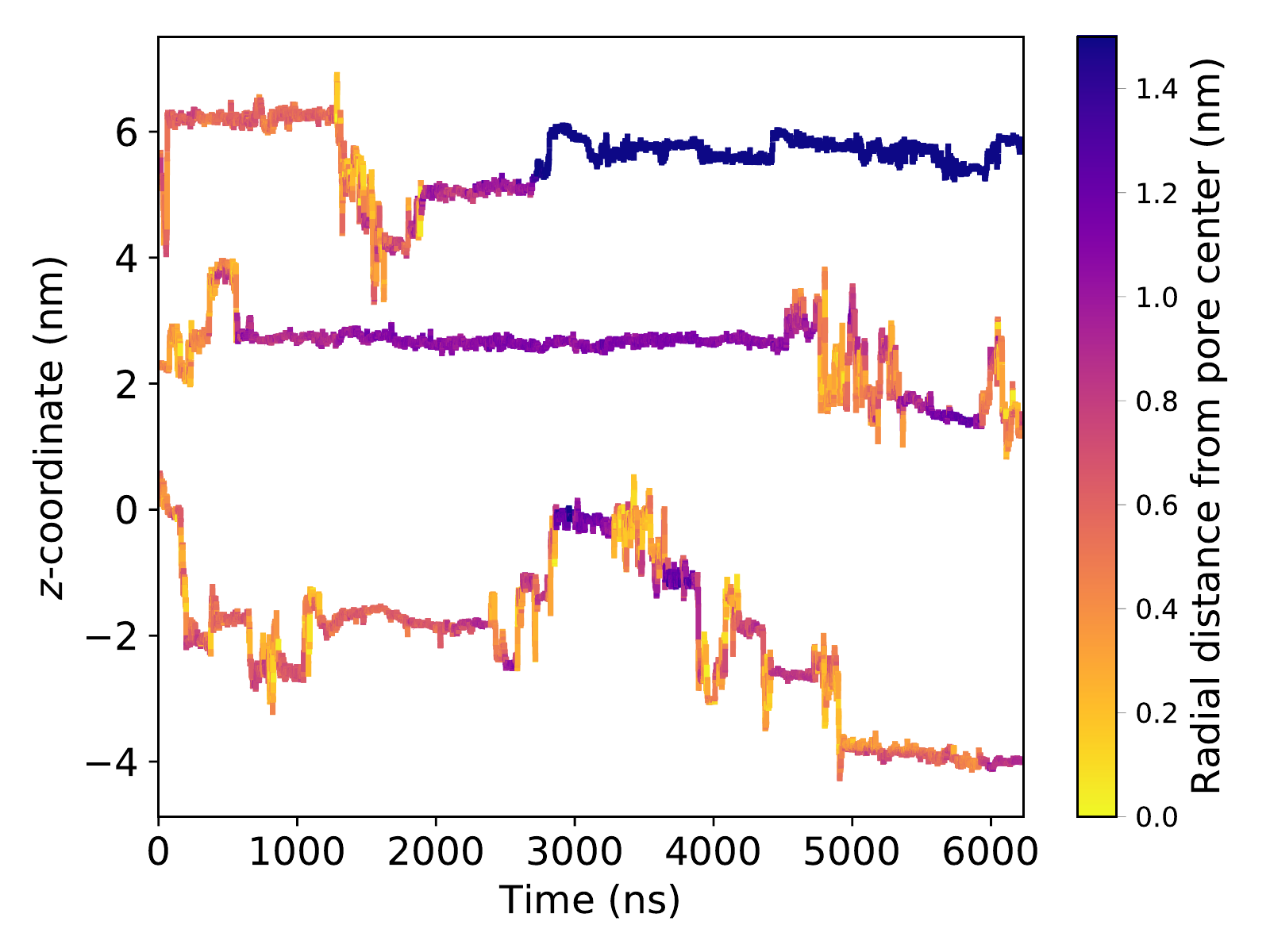}
  \caption{acetic acid}\label{fig:ACH_trajectories}
  \end{subfigure}
  \caption{Three representative trajectories generated from each solute exhibit hops
	  between periods of entrapment, characteristic of a CTRW. Solute
	  dynamics show radial dependence, represented by the color at each
	  time point. The longest periods of entrapment typically occur when
	  solutes drift far from the pore center and into the tails.
  }\label{fig:solute_trajectories}
  \end{figure*}

  We find the data suggests that solute motion in this system can be
  well-modeled by subordinated fractional Brownian motion. In
  Figure~\ref{fig:solute_trajectories}, we plotted representative trajectories
  for each solute. They are characterized by intermittent hops between periods
  of entrapment. The near-Gaussian distribution of jump lengths and power law
  distribution of dwell times are both characteristic of CTRWs
  (See Figure~\ref{S-fig:anticorrelated_hops} of the Supporting Information). 
  Anti-correlation between hops suggests a fractional diffusion process is 
  subordinate to the CTRW. Of the different models, a process subordinated by 
  FBM or FLM is best supported by the data because the analytical correlation 
  structures of hop lengths are close to those observed in our simulations 
  (See Figure~\ref{S-fig:anticorrelated_hops} of the Supporting Information). Fractional 
  motion is common in crowded viscoelastic environments where motion is highly
  influenced by the movement of surrounding components, such as monomer tails 
  in our case.~\cite{ernst_fractional_2012}
  
  We modeled the distributions of hop lengths in two ways. First, we assumed
  the distribution to be Gaussian since it is possible to exactly simulate
  realizations of fractional Brownian motion. Second, we fit the distributions
  to L\'evy stable distributions since it is more general than the Gaussian
  distribution. We plotted the MLE fits of both on top of each solute's hop length
  distribution in Figure~\ref{S-fig:anticorrelated_hops} of the
  Supporting Information. The L\'evy distribution does a better job capturing
  the somewhat heavy tails and high density near 0 of the hop length distributions. 
  However, since there are no known exact simulation techniques for generating 
  realizations of fractional L\'evy motion, this more general fit may not be 
  worthwhile. In fact, we typically observe very little difference between model
  predictions parameterized with FLM versus FBM.

  We also modeled the distribution of dwell times in two ways. First, we
  assumed pure power law behavior since it is consistent with most theoretical
  descriptions of CTRWs. The data fits well to this model at short dwell times
  but the density of long dwell times is over-estimated. In our second
  approach, we truncate the power law distribution with an exponential cut-off,
  lowering the probability of extremely long dwell times. 
  On long time scales, the density of the truncated power law drops below that
  of the pure power law and tends towards 0. 
  
\begin{figure*}
  \centering
  \begin{subfigure}{0.325\textwidth}
  \includegraphics[width=\textwidth]{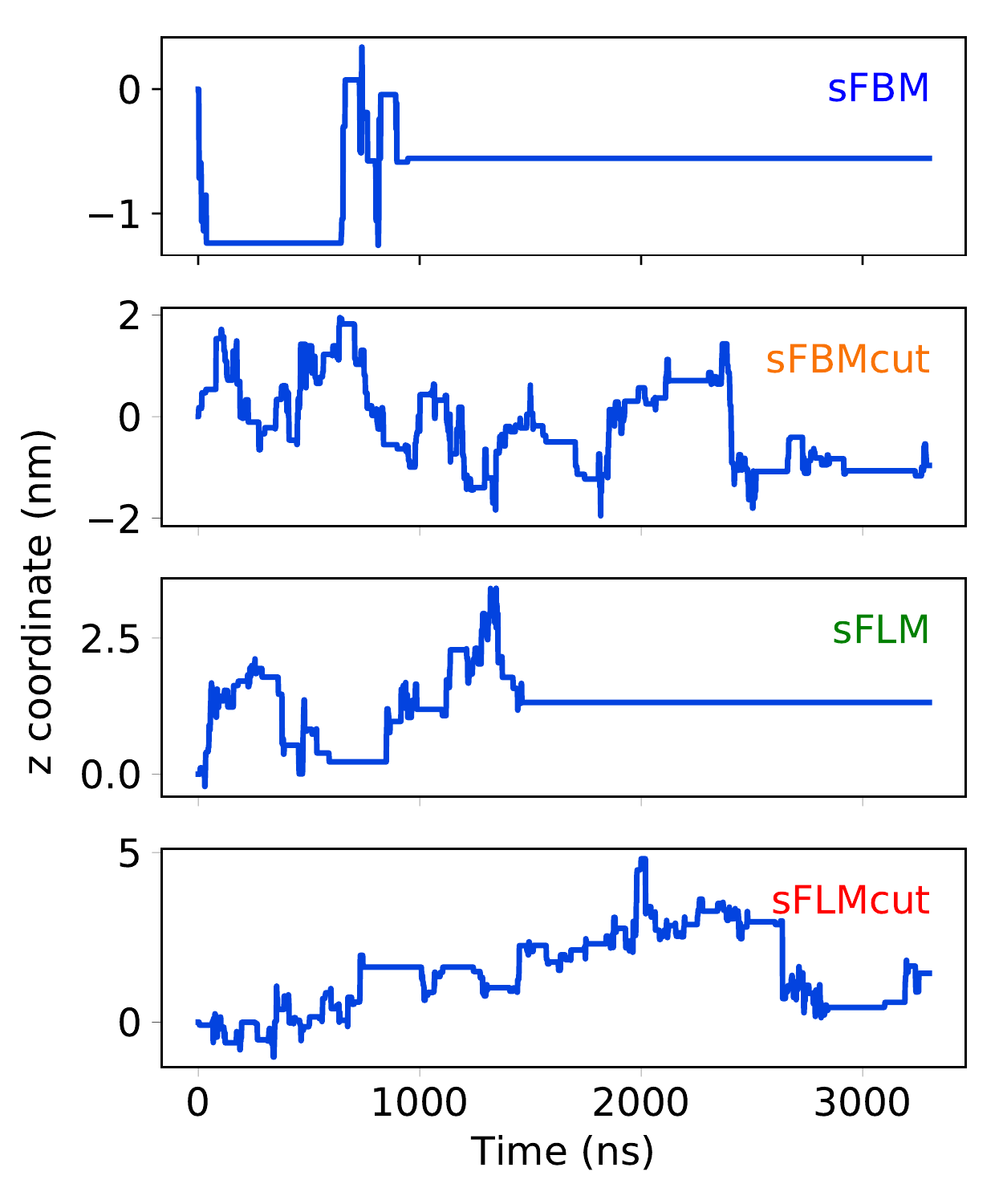}
  \caption{}\label{fig:ad_realizations_1mode}
  \end{subfigure}
  \begin{subfigure}{0.325\textwidth}
  \includegraphics[width=\textwidth]{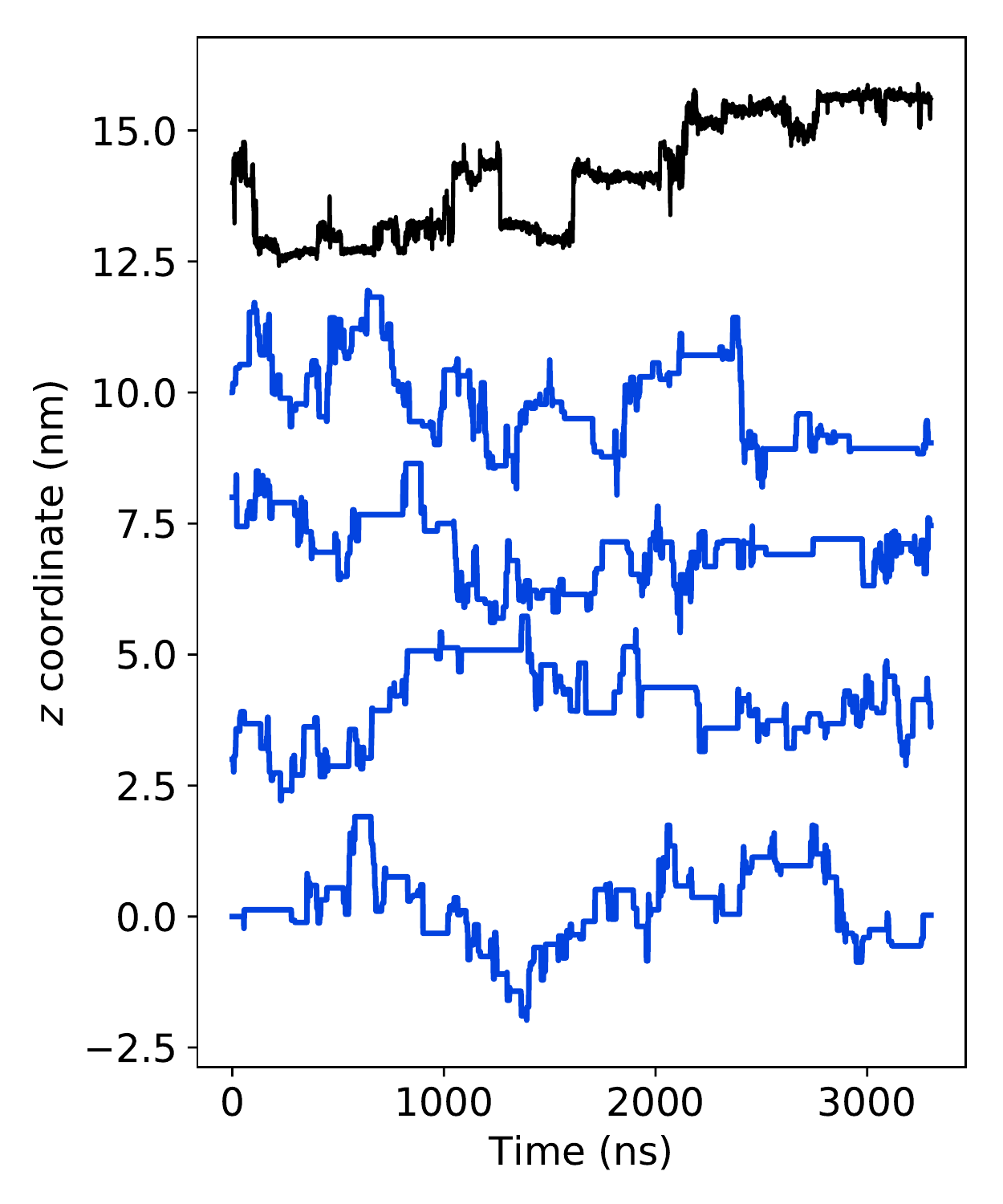}
  \caption{sFBMcut}\label{fig:ad_realizations_sFBMcut_stacked}
  \end{subfigure}
  \begin{subfigure}{0.325\textwidth}
  \includegraphics[width=\textwidth]{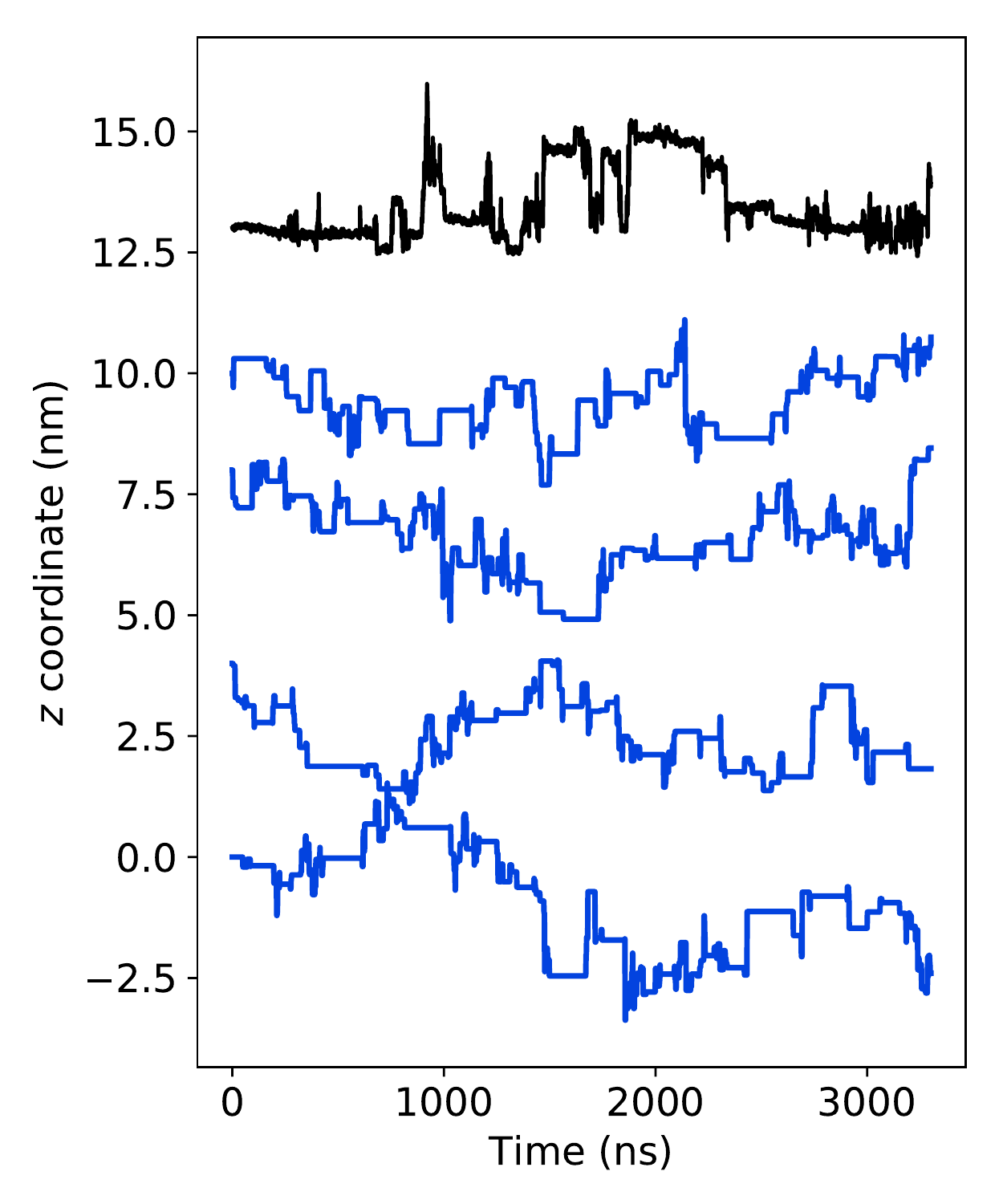}
  \caption{sFLMcut}\label{fig:ad_realizations_sFLMcut_stacked}
  \end{subfigure}
  \caption{(a) Simulated urea trajectories generated by each of the four
	  variations of the one mode AD model display qualitatively similar
	  hopping and trapping behavior to that shown in
	  Figure~\ref{fig:solute_trajectories}. Dwell times are exaggerated in
	  the sFBM and sFLM models because the power law dwell time
	  distributions are not truncated and have infinite variance. In (b)
	  and (c) we compare additional trajectories simulated with the sFBMcut
	  and sFLMcut models (blue) to MD solute trajectories (black).
	  Trajectories are vertically offset for visual clarity.
  }\label{fig:ad_eyetest}
  \end{figure*}

  Several of the choices of AD models and parameters we have described yield
  qualitatively similar trajectories to those seen in our MD simulations.  In
  Figure~\ref{fig:ad_eyetest}, we plot representative sample trajectories for
  each combination of the dwell time and hop distribution, labeled according to
  Table~\ref{table:anomalous_models}. The sFBMcut and sFLMcut in particular
  resemble the trajectories in Figure~\ref{fig:solute_trajectories}.  When we
  do not truncate the dwell time distribution, the trajectories tend to
  incorporate very long dwell times as shown by the sFBM and sFLM models.
  Predictions made with these models consistently under-predict the MD MSDs
  (see Figure~\ref{S-fig:anomalous_msds_bothmode} of the Supporting Information).
  Therefore, we will not include the pure sFBM or sFLM models in the
  remainder of our analysis.   
  
  \subsubsection{Model predictions}\label{section:AD_all_data}
  
\begin{figure*}
  \centering
  \begin{subfigure}{0.45\textwidth}
  \includegraphics[width=\textwidth]{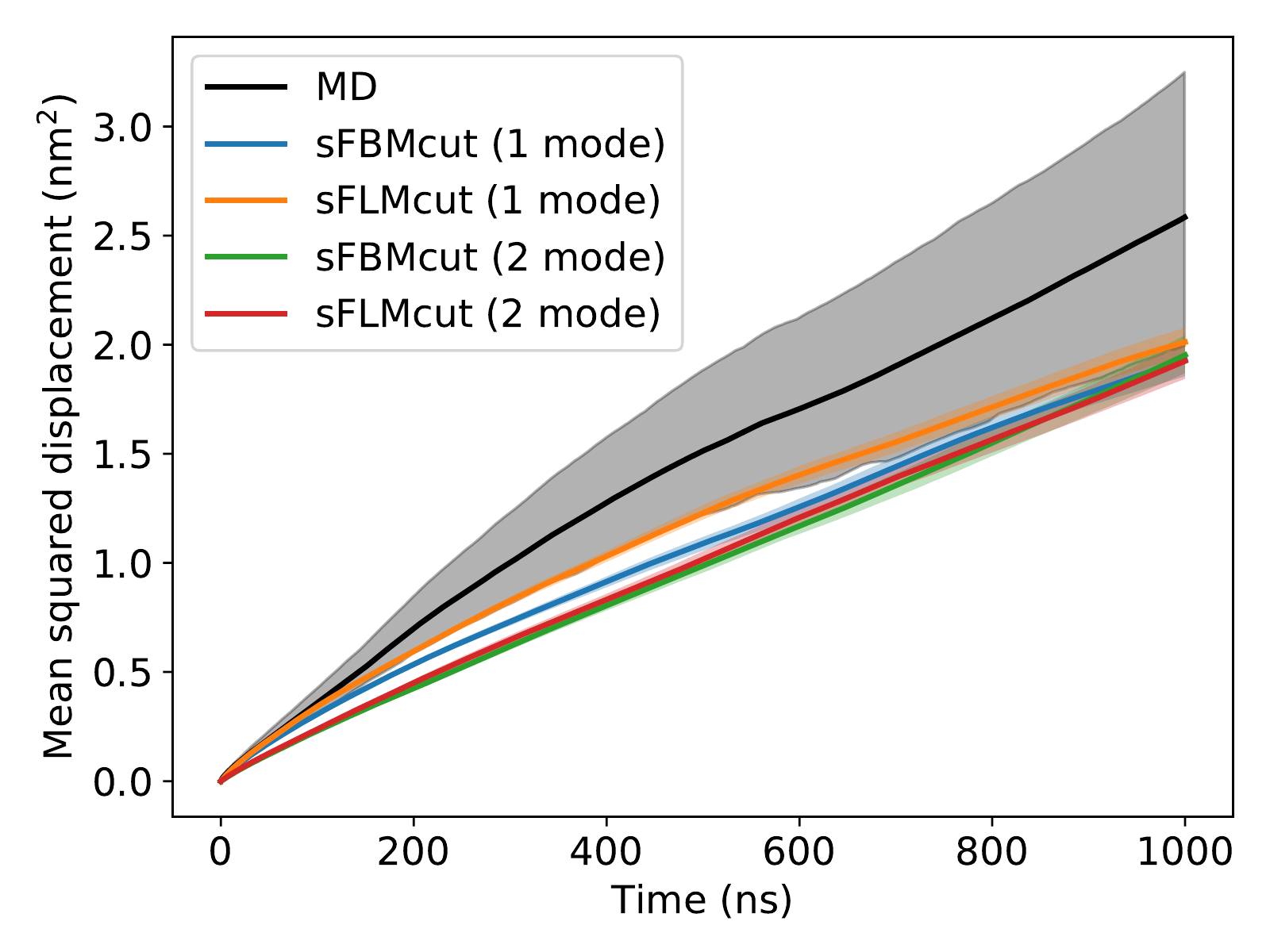}
  \caption{Urea}\label{fig:bothmode_msd_comparison_URE}
  \end{subfigure}
  \begin{subfigure}{0.45\textwidth}
  \includegraphics[width=\textwidth]{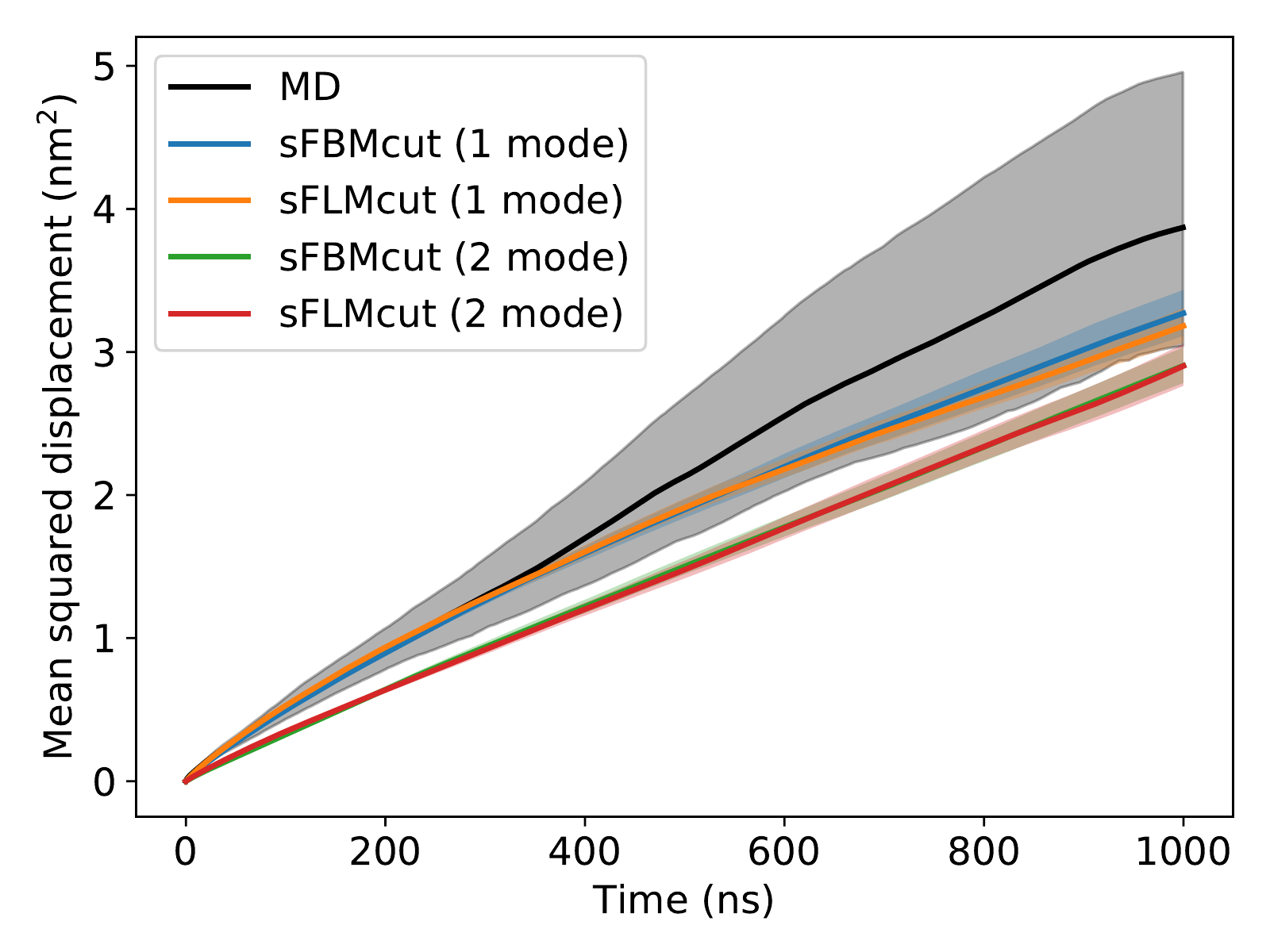}
  \caption{Ethylene Glycol}\label{fig:bothmode_msd_comparison_GCL}
  \end{subfigure}
  \begin{subfigure}{0.45\textwidth}
  \includegraphics[width=\textwidth]{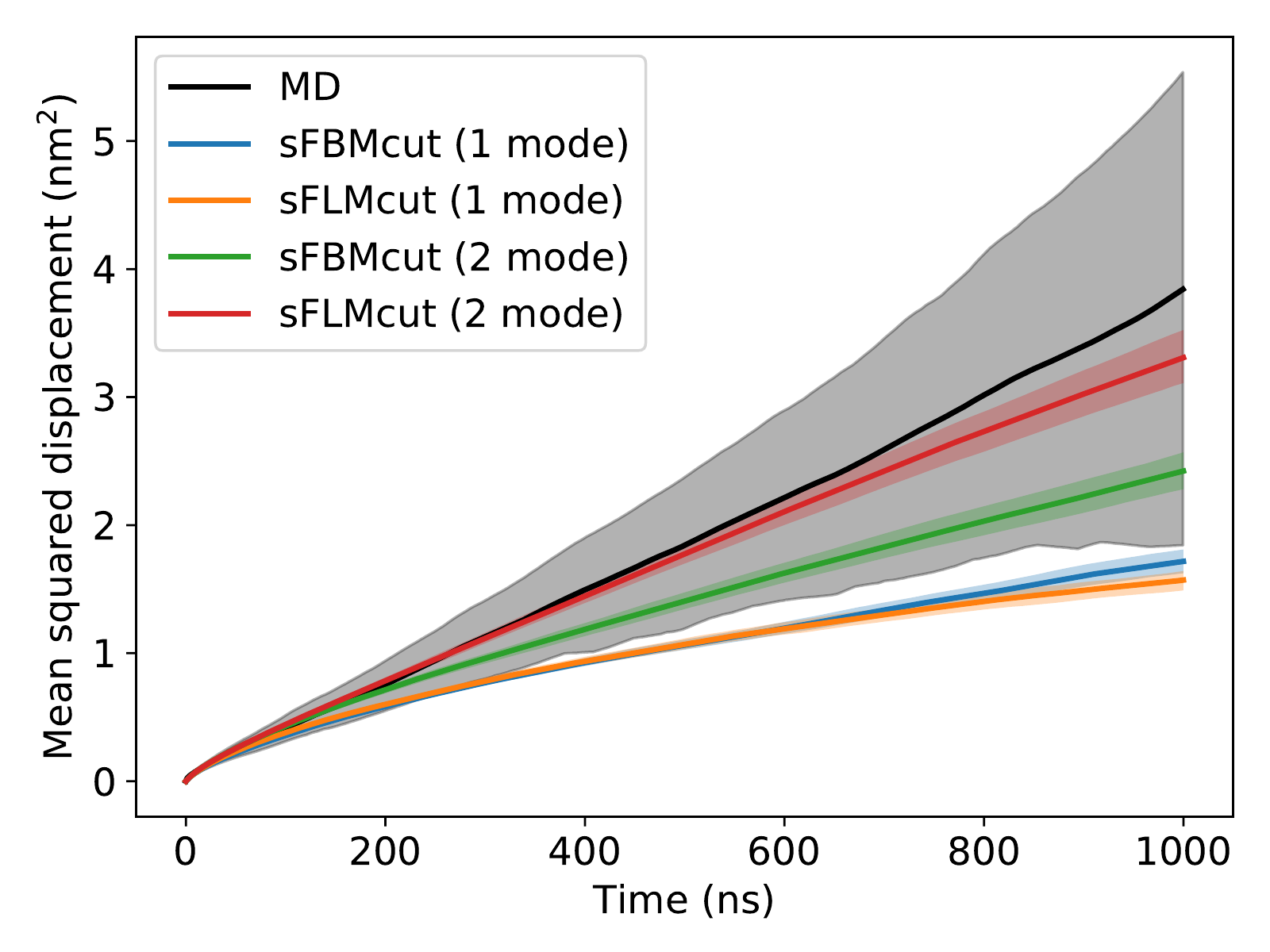}
  \caption{Methanol}\label{fig:bothmode_msd_comparison_MET}
  \end{subfigure}
  \begin{subfigure}{0.45\textwidth}
  \includegraphics[width=\textwidth]{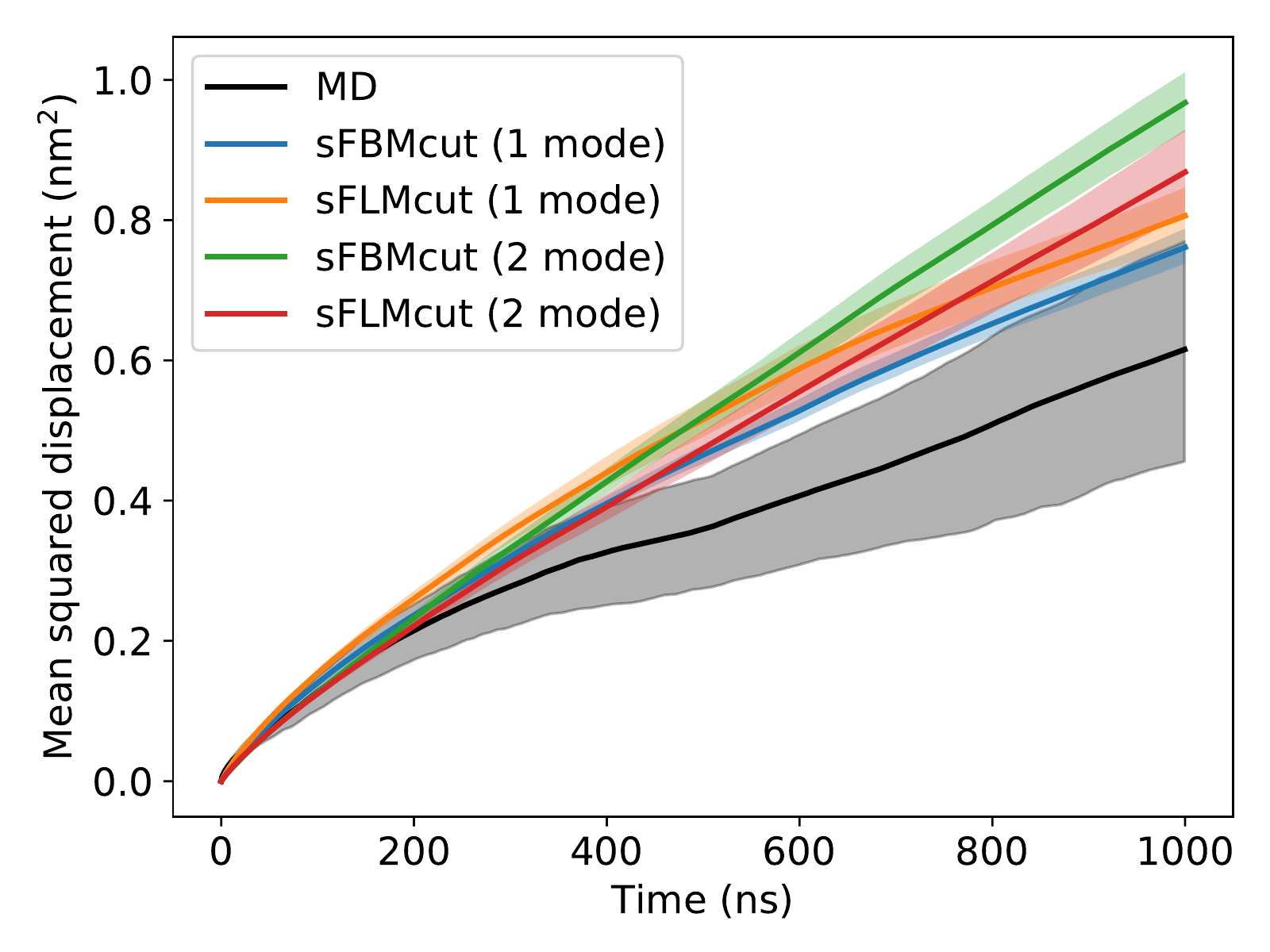}
  \caption{Acetic Acid}\label{fig:bothmode_msd_comparison_ACH}
  \end{subfigure}
  \caption{In most cases, MSDs generated from realizations of both the one and
	  two mode AD models lie within or near the 1$\sigma$ confidence
	  intervals of MD-generated data. Drawing hops from a truncated L\'evy
	  stable distribution (sFLMcut) yields MSDs similar to when hops are
	  drawn from Gaussian distributions (sFBMcut). In most cases, the one
	  mode simulated MSDs under-predicted the mean at long timescales
	  partially because they show pronounced curvature which the MD MSDs
	  lack. The two mode predictions show less curvature than the one mode
	  MSDs because the hop correlation structure is broken every time a
	  transition between tails occurs.
  }\label{fig:anomalous_msds}
  \end{figure*}

\begin{figure*}
  \centering
  \begin{subfigure}{0.325\textwidth}
  \includegraphics[width=\textwidth]{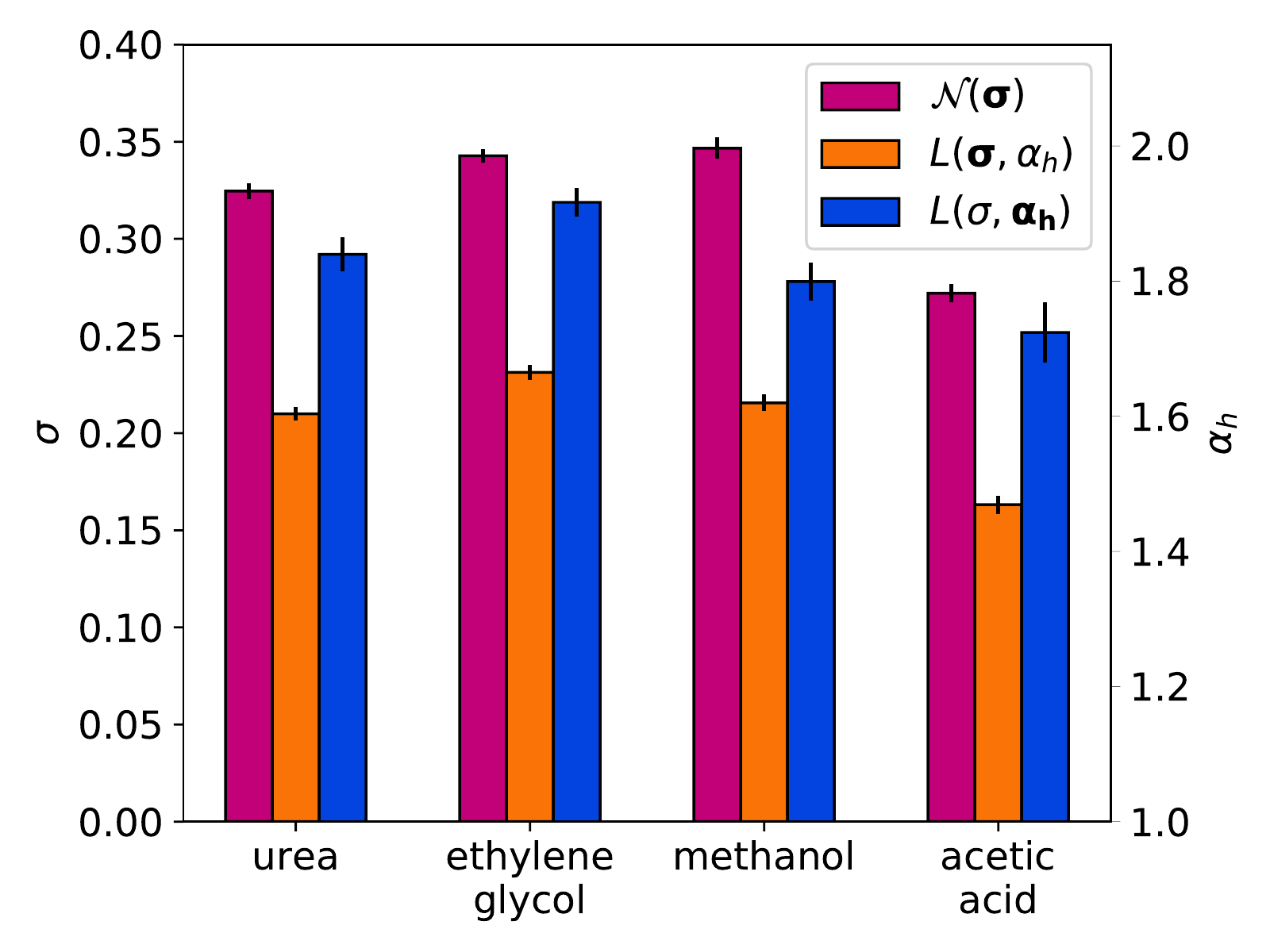}
  \caption{}\label{fig:1mode_AD_hops}
  \end{subfigure}
  \begin{subfigure}{0.325\textwidth}
  \includegraphics[width=\textwidth]{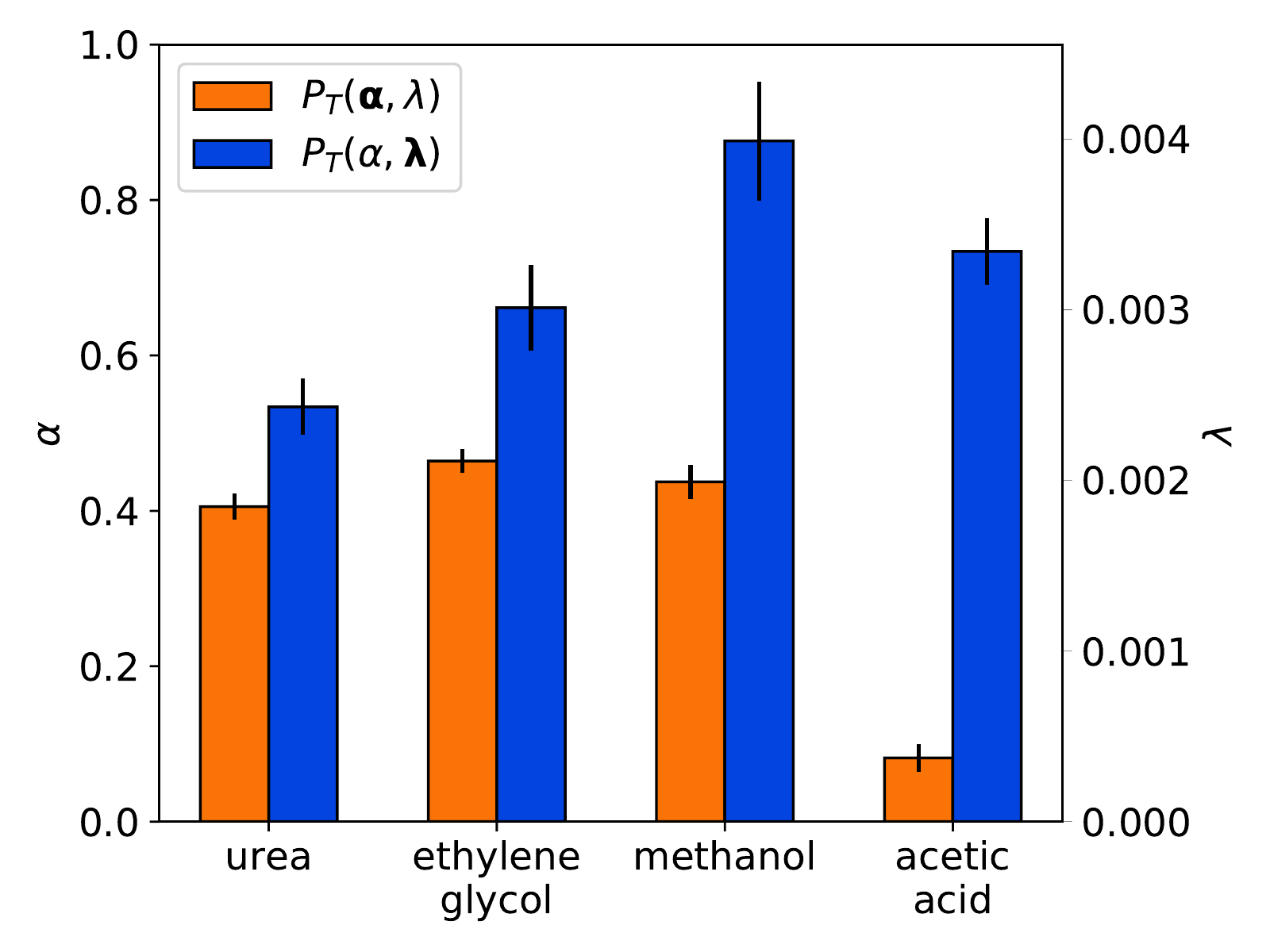}
  \caption{}\label{fig:1mode_AD_dwells}
  \end{subfigure}
  \begin{subfigure}{0.325\textwidth}
  \includegraphics[width=\textwidth]{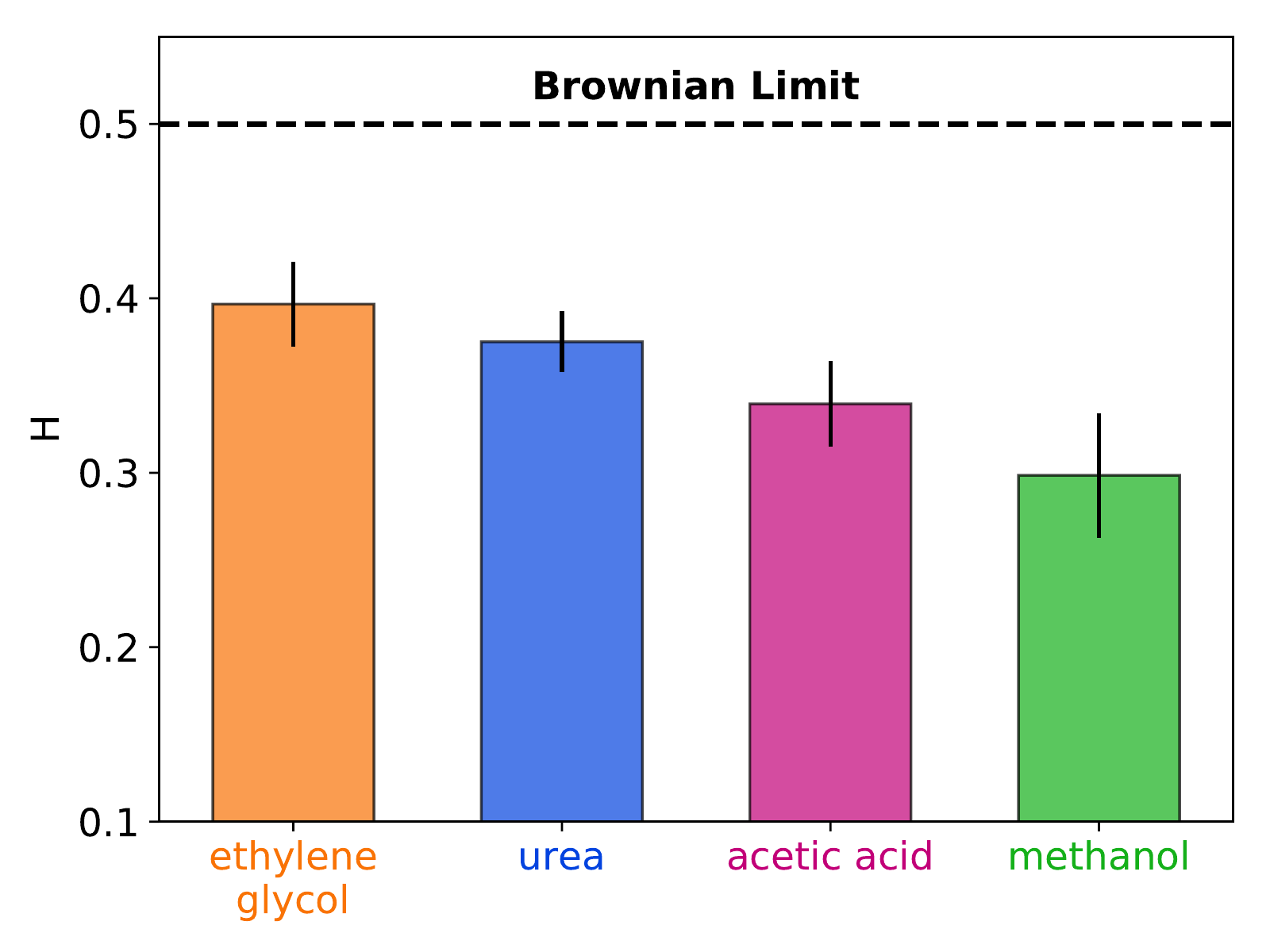}
  \caption{}\label{fig:hurst_barchart}
  \end{subfigure}
  \caption{The parameters of the one mode model reveal differences in dynamics
	  between solutes. (a) We parameterized Gaussian,
	  $\mathcal{N}(\sigma)$, and L\'evy stable, $L(\sigma, \alpha_h)$,
	  distributions to describe solute hop lengths. We assume the mean
	  ($\mu$) to be zero for these distributions and no
	  skewness ($\beta = 0$) in the L\'evy stable distributions. High
	  values of $\sigma$ and lower values of $\alpha_h$ result in larger
	  hops. (b) We parameterized a pure power law, $P(\alpha)$, and a
	  truncated power law, $P_T(\alpha, \lambda)$, distribution to describe
	  solute dwell times. Lower values of $\alpha$ lead to heavier power
	  law tails and higher values of $\lambda$ truncate the distribution at
	  lower dwell times. (c) Finally, we parameterized the hop
	  autocorrelation function, $\gamma(H)$, to describe the degree of
	  correlation between hops. Simulations with higher values of $H$ display
	  behavior closer to the Brownian limit.}\label{fig:1mode_parameters}
  \end{figure*}

\begin{figure*}
  \centering
  \begin{subfigure}{0.325\textwidth}
  \includegraphics[width=\textwidth]{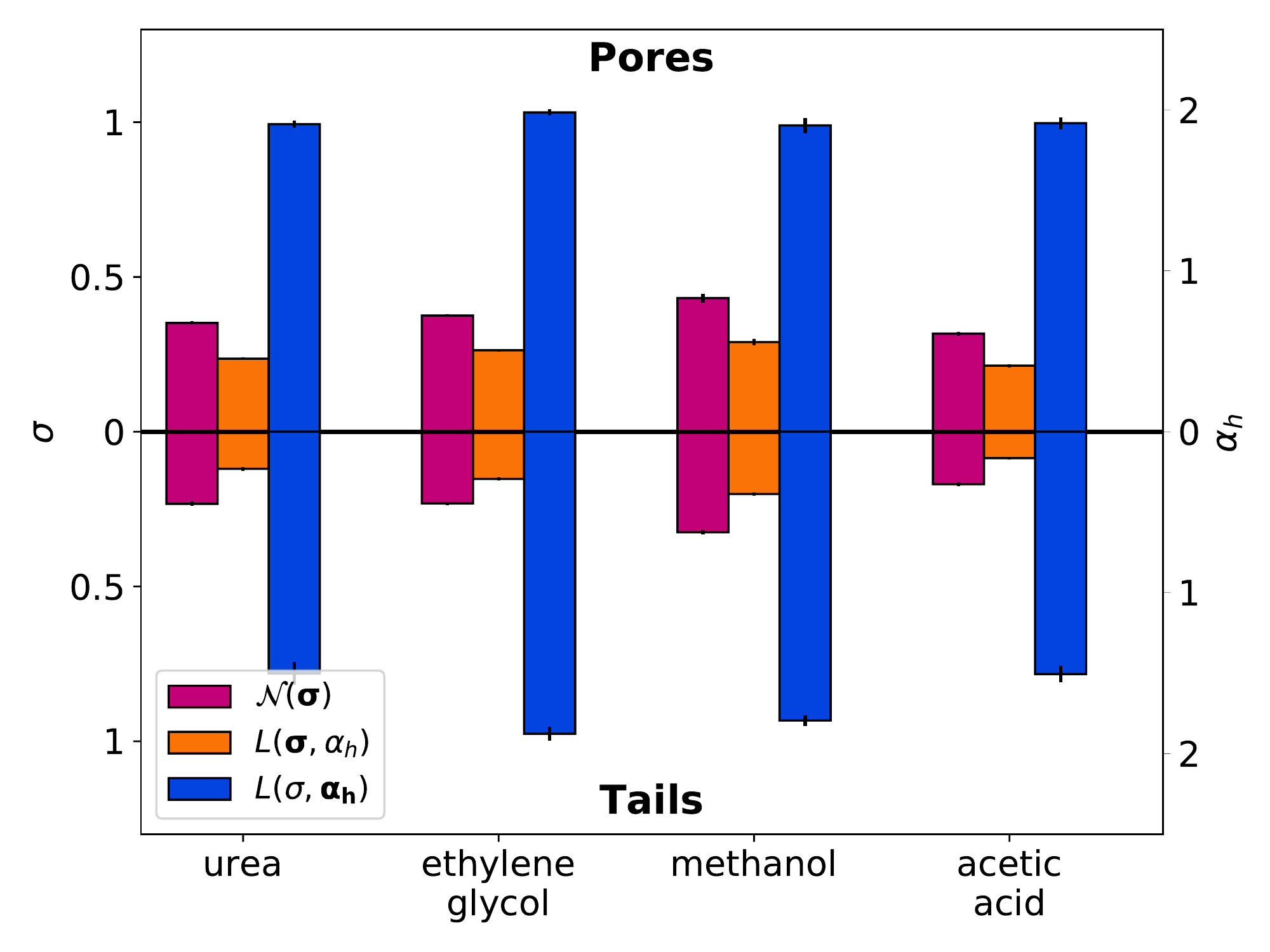}
  \caption{}\label{fig:2mode_AD_hops}
  \end{subfigure}
  \begin{subfigure}{0.325\textwidth}
  \includegraphics[width=\textwidth]{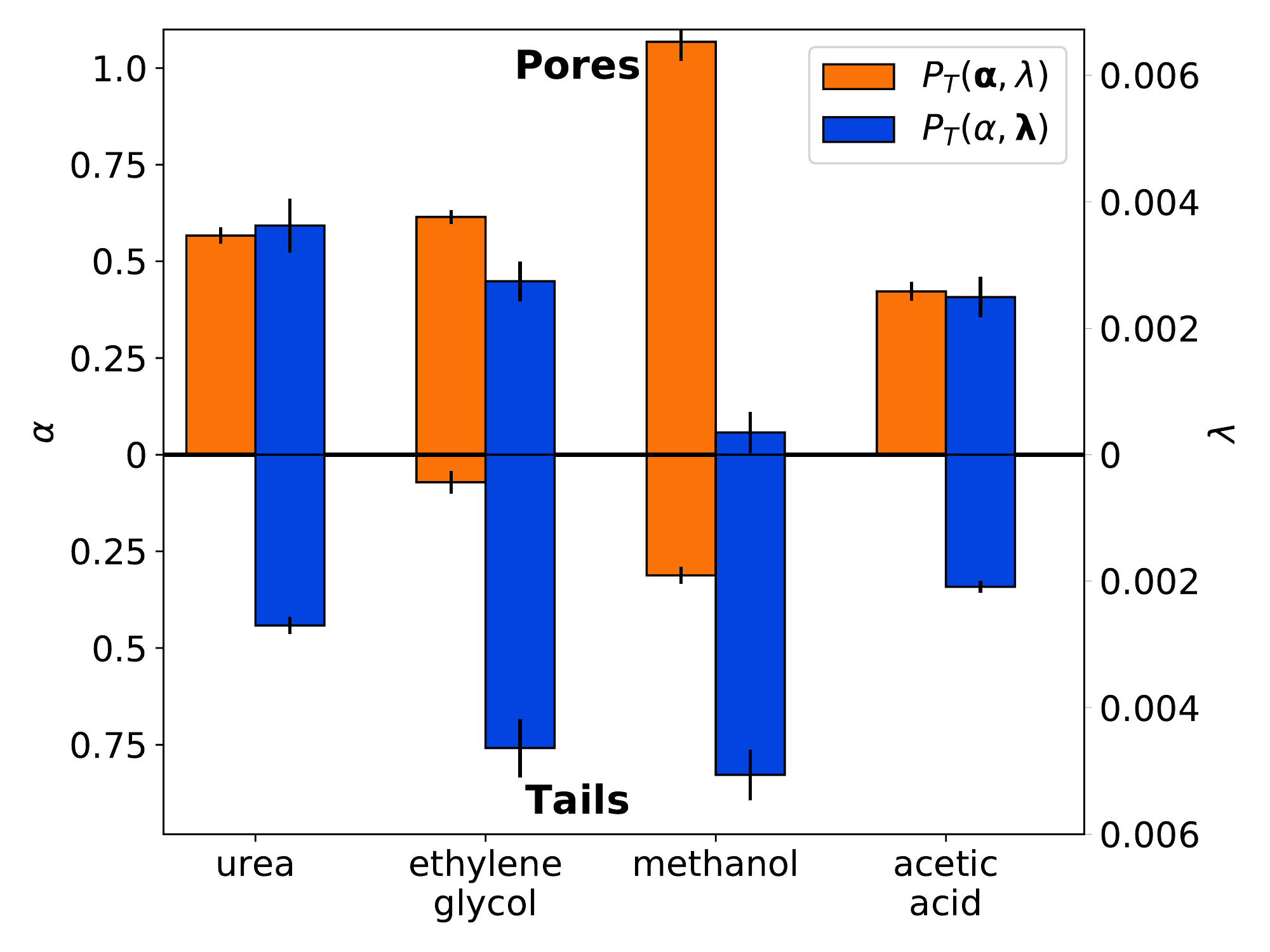}
  \caption{}\label{fig:2mode_AD_dwells}
  \end{subfigure}
  \begin{subfigure}{0.325\textwidth}
  \includegraphics[width=\textwidth]{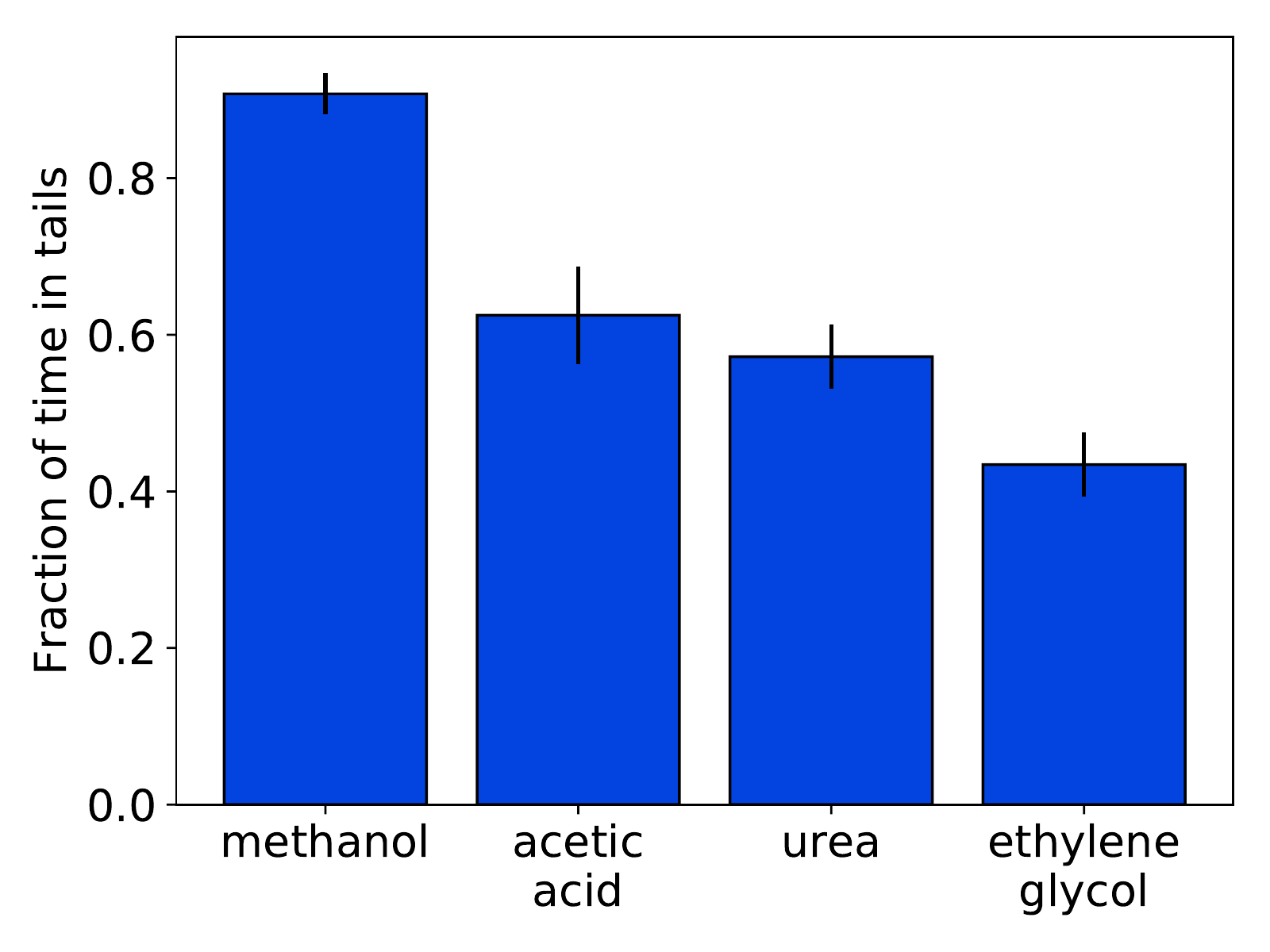}
  \caption{}\label{fig:AD_mode_occupation}
  \end{subfigure}
  \caption{The two mode model parameterizes solute behavior in the pore and tails
	  separately. We consider solutes to be within the pore region if they
	  are 0.75 nm from a given pore center, otherwise they are in the
	  tails. (a) Generally, movement is much more restricted in the tail
	  region, parameterized by lower $\sigma$ values (smaller hops) for the
	  Gaussian and L\'evy stable distributions. Values of $\alpha_h$ are
	  significantly lower for urea and acetic acid meaning there is a
	  larger probability that they will take large hops. (b) Dwell times
	  are longer in the tails. Lower values of $\alpha$ correspond to power
	  laws with heavier tails and thus higher probabilities of long dwell
	  times. There is no easily discernible trend in $\lambda$ of the
	  truncated power law distribution. Note that we used the same Hurst
	  parameter for both modes (shown in Figure~\ref{fig:hurst_barchart})
	  due to a low number of sufficiently long sequences of hops in each
	  mode. (c) Solutes spend various amounts of time in the tail and pore
	  region dependent on their size, shape and chemical functionality.
	  Methanol's small size favors occupation of the much larger accessible
	  volume in the tails. Urea and acetic acid are fairly stable in both
	  regions since they are small and hydrophilic. Ethylene glycol has a
	  slight preference for the pores likely because it is a larger
	  molecule with two hydrophilic hydroxyl groups.
  }\label{fig:2mode_parameters}
  \end{figure*}

  We obtain reasonable predictions of the MD simulated MSDs when we
  parameterize AD models with all available data after the perceived
  equilibration time.   
  We observe that in some cases, solute trajectories extracted from our MD
  simulations may display non-stationary behavior. Our brief analysis 
  in Section~\ref{S-section:msd_comparison} of the Supporting Information 
  suggests that we may be operating on the border of the minimum amount of
  data required to accurately parameterize AD approach models.
  The MSD curves generated from both the one and two mode models are overlaid
  with the MD simulated MSDs for comparison in Figure~\ref{fig:anomalous_msds}.
  The associated parameters for the one and two mode models are presented in
  Figures~\ref{fig:1mode_parameters} and~\ref{fig:2mode_parameters}.
  
  The one and two mode AD models do a fairly good job of predicting the
  magnitude of the MD MSD curves examined up to a 1000 ns time lag. In most
  cases, both the Brownian (sFBMcut) and L\'evy (sFLMcut) versions of both the
  one and two mode AD models give very similar results, 
  because  
  the hop length distributions are nearly Gaussian even when fit to a more general
  L\'evy distribution ($\alpha_h \sim 2$). 
  
  The deviation between the one mode MSD predictions and MD are primarily due
  to differences in their curvature at long time lags. The shape and magnitude
  of the predicted curves appears accurate relative to MD at short time lags.
  However, the modeled trajectories undershoot the mean MD MSD as the time lag
  increases. Long time positional anti-correlation, on the order of hundreds of
  ns, may not exist in the MD system. Eventual loss of correlation would result 
  in a shift from sub-linear, or subdiffusive, to linear, or diffusive, MSD behavior, as observed in the MD 
  trajectories. Acetic acid exemplifies this point. At first glance, acetic acid's
  predicted MSD appears to match the curvature of MD quite well, but closer 
  examination reveals that
  the MD MSD curve may actually shift from a sub-linear to a linear regime around
  500 ns.
  
  The two mode models display curvature more consistent with MD but for
  non-physical reasons. Every time a switch between the pores and tails occurs,
  the width of the distribution used to model hops changes, and the correlation structure is broken. 
  Solutes that switch between modes the least show
  predicted MSDs with the greatest curvature. Due to the much larger
  accessible volume that a smaller molecule has, methanol spends $>90\%$ of its
  time in the hydrophobic tails (see Figure~\ref{fig:AD_mode_occupation}), so
  mode transitions are relatively rare and the predicted MSDs have significant
  curvature. This artificially resolves the problem of long timescale
  correlation, however, it has no physical basis.
  
  The model parameters for the one and two mode models tell stories about each
  solute's behavior that help explain the difference between the MSDs of
  different solutes. Higher values of $\sigma$ and lower values of $\alpha_h$
  indicate larger average hop length magnitudes by increasing the hop length
  distribution's width and tail density respectively. Higher values of $\alpha$
  indicate a lower probability of long dwell times. Higher values of $\lambda$
  truncate the power law distribution earlier preventing extremely long dwell
  times. Values of $H$ closer to the Brownian limit of 0.5 indicate a lower
  degree of negative correlation between hops. All of these changes in physical
  behavior contribute to an overall increase in the predicted MSD.

  Examining first the parameters of the one mode model, we can begin to break
  down the trends in solute MSDs. The parameters belonging to ethylene glycol
  and methanol are relatively similar, which is consistent with their similar MD
  MSDs and qualitatively similar MD time series.
  Relative to ethylene glycol, methanol tends to stay trapped for less
  time and takes larger hops but the most substantial difference is with
  respect to their Hurst parameters. Methanol has the lowest $H$ of all the
  solutes studied because it spends the majority of its time outside the pore
  region where collisions with tails are frequent.
  Urea has the third highest MSD which is primarily a consequence of more
  frequent and longer dwell times (lower $\alpha$ and $\lambda$). Urea's hop
  lengths ($\sigma$) and correlation ($H$) are comparable to ethylene glycol
  and methanol. Acetic acid has the smallest MSD among the solutes studied due
  to longer periods of entrapment and shorter hops. Its trapping behavior is
  parameterized by an $\alpha$ value significantly lower than other solutes,
  but an intermediate $\lambda$ value, suggesting it experiences many
  medium-length periods of entrapment. Its hops are smaller but are slightly
  compensated by a heavier tailed distribution (lower $\alpha_h$) than the
  other solutes. 
  
  We can use the two mode model to gain an even deeper understanding of solute
  behavior in the pore versus in the tails. It is clear that solutes are
  significantly slowed while they are in the tail region where long dwell times
  are more probable (smaller $\alpha$) and hops are smaller (smaller $\sigma$).
  Each solute spends a different amount of time in the tails (see
  Figure~\ref{fig:AD_mode_occupation}). Urea and acetic acid spend slightly
  more than half of their time in the tails (56\% and 62\% of their time
  respectively) while ethylene glycol spends about 44\% of its time in the
  tails. Urea and acetic acid's compact, flat structure allows it to more
  easily partition into the tails while ethylene glycol prefers the pore region
  due to its two hydrophilic hydroxyl groups. In contrast, methanol spends 91\%
  of its time in the tails, likely due to its small size. The value of
  $\alpha_h$ for urea and acetic acid in the tails is 1.50, meaning its hop
  distribution is heavy tailed relative to ethylene glycol and methanol, whose
  $\alpha_h$ values are 1.90 and 1.85 respectively, which is more consistent
  with a Gaussian distribution ($\alpha$=2). Acetic acid and urea are
  structurally similar molecules, both planar with two heavy atoms attached to
  a carbonyl group. Their small size and rigid shape may allow them to
  occasionally slip through gaps in the tails. Meanwhile, methanol is small
  enough that it does not need to make larger jumps to escape traps.

  Overall, the AD approach does a reasonable job of predicting solute MSDs and
  its parameters can help us further understand solute dynamics. Hops appear to
  be well modeled as anti-correlated draws from either Gaussian or L\'evy
  stable distributions. The data strongly suggests that one must truncate the
  power law dwell time distributions in order to obtain accurate MSD estimates.
  Trajectories generated by pure power laws are qualitatively non-physical. We
  can further understand solute dynamics by adding radially dependent parameter
  distributions as in the 2 mode model. A significant amount of solute
  trajectory data is necessary in order to achieve good parameter estimates. 

  \subsection{The Markov State-Dependent Dynamical Model}\label{section:msm_results}
  
  The AD model is useful if one does not know exact transport mechanisms 
  in a system since it only requires time series data. However, since we have
  already studied transport mechanisms in detail in our previous work, we can
  attempt to model transport as transitions between known discrete
  states, defined in Table~\ref{table:states}, with state-dependent positional
  fluctuations, which we refer to as the Markov state-dependent dynamical model (MSDDM).

  \subsubsection{Solute state preferences}\label{section:state_preferences}

\begin{figure}
  \centering
  \includegraphics[width=0.475\textwidth]{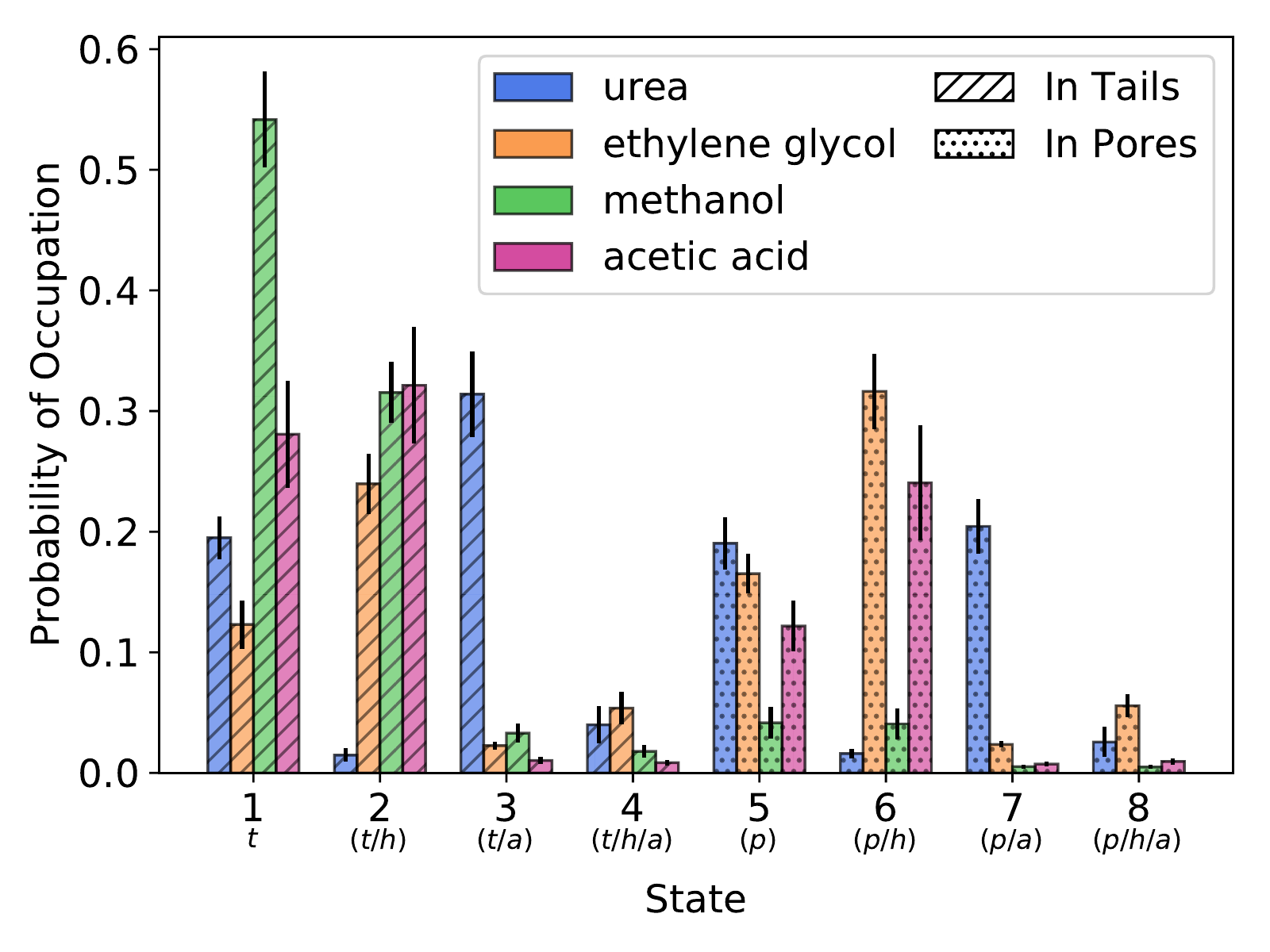}
  \caption{Solutes spend varying amounts of time under the influence of each
	  trapping mechanism. To aid the reader, we labeled each state with an
	  abbreviation which identifies the combination of conditions to which
	  each solute is subject in each state: t - tails, p - pores, h -
	  hydrogen bonded, a - associated with sodium. Solutes tend to favor
	  the same types of interactions (e.g. hydrogen bonding and/or
	  associating with sodium ions) independent of whether they are in the
	  pores or the tails.}\label{fig:state_probabilities}
  \end{figure}

  Solute size and chemical functionality influence which states are visited
  most frequently.  In Figure~\ref{fig:state_probabilities}, we plotted the
  probabilities of occupying a given state at any time. Solutes tend to favor
  the same types of interactions independent of which region they are in. We
  can relate these interactions to solute chemical functionality and use that
  intuition to hypothesize new designs for LLC monomers which control the
  transport rate of specific solutes.
  
  \begin{itemize}
  \item \textit{Urea} spends the largest fraction of its time trapped via association with
  sodium ions. It does so 31\% of the total time while in the tails (state 3)
  and 21\% of the total time while in the pores (state 7). Note that sodium
  does not drift significantly into the tails but sits close to the pore/tail
  region boundary. The electron-dense and unshielded oxygen atom of urea's
  carbonyl group is prone to associate with positively charged sodium ions. The
  nitrogen atoms of urea are only weak hydrogen bond donors. Therefore, the
  transport rate of urea is likely to be most significantly modified by
  removing or changing the identity of the counter-ion.
  
  \item \textit{Ethylene glycol} spends the largest fraction of its time trapped in a hydrogen
  bonded state. It does so 24\% of the total time while in the tails (state 2)
  and 32\% of the total time while in the pores (state 6). The two hydroxyl
  groups of ethylene glycol readily donate their hydrogen atoms to the
  carboxylate head groups and the ether linkages between the head groups and
  monomer tails. The transport rate of ethylene glycol might be modified by
  removing hydrogen bond accepting groups from the LLC monomers, especially
  those which stabilize the solute in the tails (i.e. the ether linkages). 
  
  \item \textit{Methanol} spends most of its time unbound in the tail region (state 1) and
  spends a significant portion of time hydrogen bonded while in the tail
  region. Tail region hydrogen bonds are donated from methanol to the ether
  linkages between the monomer head groups and tails, as well as to the ester groups
  at the ends of the tails. One might modify the
  rate of methanol transport by controlling its partition into the monomer
  tails. This might be achieved by adding cross-linkable groups near the
  monomer head groups.
  
  \item \textit{Acetic acid} spends the majority of its time hydrogen bonding both in
  and out of the pore (states 2 and 6). Although it has an unshielded carbonyl
  group in its structure, association with sodium ions in this environment is
  apparently a much weaker interaction than hydrogen bonding.  As with ethylene
  glycol, one might modify the transport rate of acetic acid by removing
  hydrogen bond accepting constituents of the LLC monomer. With this
  modification, we hypothesize that acetic acid might show similar transport
  rates to urea given their structural similarity.
  \end{itemize}
  
  \subsubsection{Parameters of the MSDDM}\label{section:msddm_parameterization}

  To create an MSDDM for each solute, we determined the state sequence associated
  with each solute trajectory based on the geometric indicators of the states 
  indicated in Table~\ref{table:states}. We then generated emission distributions of fluctuations
  within each state as well as transitions between states. In theory, one could 
  parameterize separate transition distributions for those which occur in the tails 
  versus in the pores, however this would lead to a broken correlation structure 
  similar to that seen in the two mode AD models.

\begin{figure*}
  \centering
  \begin{subfigure}{0.325\textwidth}
  \includegraphics[width=\textwidth]{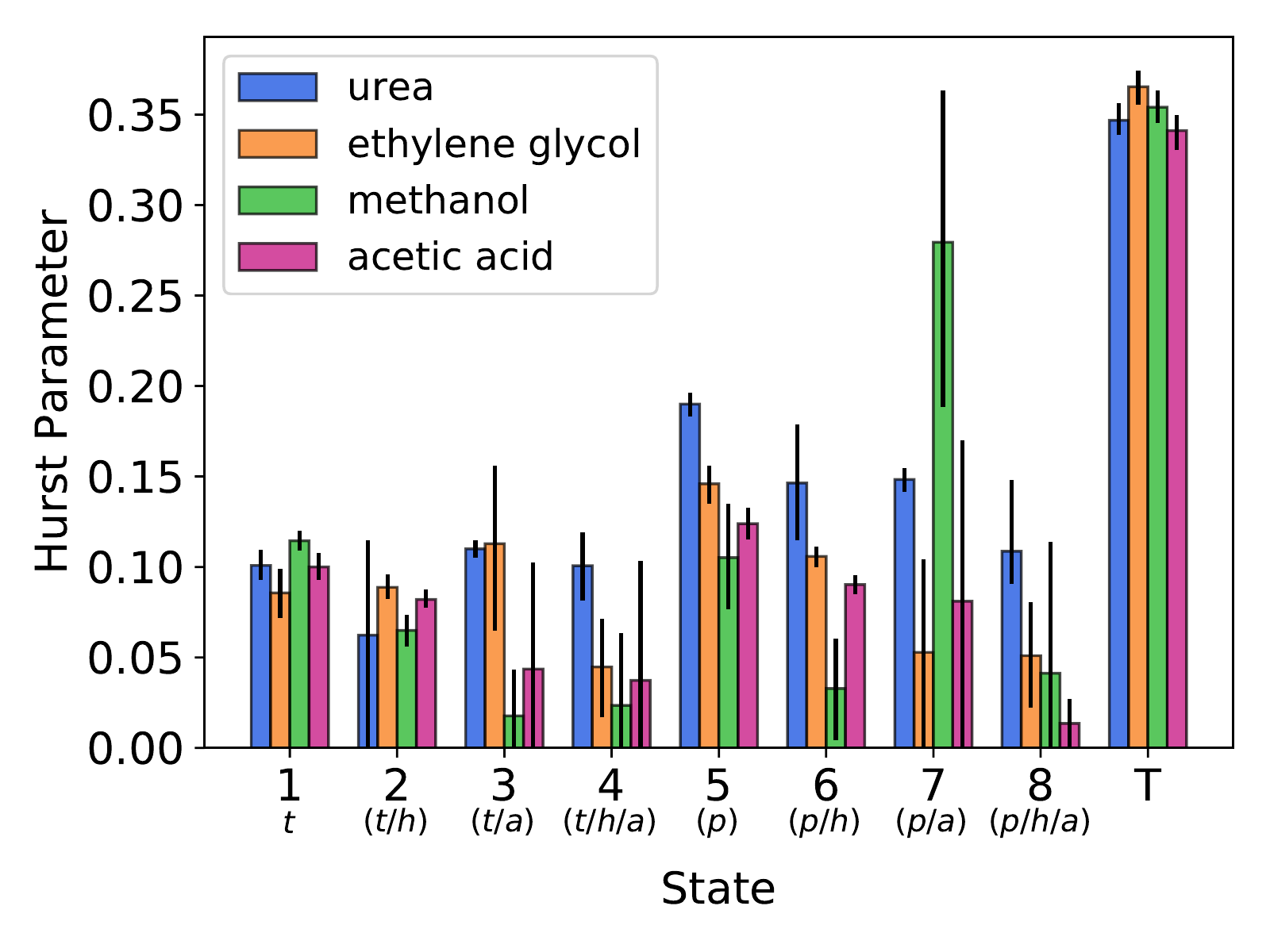}
  \caption{}\label{fig:H_v_state}
  \end{subfigure}
  \begin{subfigure}{0.325\textwidth}
  \includegraphics[width=\textwidth]{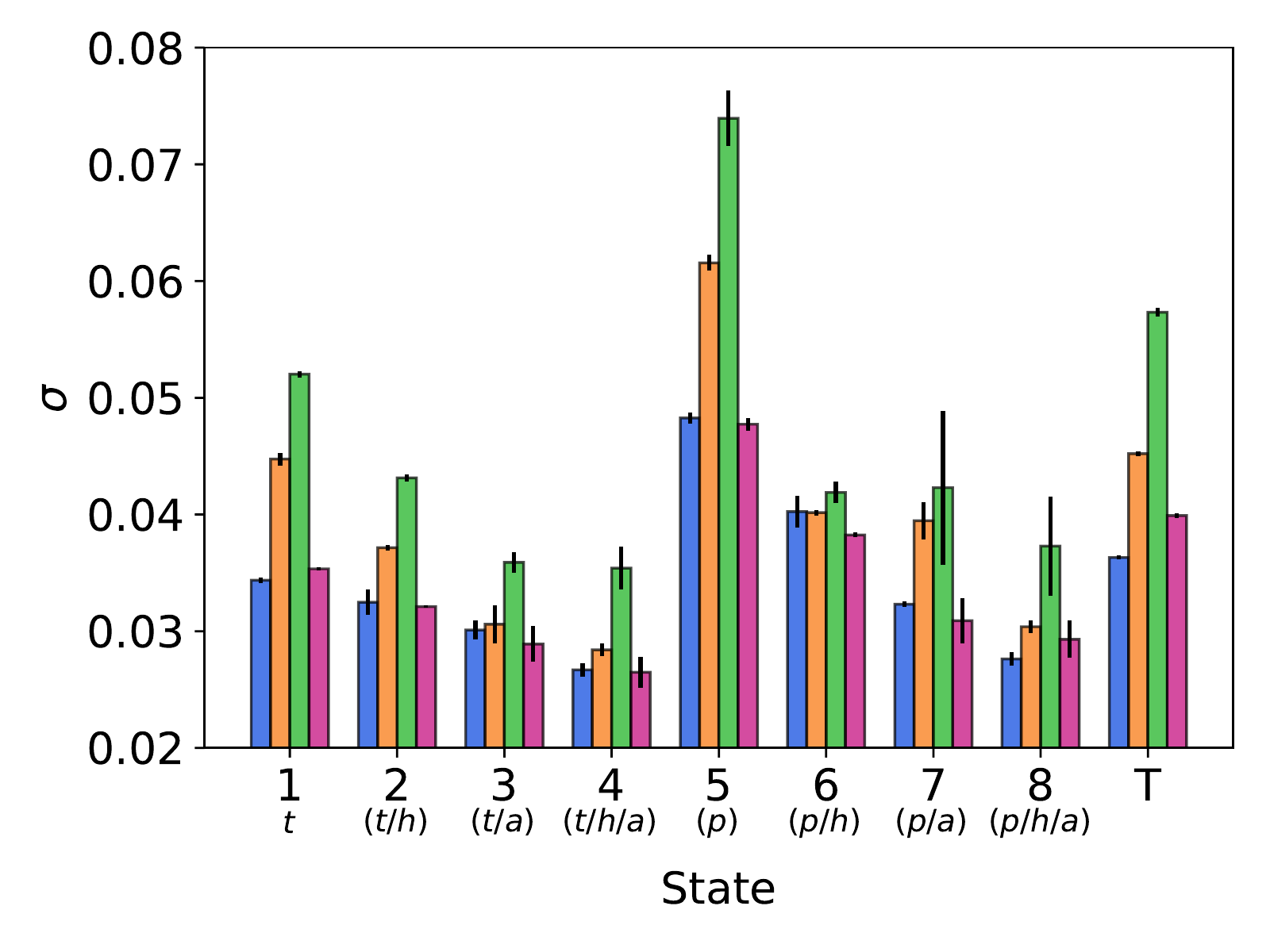}
  \caption{}\label{fig:sigma_v_state}
  \end{subfigure}
  \begin{subfigure}{0.325\textwidth}
  \includegraphics[width=\textwidth]{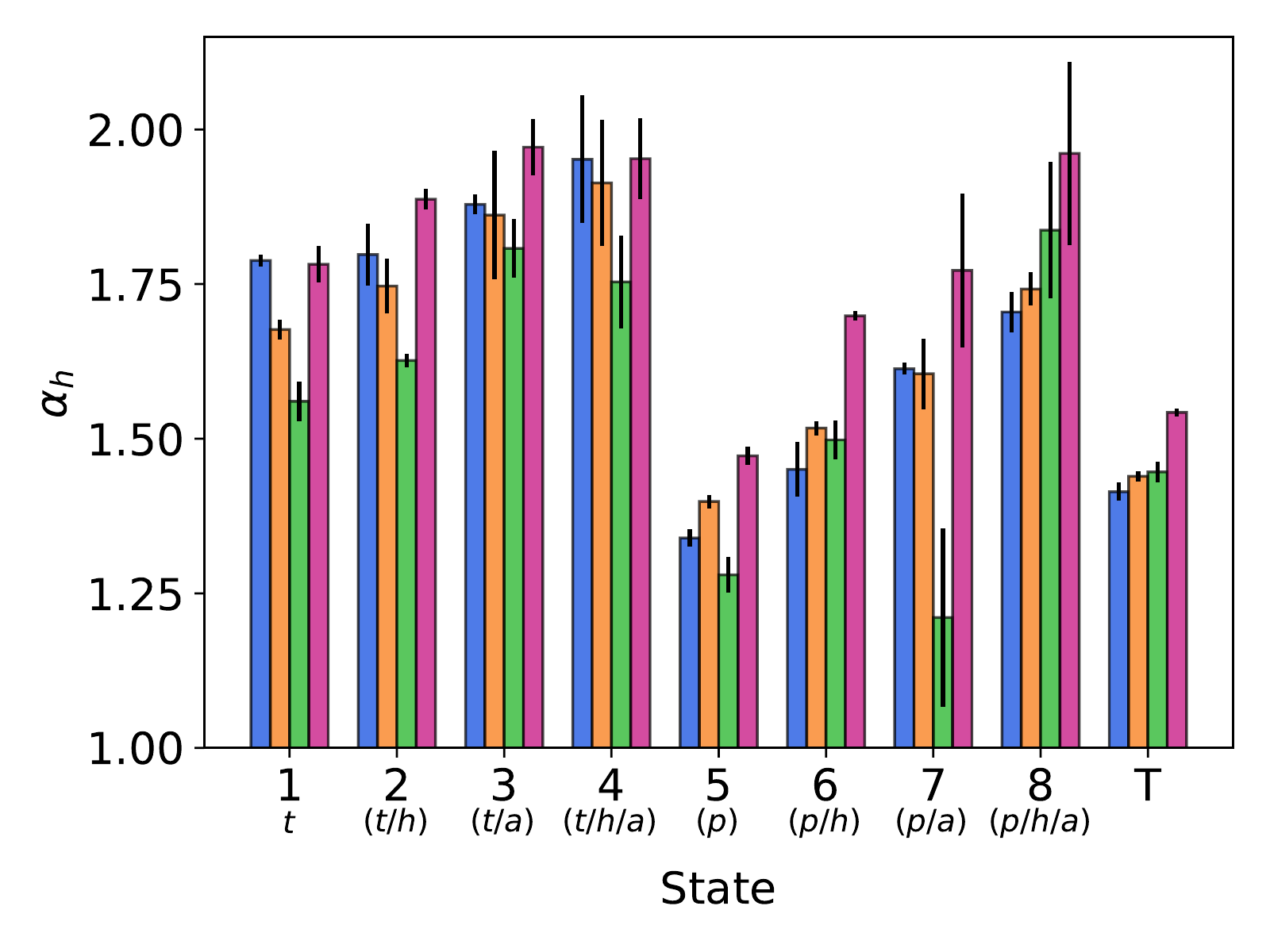}
  \caption{}\label{fig:alpha_v_state}
  \end{subfigure}
  \caption{The parameters of the MSDDM are strong functions of trapping
	  mechanisms. We observe different parameters but with similar trends
	  between the tail and pore region. The states are defined in
	  Table~\ref{table:states}. See Figure~\ref{fig:state_probabilities}
	  for a description of the abbreviation under each state number. The
	  legend in (a) applies to all subplots. (a) Motion is highly
	  anti-correlated in trapped states. As the number of simultaneously
	  influencing trapping mechanisms increases, the Hurst parameter, $H$,
	  decreases. $H$ is highest (closest to Brownian) during transitions 
	  between states (state T). (b) As more trapping mechanisms simultaneously influence solutes,
	  the width of the hop length distribution ($\sigma$) decreases. The
	  largest hops occur when solutes are unbound in the pores. (c) The
	  weight of the hop length distribution's tails, parameterized by
	  $\alpha_h$, increases as more trapping mechanisms influence solutes
	  simultaneously. The transitional hop distributions have among the
	  heaviest tails.}\label{fig:msddm_parameters}
  \end{figure*}

  We observe correlated emissions drawn from L\'evy stable distributions. The
  deviation of the emission distributions from Gaussian behavior is far more
  pronounced than that seen in the hop length distributions of the previous
  section (see Figure~\ref{S-fig:gaussian_levy_comparison} of the Supporting
  Information). We therefore did not consider the Gaussian case. The correlation
  structure between hops is consistent with that of FLM (see 
  Figure~\ref{S-fig:msddm_acf}). The parameters of the L\'evy stable distributions
  along with their Hurst parameters are visualized in Figure~\ref{fig:msddm_parameters}
  (and tabulated in the Supporting Information, Table~\ref{S-table:msddm_params}). 
  

  Most of a solute's MSD is a consequence of transitions between the 8 states
  in Table~\ref{table:states}. Perfectly anti-correlated motion ($H$=0) results
  in no contribution to the solute's MSD. Motion while trapped in a state is
  highly anti-correlated as indicated by their consistently low Hurst
  parameters. There is a weak negative trend in the Hurst parameter values as
  the number of simultaneously influencing trapping mechanisms increases
  (Figure~\ref{fig:H_v_state}). The Hurst parameters for transitional (T)
  emissions are up to 18 times higher than emissions from trapped states. The
  value of $\alpha_h$ for transition emissions is also relatively low giving
  higher probabilities to larger hops.
  
  As solutes are influenced by more trapping interactions simultaneously
  (e.g.~hydrogen bonding \textit{and} association with sodium versus just
  hydrogen bonding), the width of the hop length distribution, $\sigma$,
  decreases while its L\'evy index, $\alpha_h$, increases. Treating states in
  the tail and pore regions independently, $\sigma$ is largest and $\alpha_h$
  is smallest when solutes are not hydrogen bonding or associating with sodium
  (states 1 and 5). Solutes are free to move and take occasionally large hops.
  The smallest $\sigma$ and highest $\alpha_h$ values are measured when solutes
  are hydrogen bonding and associating with sodium at the same time (states 4
  and 8). Motion is restricted by multiple stabilizing forces which maintains a
  relatively narrow distribution of hop lengths.
  
  \subsubsection{Application of the MSDDM}\label{section:msddm_application}
  
\begin{figure*}
  \centering
  \begin{subfigure}{0.45\textwidth}
  \includegraphics[width=\textwidth]{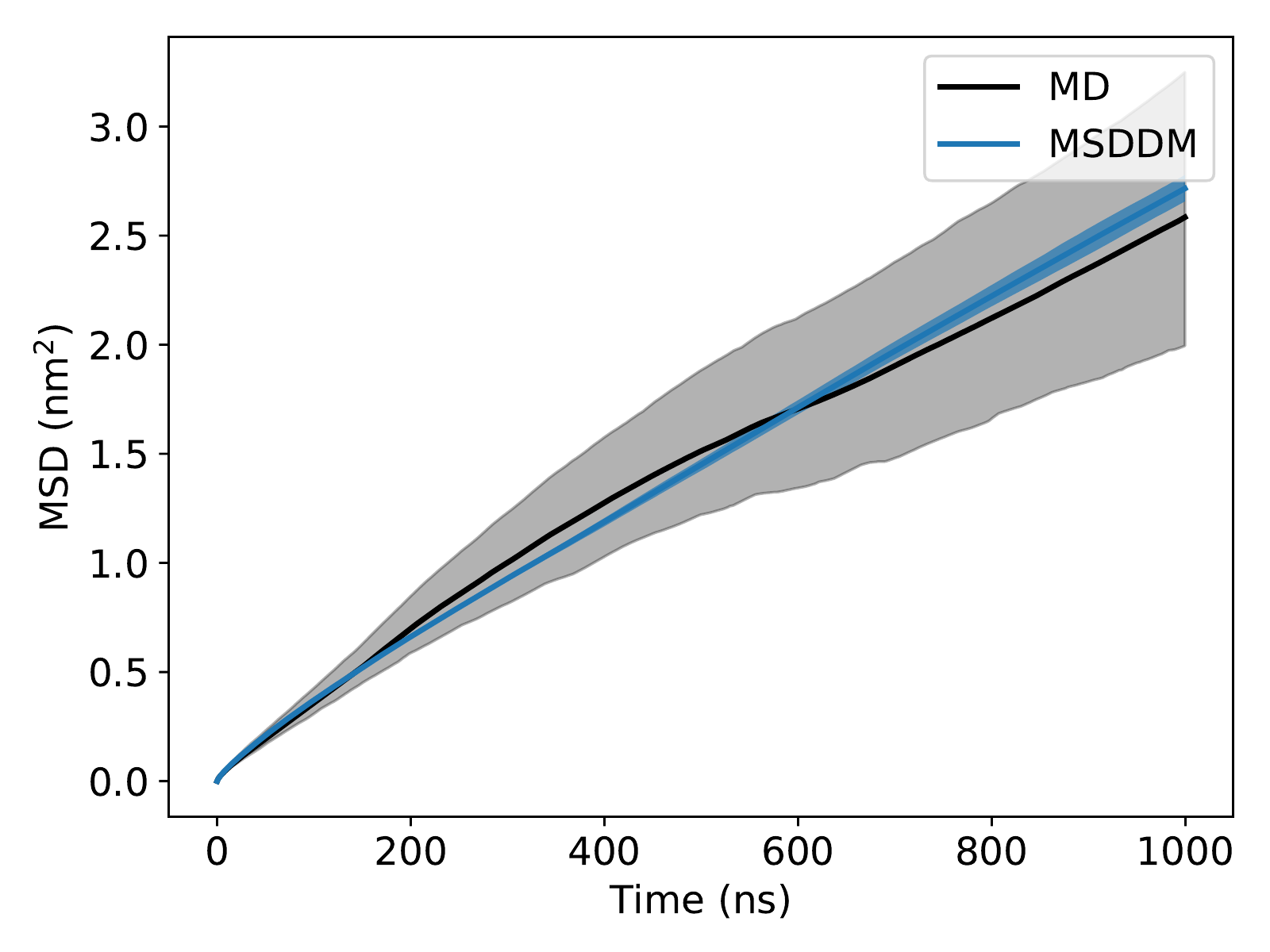}
  \caption{Urea}\label{fig:URE_msddm}
  \end{subfigure}
  \begin{subfigure}{0.45\textwidth}
  \includegraphics[width=\textwidth]{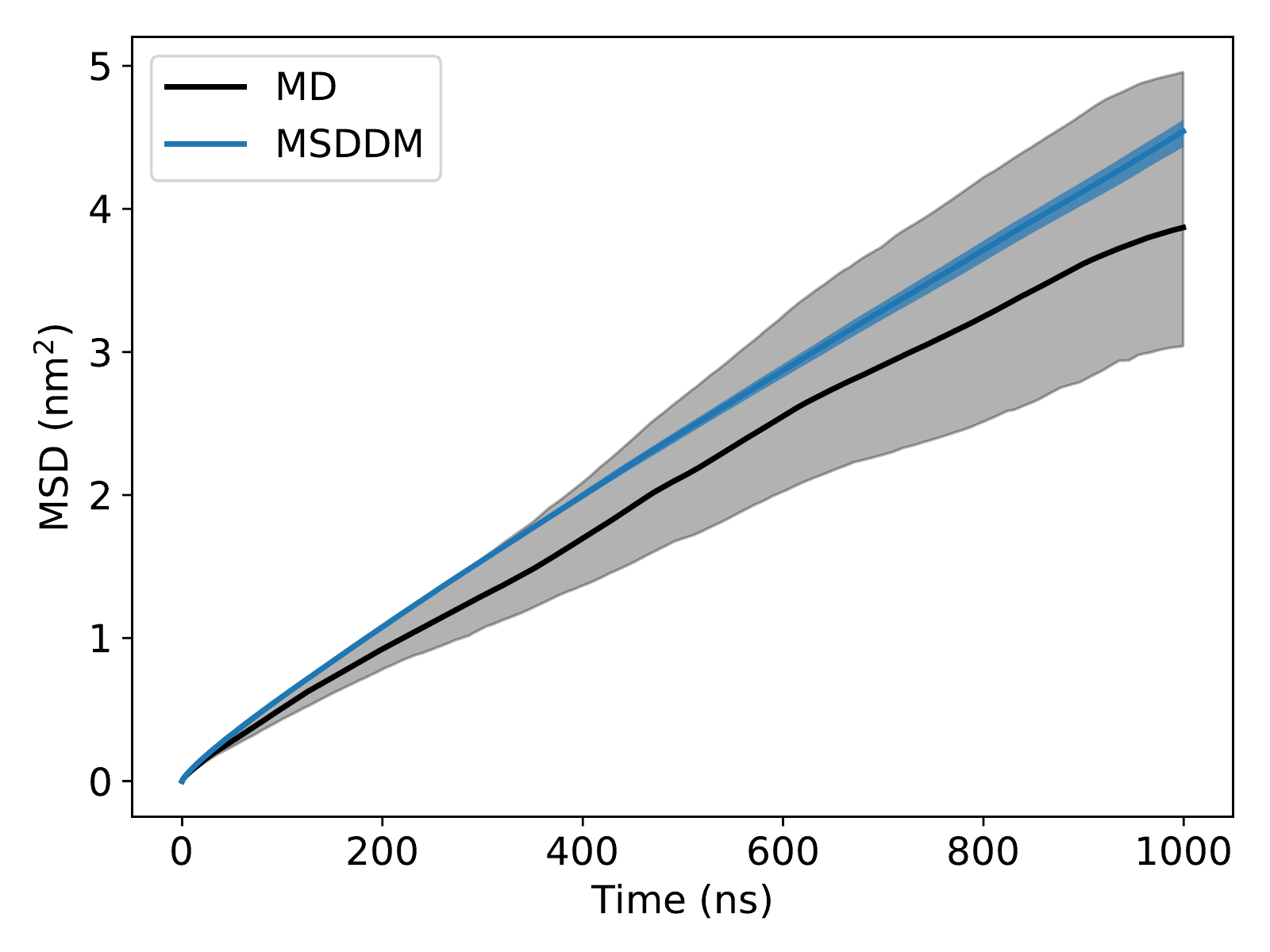}
  \caption{Ethylene Glycol}\label{fig:GCL_msddm}
  \end{subfigure}
  \begin{subfigure}{0.45\textwidth}
  \includegraphics[width=\textwidth]{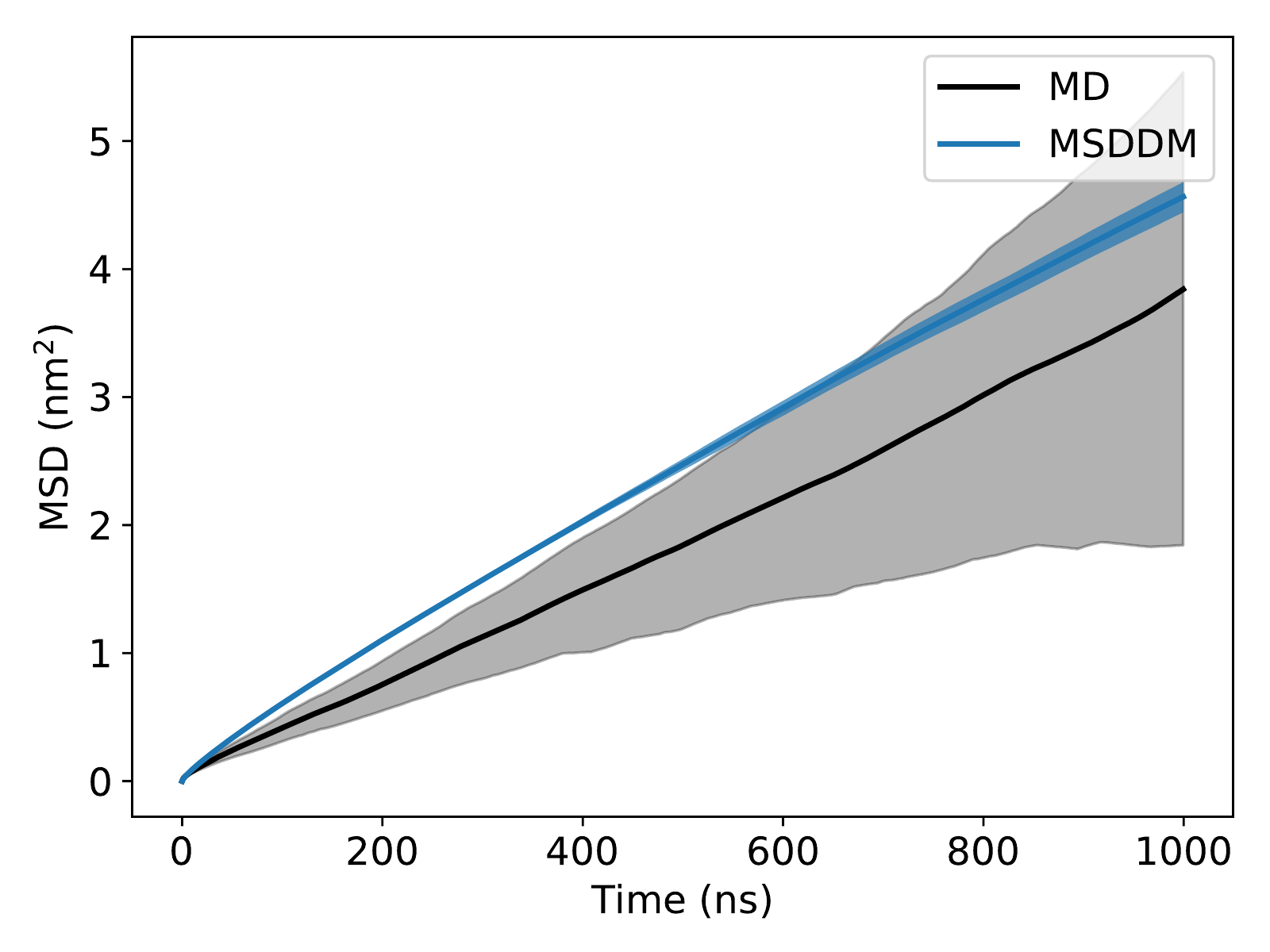}
  \caption{Methanol}\label{fig:MET_msddm}
  \end{subfigure}
  \begin{subfigure}{0.45\textwidth}
  \includegraphics[width=\textwidth]{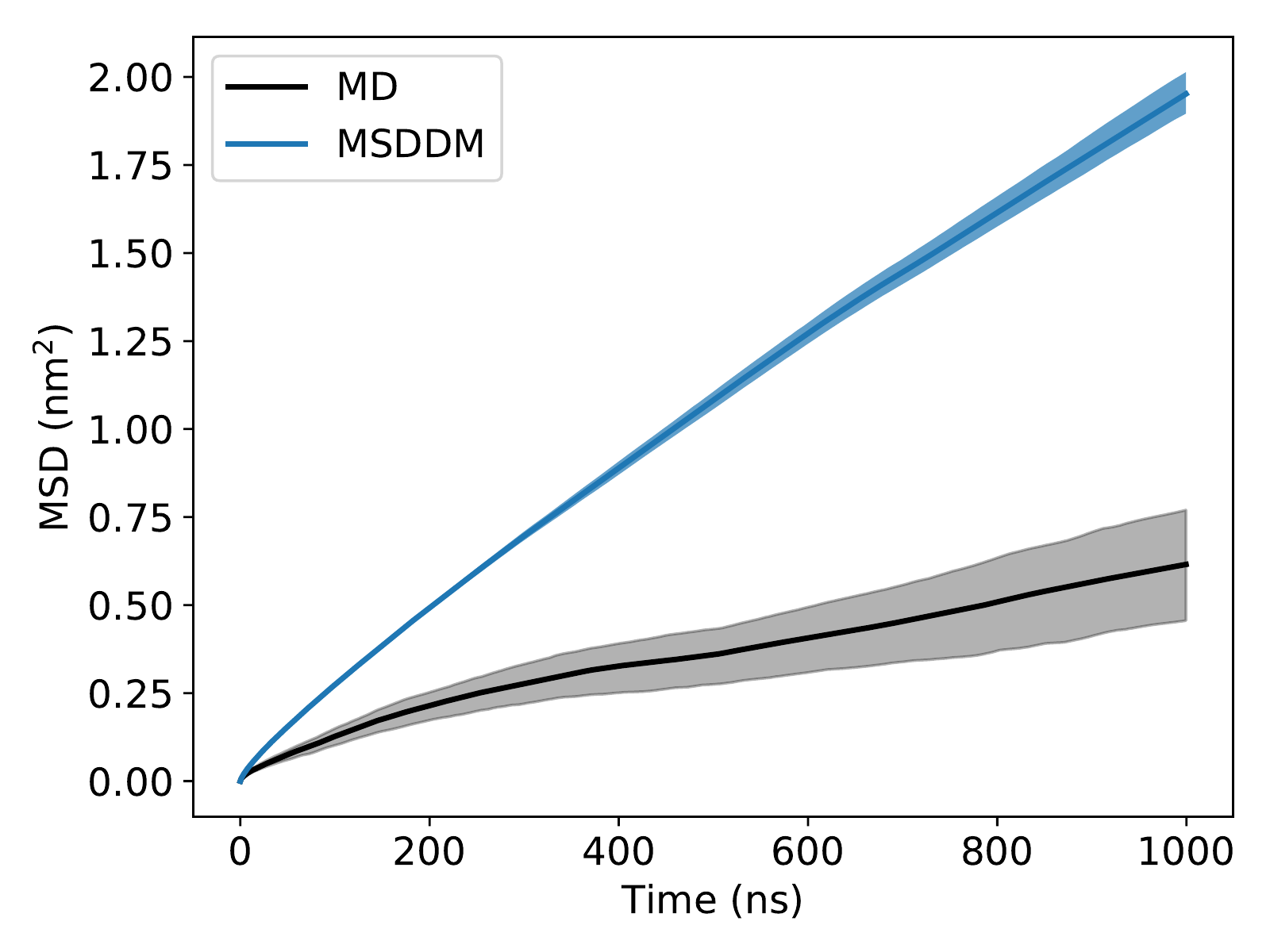}
  \caption{Acetic Acid}\label{fig:ACH_msddm}
  \end{subfigure}
  \caption{In most cases, the magnitude of the MSD curves predicted by the
	  MSDDM agree well with those generated from MD simulations. The
	  predicted MSD curves of urea and ethylene glycol lie within the
	  1$\sigma$ confidence intervals of MD for all time lags. Methanol
	  over-predicts the MSD at small time lags and acetic acid grossly
	  over-predicts the MSD at all time lags.  Like the AD approach models,
	  the MSDDM doesn't fully capture the curvature of the MD MSD curves.
	  }\label{fig:msddm_performance}
  \end{figure*}

  Quantitatively, the dynamics of urea, ethylene glycol and, to a lesser extent, 
  methanol appear to be well-captured by the MSDDM. We simulated 1000 MSDDM trajectory realizations
  for each solute, as described in Section~\ref{method:MSMs}, then calculated their
  MSDs (see Figure~\ref{fig:msddm_performance}). In most cases, the MSDDM predicts
  the magnitude of the MD MSDs within their 1$\sigma$ confidence intervals. 
 
  The predicted MSD of acetic acid is severely over-estimated primarily due to
  under-estimation of hop anti-correlation ($H$ too high) and an over-estimate
  of its hop lengths ($\sigma$ and $\alpha_h$ too high). Acetic acid's
  predicted MSD actually lies just within the lower bound of urea's MD MSD
  1$\sigma$ confidence interval, but lower than the prediction of the MSDDM for
  urea. As stated earlier, most of a solute’s motion modeled by the MSDDM is
  due to displacements during state transitions. Acetic acid has a transitional
  Hurst parameter close to urea’s and transitional $\sigma$ and $\alpha_h$
  values that are higher than urea’s. The primary reason that the MSDDM
  predictions for acetic acid's MSD are lower than urea's appears to be because
  of higher hop anti-correlation (lower $H$) in all states, including
  transitions. There is also the strong possibility that the over-estimate is
  a consequence of lumping together all of acetic acid’s transitional hops into
  a single correlated distribution.

\begin{figure*}
  \centering
  \begin{subfigure}{0.24\textwidth}
  \includegraphics[width=\textwidth]{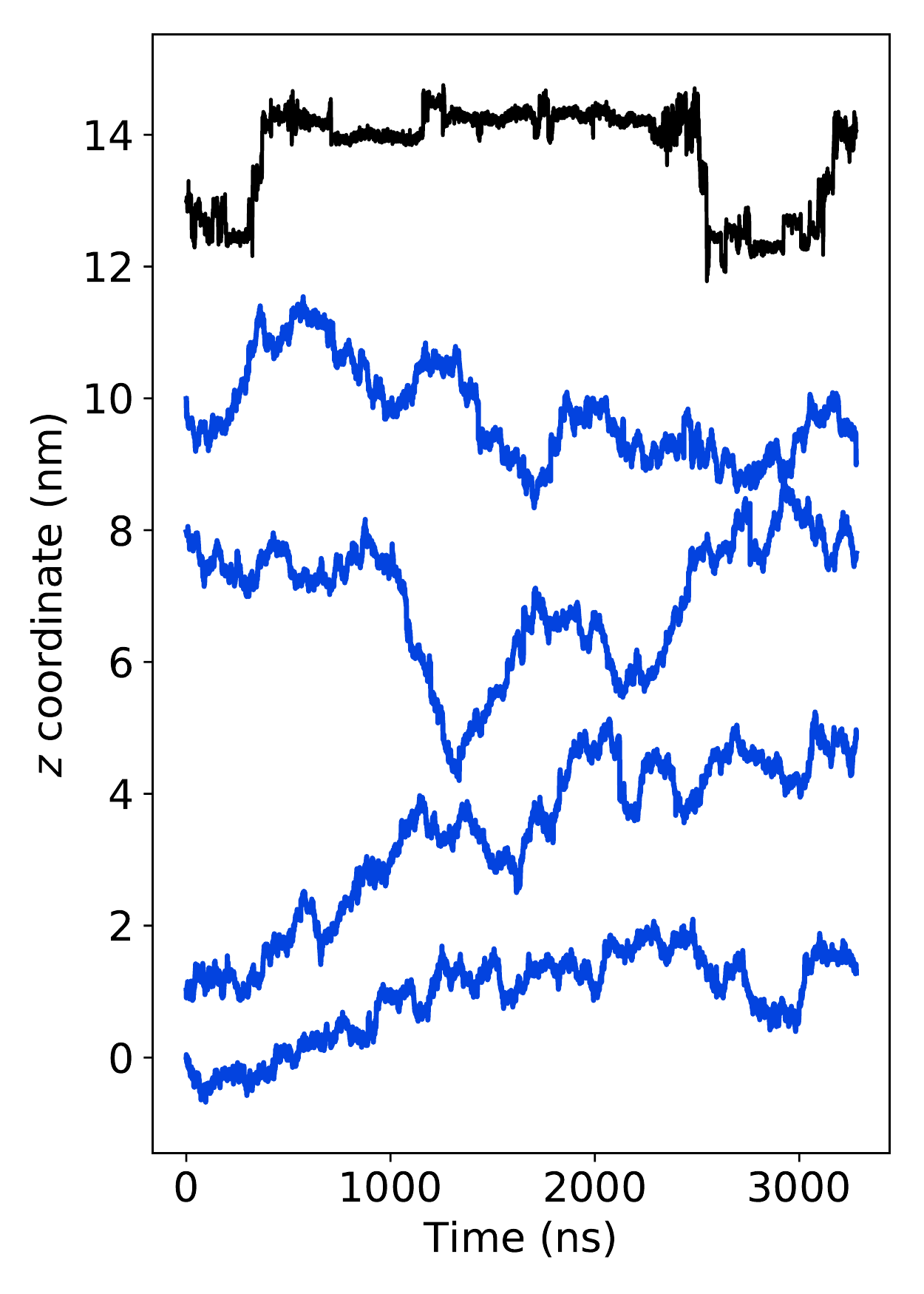}
  \caption{Urea}\label{fig:stacked_msddm_realizations_URE}
  \end{subfigure}
  \begin{subfigure}{0.24\textwidth}
  \includegraphics[width=\textwidth]{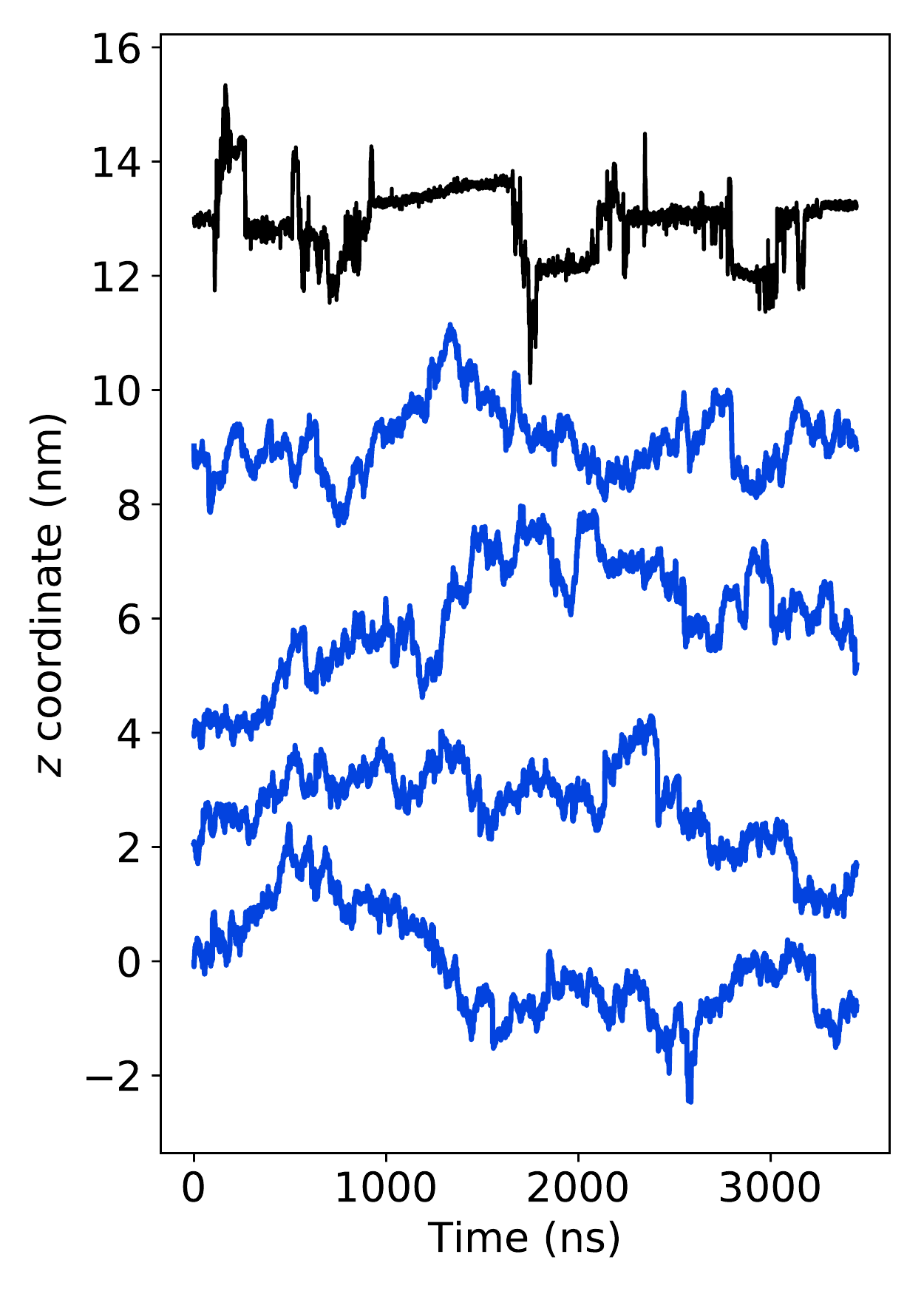}
  \caption{Ethylene glycol}\label{fig:stacked_msddm_realizations_GCL}
  \end{subfigure}
  \begin{subfigure}{0.24\textwidth}
  \includegraphics[width=\textwidth]{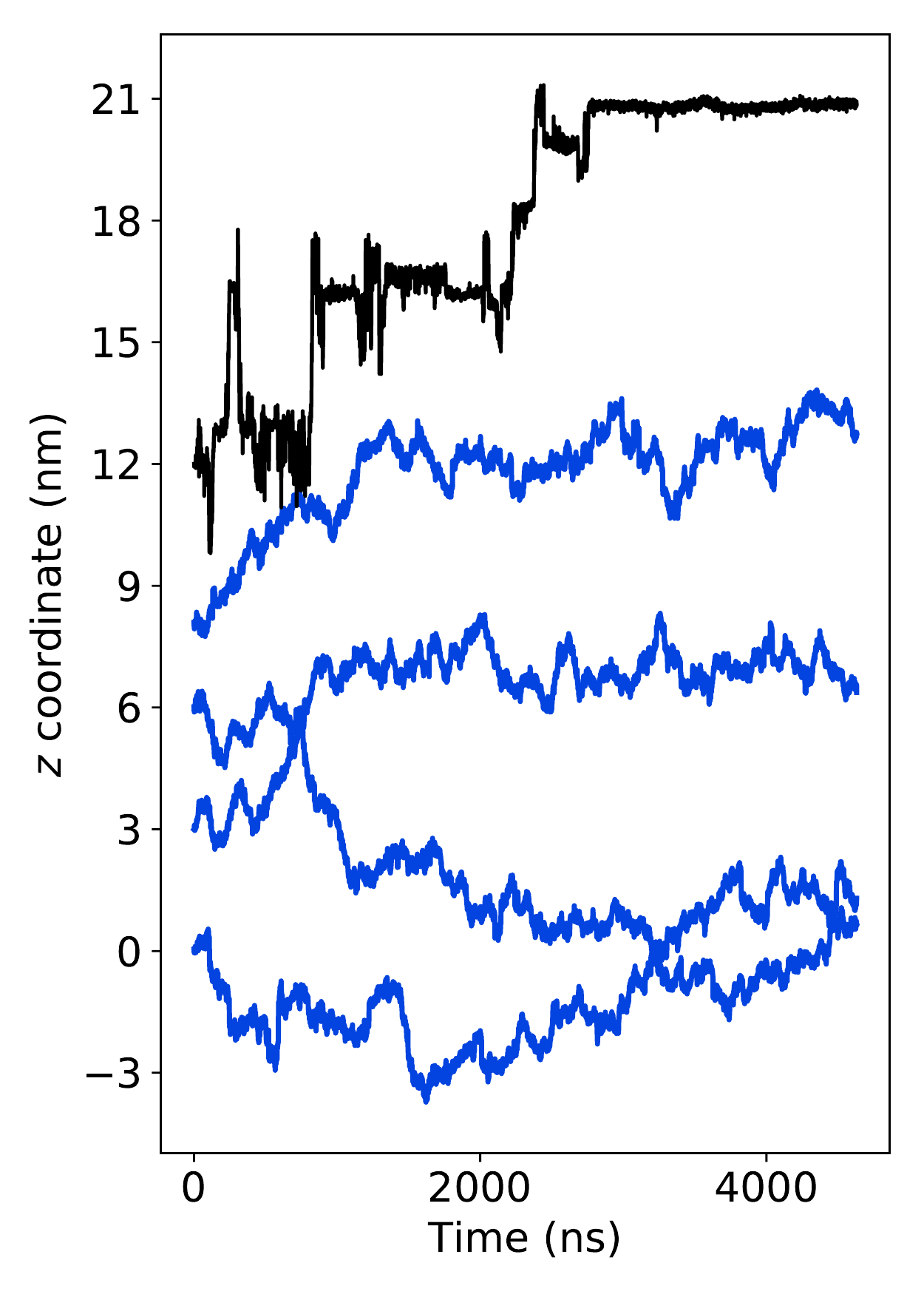}
  \caption{Methanol}\label{fig:stacked_msddm_realizations_MET}
  \end{subfigure}
  \begin{subfigure}{0.24\textwidth}
  \includegraphics[width=\textwidth]{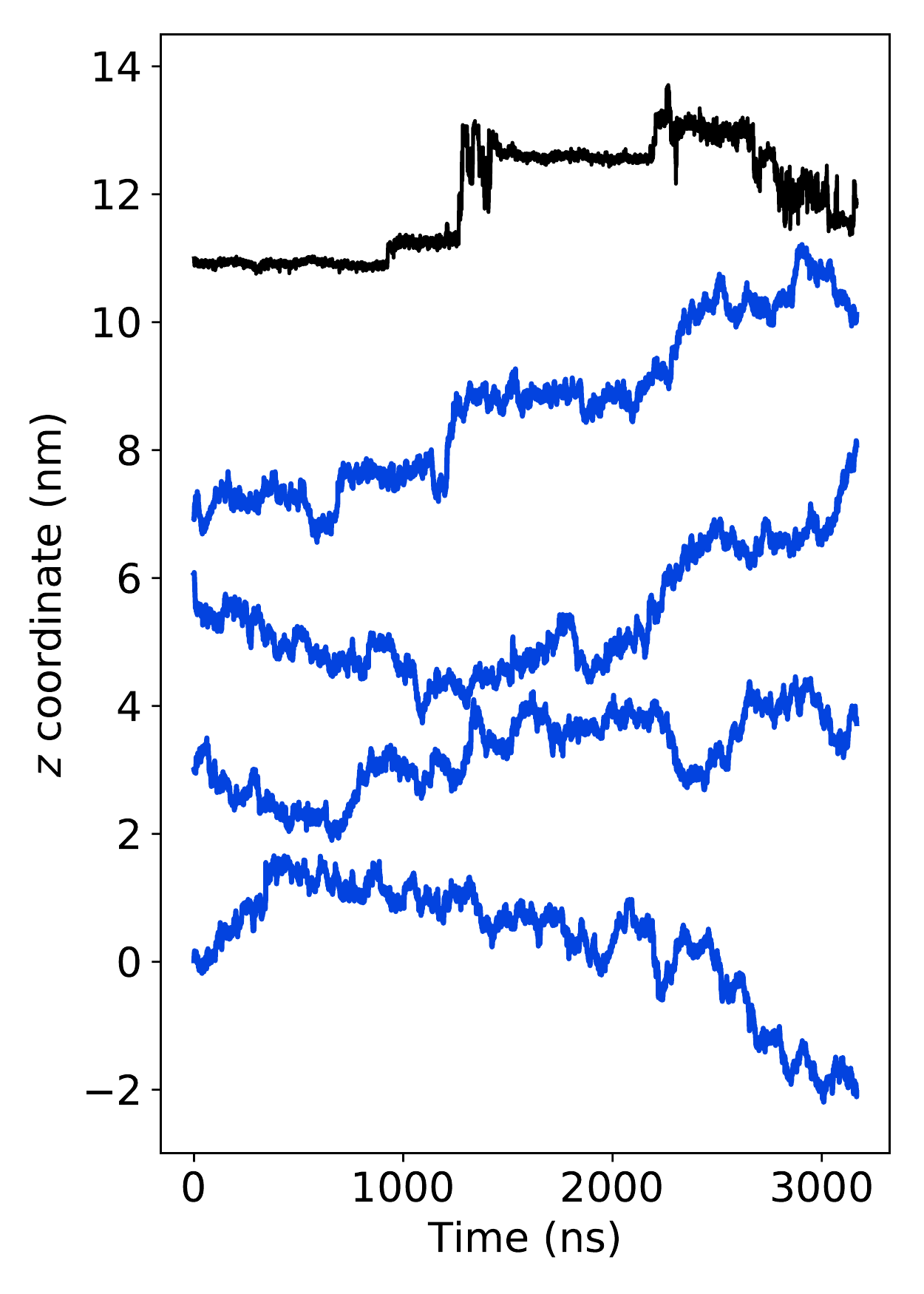}
  \caption{Acetic Acid}\label{fig:stacked_msddm_realizations_ACH}
  \end{subfigure}
  \caption{Realizations of the MSDDM for each solute (blue) do not reproduce
	  the hopping and trapping behavior observed in our MD simulations
	  (black). The trajectories are qualitatively similar to what one
	  might expect for Brownian motion even though the MSDs are often similar
	  to the atomistic systems.
  	  }\label{fig:msddm_eyetest}
  \end{figure*}

  Despite relatively good predictions of the MD MSD, qualitative mismatch
  between simulated MSDDM and MD trajectories suggest that the MSDDM may be
  getting the right answers for the wrong reasons. We plotted typical
  realizations of the MSDDM for each solute and compared them to MD in
  Figure~\ref{fig:msddm_eyetest}. There is little evidence of trapping behavior
  or large hops. There are two reasons for this behavior. First, the width of
  hop length distributions are much smaller than those of the AD model. Closer
  examination of the characteristic MD trajectories shown in
  Figure~\ref{fig:solute_trajectories} reveal that hops tend to be an
  accumulation of a series of hops in the same direction. All of the hops in
  the MSDDM are negatively correlated which prevents this from happening. The
  second reason is a consequence of using a single hop length distribution for
  transitions. This was necessary because we could not collect enough data to
  fit all of the possible transition distributions and because we could not
  correlate emissions coming from different hop distributions. Many
  transitions occur between two trapped states where the transitional hops are
  actually very small. Our model ignores this physical restriction which can
  cause the solute to drift rather than stay trapped.
  
  \subsection{Solute Flux and Selectivity}\label{section:mfpt}
  
  We used the one mode sFBMcut (see Table~\ref{table:anomalous_models}) AD
  model in order to demonstrate how one can use its realizations in order to
  calculate the flux (see section~\ref{method:mfpt}) of solutes given model
  parameters extracted from MD simulations. The one mode sFBMcut model
  generates predictions similar to the one mode sFLMcut model at a lower
  computational cost. We do not consider the two mode AD model because it has a
  broken correlation structure and we do not consider the MSDDM because its
  realizations do not display the expected hopping and trapping behavior. 

  It is computationally infeasible to simulate trajectories long enough that
  they traverse the length of a macroscopic pore. To date, the thinnest
  H\textsubscript{II} LLC membrane synthesized with the monomer in this work
  was 7$\mu$m thick.~\cite{feng_thin_2016}
  Using 24 cores to simulate trajectory realizations in
  parallel, it takes on the order of 1 day to simulate 10000 sFBMcut
  realizations of solutes traversing a 50 nm pore. The RAM requirements and
  performance scales greater than linearly and thus would take an infeasible
  amount of memory and time to simulate transport through a pore over 100 times
  longer. One could improve performance significantly by simulating less
  trajectories. In Figure~\ref{S-fig:flux_curve_sensitivity} of the Supporting
  Information, we determined that one can simulate as few as 100 sFBMcut
  realizations in order to parameterize Equation~\ref{eqn:flux_decay}. For
  better precision, we recommend simulating at least 1000 realizations.
  However, even with an order of magnitude decrease in number of trajectories,
  it is still infeasible to simulate experimental-length pores.
  
\begin{figure*}
  \centering
  \begin{subfigure}{0.485\textwidth}
  \includegraphics[width=\textwidth]{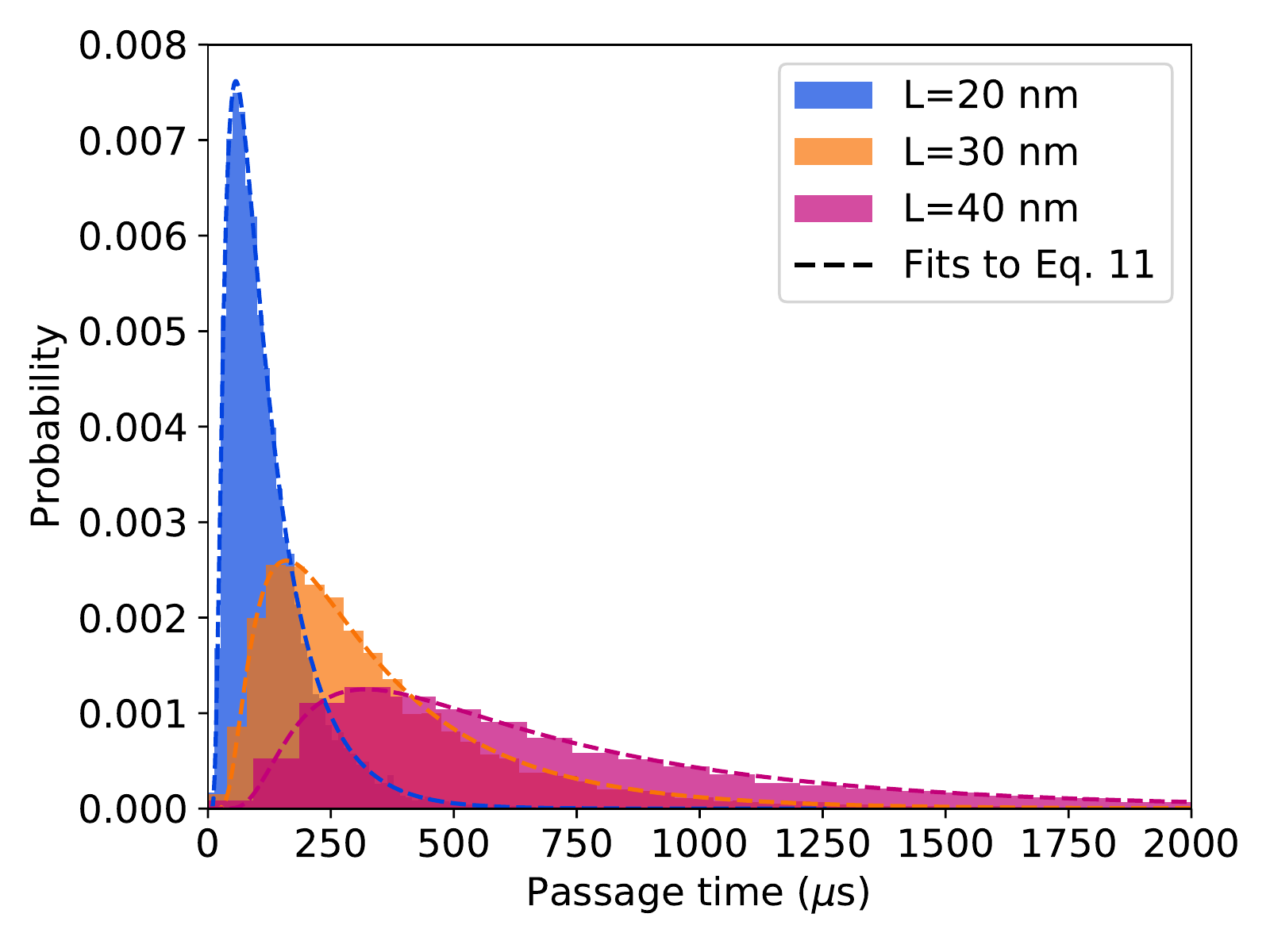}
  \caption{}\label{fig:fpt_distributions}
  \end{subfigure}
  \begin{subfigure}{0.485\textwidth}
  \includegraphics[width=\textwidth]{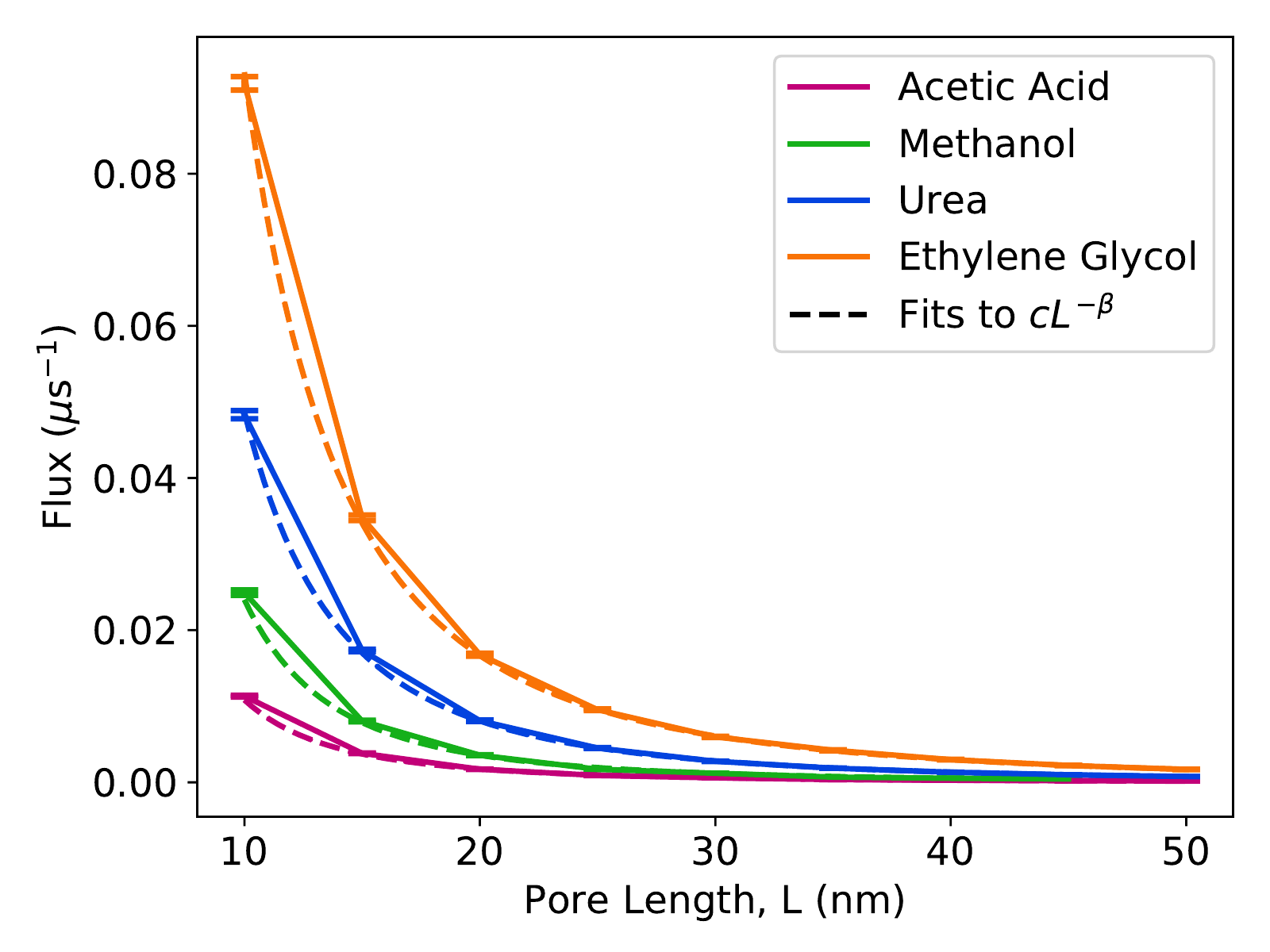}
  \caption{}\label{fig:flux_curves_ad}
  \end{subfigure}
  \caption{(a) The distributions of first passage times generated from the
	  sFBMcut model fit well to Equation~\ref{eqn:passage_times}. We show
	  similar fits for the remaining solutes in
	  Figure~\ref{S-fig:ad_fpt_fits} of the Supporting Information. (b) The
	  single particle flux measured by the sFBMcut AD model decays with
	  increasing pore length. The rankings of solute fluxes are consistent
	  with the MSDs predicted by each model. We fit the single particle
	  solute flux versus pore length, $L$, to a power law function of the
	  form $AL^{-\beta}$ (dashed lines). 
  }\label{fig:flux_curves}
  \end{figure*}

  We used simulated trajectories which traverse computationally-reasonable
  length pores in order to construct an empirical model which one can use to
  estimate particle flux for arbitrary length pores. We fit
  Equation~\ref{eqn:passage_times} to the empirical distribution of first
  passage times in Figure~\ref{fig:fpt_distributions} and used the expected
  value of the analytical equation to calculate flux from
  Equation~\ref{eqn:hill_relation}. As shown in
  Figure~\ref{fig:flux_curves_ad}, the flux appears to scale according to a
  power law of the form:
  \begin{equation}
  J(L) = AL^{-\beta} 
  \label{eqn:flux_decay}
  \end{equation}

\begin{figure*}
  \centering
  \begin{subfigure}{0.45\textwidth}
  \includegraphics[width=\textwidth]{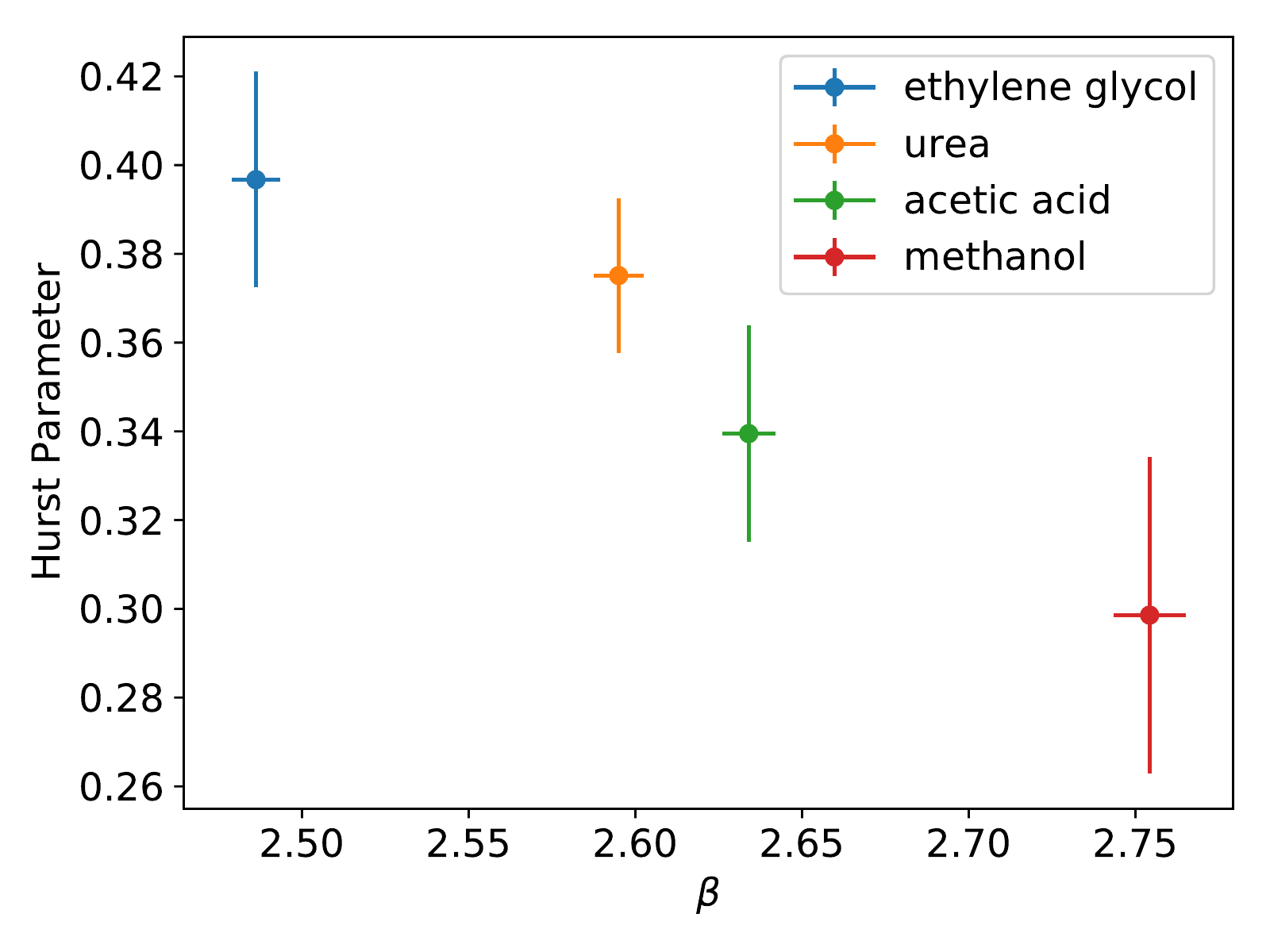}
  \caption{}\label{fig:beta_hurst}
  \end{subfigure}
  \begin{subfigure}{0.45\textwidth}
  \includegraphics[width=\textwidth]{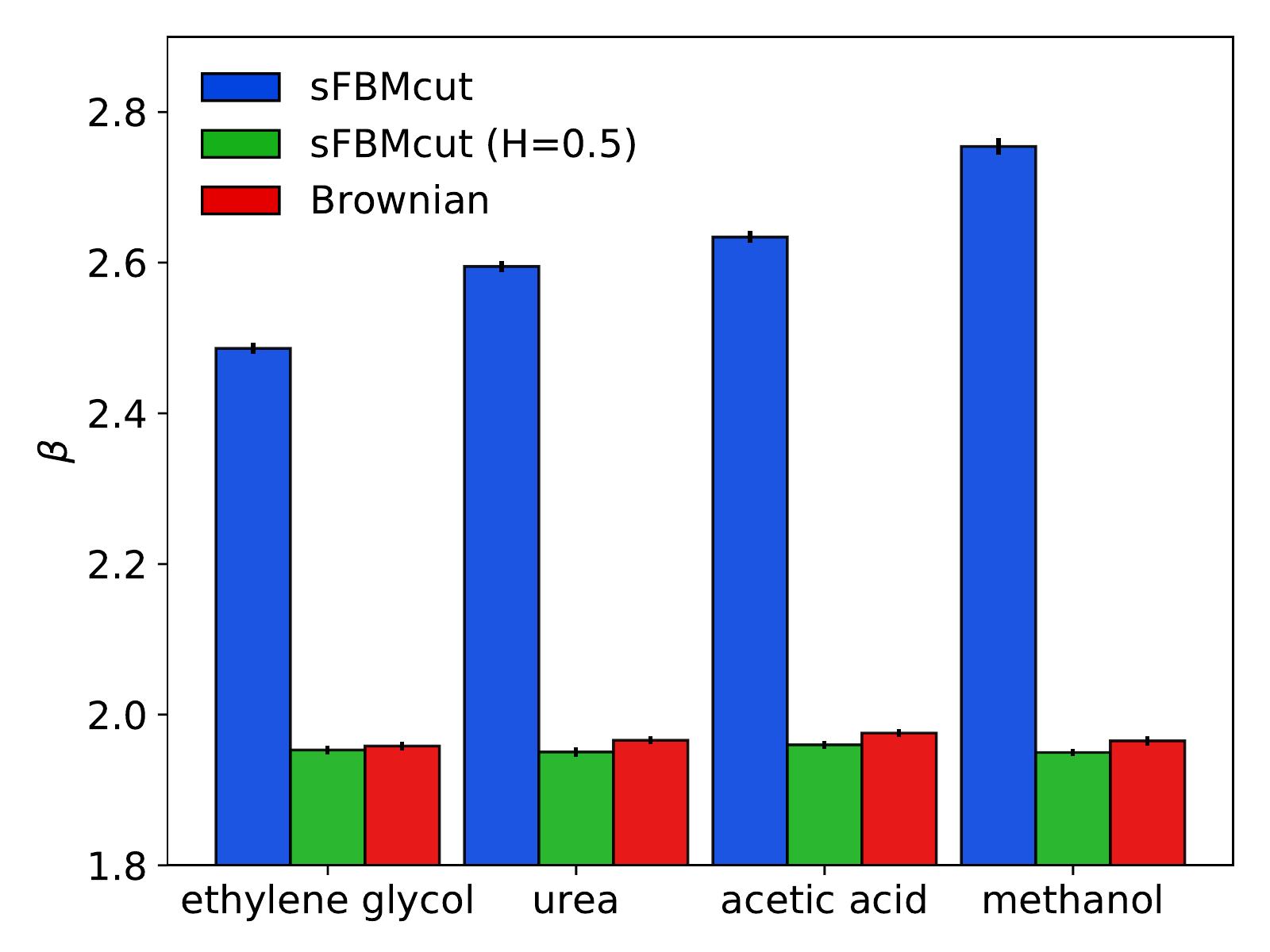}
  \caption{}\label{fig:beta_variations}
  \end{subfigure}
  \begin{subfigure}{0.31\textwidth}
  \vspace{-0.25cm}
  \includegraphics[width=\textwidth]{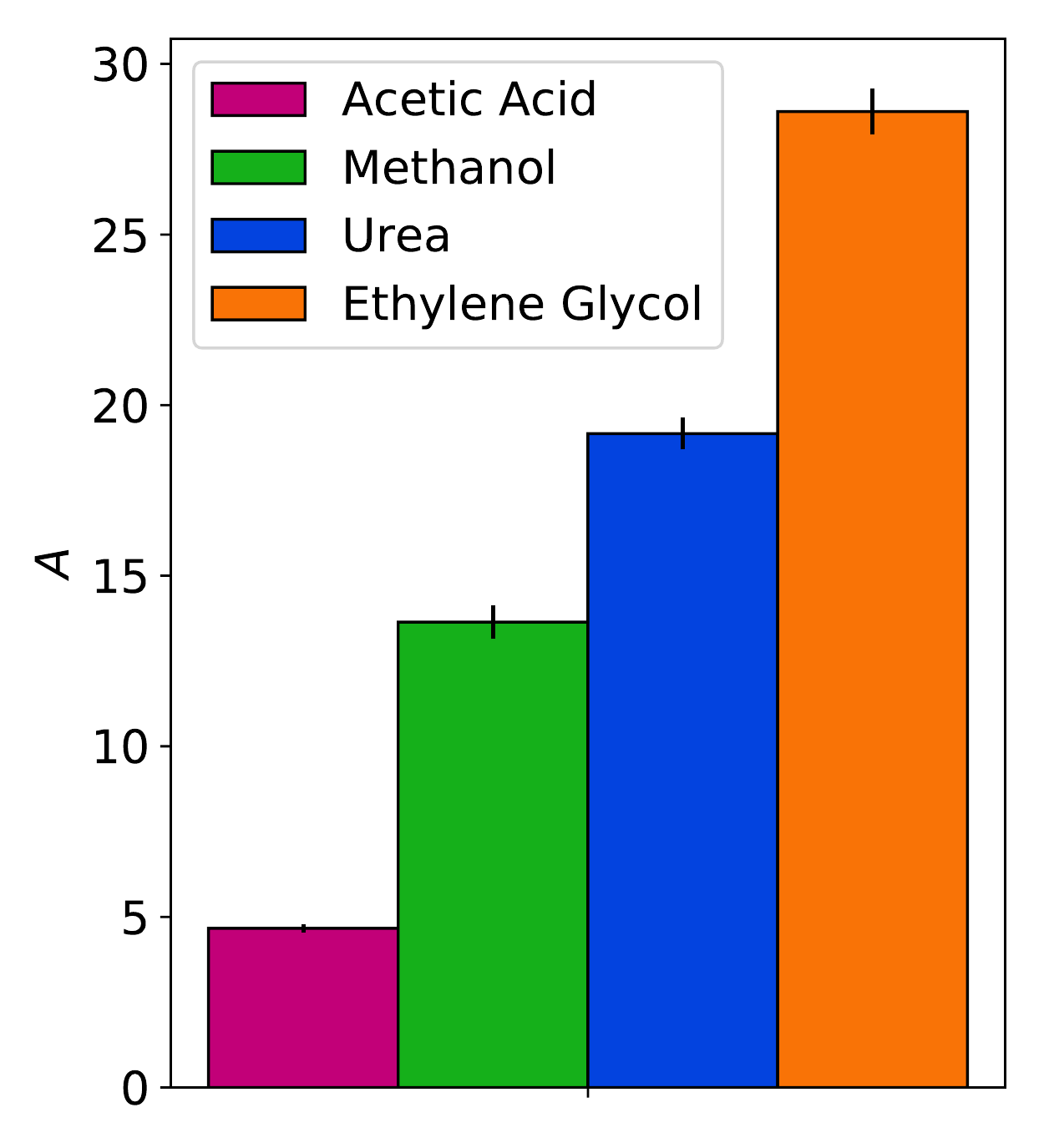}
  \caption{}\label{fig:c_parameters}
  \end{subfigure}
  \begin{subfigure}{0.66\textwidth}
  \includegraphics[width=\textwidth]{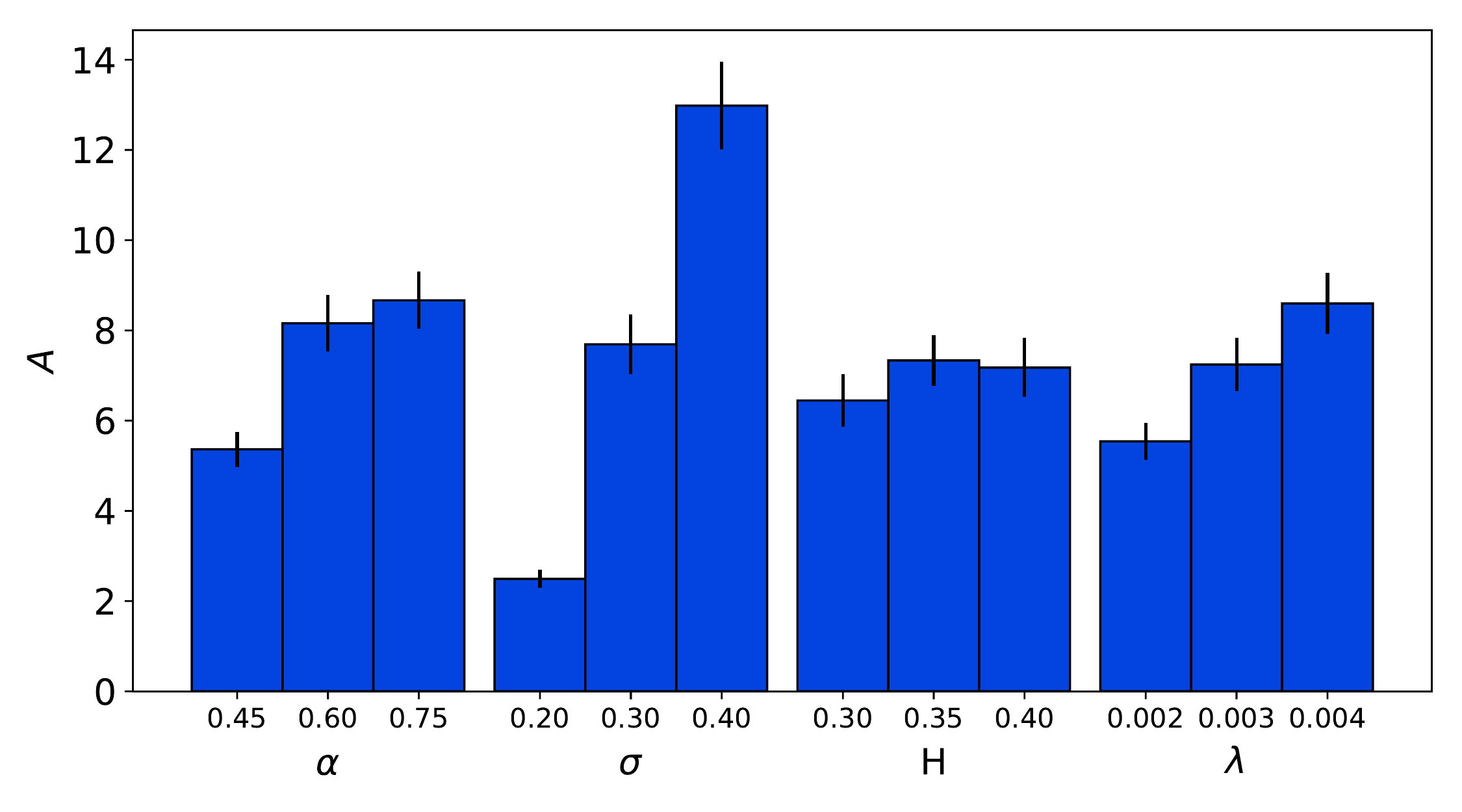}
  \caption{}\label{fig:c_influence}
  \end{subfigure}
  \vspace{-.5cm}
  \caption{(a) Increased anti-correlation between hops (decreased $H$) appears
	  to increase the length-dependence of solute flux (higher $\beta$) for
	  the sFBMcut model.  (b) If we remove anti-correlation, by setting
	  $H$=0.5, the length dependence parameter drops to a value near 2.
	  When we remove hop anti-correlation and dwell times between hops,
	  effectively simulating Brownian motion, the $\beta$ parameter stays
	  near 2.  This suggests that $\beta$ does not depend on dwell times
	  (parameterized by $\alpha$ and $\lambda$). (c) As solute flux
	  increases, the scaling of the flux curves, $A$, increases (compare
	  ranking with Figure~\ref{fig:flux_curves_ad}).  (d) Physical
	  processes which increase the rate of solute displacement result in
	  larger values of $A$. To test the dependence of $A$ on $\alpha$,
	  $\sigma$, $H$ and $\lambda$, we chose a single set of parameters,
	  representative of solutes parameterized by the sFBMcut AD model, and
	  generated realizations of the model by varying each parameter
	  independently about the same base parameter set. 
	  Decreased dwell times (increased $\alpha$), increased hop lengths (increased
	  $\sigma$), and a lower cut-off to the dwell time distribution
	  (increased $\lambda$) lead to increases in $A$. 
	  and non-linearly to $\alpha$.  The data suggests that the $A$
	  parameters do not depend on hop anti-correlation ($H$).
 	  }\label{fig:beta}
  \end{figure*}

  The scaling of solute flux with pore length is primarily influenced by
  anti-correlation between solute hops. In Figure~\ref{fig:beta}a, we show that
  $\beta$ is inversely related to the Hurst parameter. This makes intuitive
  sense since higher degrees of anti-correlation should slow the rate at which
  solutes cross a membrane pore of a given length. When we remove anti-correlation between
  hops (set $H$=0.5), the length dependence becomes the same for all solutes,
  dropping to a value just below 2. This also implies that hop lengths and
  dwell times do not affect length dependence since each solute exhibits
  different hopping and trapping behavior yet are all parameterized by the same
  value of $\beta$ when $H$=0.5. We further verified this claim by removing
  hop anti-correlation and setting all dwell times equal to one timestep,
  effectively simulating Brownian motion.  The length dependence remains
  unchanged. 

  Dwell times and hop lengths directly modify the rate at which solutes move 
  through the membrane pores and are reflected in the scaling pre-factor of
  the solute flux curves, $A$. Comparison of Figure~\ref{fig:c_parameters} with 
  Figure~\ref{fig:flux_curves_ad} reveals that the ranking of the $A$ parameters
  is consistent with the ranking of solute flux. In Figure~\ref{fig:c_influence} we
  demonstrate that decreasing dwell times (increasing $\alpha$), cutting off 
  the dwell time distribution at shorter times (increasing $\lambda$), and increasing 
  hop lengths (increasing $\sigma$) independently lead to an increase in $A$. 
  Figure~\ref{fig:c_influence} also suggests that $A$ is not dependent on hop 
  anti-correlation ($H$).
  
  The power law decay of the flux with pore length implies the following
  relationship for selectivity via substitution of
  Equation~\ref{eqn:flux_decay} into Equation~\ref{eqn:selectivity}:
  \begin{equation}
  S_{ij}(L) = \bigg(\frac{A_i}{A_j}\bigg)L^{(\beta_i - \beta_j)}
  \label{eqn:selectivity_ratio}
  \end{equation}

  \begin{figure}
  \centering
  \includegraphics[width=0.5\textwidth]{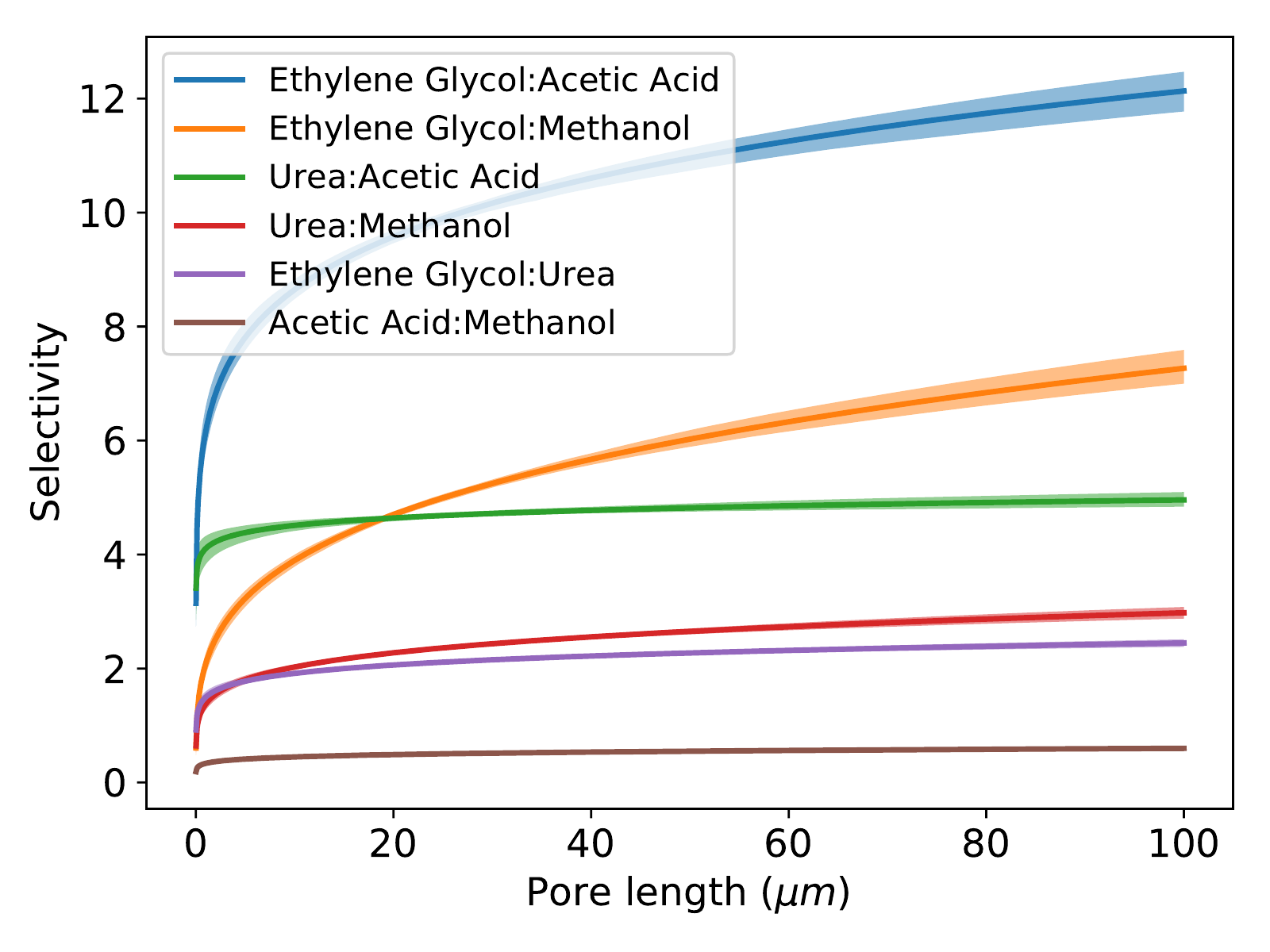}
  \caption{The selectivity between pairs of species changes monotonically with
	  pore length. The strength of dependence on pore length depends on
	  the difference between $\beta$ values. The largest differences in solute
	  flux result in high selectivities at any pore length. This membrane
	  may be a good candidate for the separation of ethylene glycol from
	  acetic acid. Ethylene glycol has the lowest $\beta$ value while
	  acetic acid has the second highest, leading to strong length
	  dependence. Ethylene glycol also has the highest flux and acetic acid
	  has the lowest resulting in relatively high selectivities independent
	  of pore length.}
  \label{fig:selectivity}
  \end{figure}
   
  In Figure~\ref{fig:selectivity}, we plot Equation~\ref{eqn:selectivity_ratio}
  for pore lengths ranging from those studied in
  Figure~\ref{fig:flux_curves_ad} to macroscopic-length pores. For the same
  degree of hop anti-correlation, as is the case for uncorrelated motion,
  selectivity depends on solute hop lengths and dwell times ($A$). For $(A_i /
  A_j)=1$, LLC membranes will be more selective towards passage of solutes with
  less anti-correlated hopping behavior (lower $\beta$). The selectivity
  towards solutes with less length dependent flux increases with membrane
  thickness. In most cases, the length dependence of selectivity plateaus near
  or within the range of experimentally-accessible membrane thicknesses.
  Therefore, from a practical standpoint, one should not expect significant
  changes in selectivity by varying LLC membrane thickness. 
  
  Of the solutes studied, the data suggests that this particular LLC membrane 
  might be most useful for selectively separating ethylene glycol from methanol
  and acetic acid. Relative to acetic acid, ethylene glycol takes larger hops 
  and is trapped longer, leading to a larger $A$. There is also an appreciable 
  difference in the scaling of each solute's flux with pore length ($\beta$) 
  which gives the selectivity significant length dependence. Relative to methanol,
  ethylene glycol takes similarly-sized hops and dwells which is why selectivity
  is relatively low for very short pore lengths. However, the large difference in
  hop correlation between the two solutes leads to the strongest pore length 
  dependence of selectivity. 
  
  The insights into flux and selectivity provided by our time series modeling
  approach could not be drawn easily by simply observing solute motion, the
  structure of the membrane nanopores, or even the solute trajectories
  extracted from the MD simulations. Differences in solute MSDs alone do not
  fully explain the trends and magnitudes of the selectivities shown in
  Figure~\ref{fig:selectivity}. Complex interplay between membrane
  constituents and solutes with varying chemical functionality lead to diverse
  solute behavior. Even if our model is not perfect, it provides clear logic
  behind the mechanisms leading to selective behavior which could significantly
  help illuminate any design choices. We hope this type of analysis can be 
  leveraged to explore new, interesting and complex separations problems.
  
  \section{Conclusions}
  
  We have tested two different mathematical frameworks for
  describing complex solute dynamics
  by applying them to an H\textsubscript{II} phase LLC membrane. The
  values obtained for the parameters when fitting the models to the time series
  data offer important mechanistic insight on the molecular details of
  transport. Subordinated fractional Brownian and L\'evy motion have a strong
  theoretical foundation in the anomalous diffusion literature. Our single mode
  AD model quantifies and allows comparison of the hopping and trapping behavior
  among solutes. A two mode model that describes dynamics based on whether a
  solute is in or out of the pore region allows us to break down individual
  solute motion into two distinct regimes and we showed that solute motion
  is clearly restricted while in the tail region. Our Markov state-dependent
  dynamical model uses explicitly defined trapping mechanisms and gives a nice
  description of transitions between these observed states, the equilibrium
  distribution of solutes among states as well as the type of stochastic
  behavior shown in each state. 
  
  Although large portions of the MSDs predicted by our models fall close to or
  within the 1$\sigma$ confidence intervals of MD, it is not always for
  physically accurate reasons. Qualitatively, MSDDM trajectories do not display
  the same hopping and trapping behavior shown by MD trajectories. Frequent
  transitions drawn from a single, relatively broad transition emission
  distribution prevent long periods of immobility. The most obvious solution to
  this would be to generate individual emission distributions for each type of
  transition, but this would require orders of magnitude more data or more
  prior assumptions about the distributions. This approach is further
  complicated by the need to make correlated draws from a fractional L\'evy
  process with a frequently changing distribution width. A possible
  simplification could be to assume that correlation is lost every time a state
  transition occurs.

  Although the AD approach generates qualitatively accurate trajectories and
  predicts MSDs near or within the 1$\sigma$ confidence interval of our MD
  simulations, the curvature of the predicted MSDs does not appear to be
  consistent with the MD simulations. The MSD curves calculated from MD
  simulations appear to straighten out on long timescales while those predicted
  from the AD models continuously curve. This is because pure fractional Brownian motion
  features hop correlation that persists indefinitely.
  However, it is likely that on longer timescales, hops become decorrelated, 
  causing solute dynamics to transition from subdiffusive to diffusive behavior.
  We may be able to incorporate this transition by truncating the positional 
  autocorrelation function, allowing correlation to diminish on the 100 ns 
  time scale as suggested by the physical trajectories.

  We demonstrated how one could use the one mode AD model,
  or any stochastic time series model,
  in order to to
  determine macroscopic flux and selectivity. We showed that, when using the AD
  model, solute flux decreases with pore length at a rate faster than pure
  Brownian motion due to anti-correlation between hops.
  Due to differences in hop anti-correlation, we observe length dependent
  selectivity. Based on these calculations we can hypothesize that this
  particular LLC membrane may be a good candidate for the selective separation
  of ethylene glycol from acetic acid or methanol.

  
  In a broader context, it is clear that time series modeling may be a useful way
  to form a clear connection between complex solute dynamics on the nanosecond 
  timescale and macroscopic observables. On the nanoscopic scale, one can both 
  learn and characterize dynamical modes in order to identify areas of potentially
  impactful molecular-level design changes. For example, it may be possible to 
  modify solute transport rates and selectivities in LLC membranes by redesigning 
  the LLC monomers to mitigate or enhance their chemically dependent interactions
  with different solutes. With these models, dynamics can be propagated onto much 
  longer time with computational ease. This allows one not only to predict a 
  macroscopic observable for a specific system, but to explicitly calculate the 
  effect of perturbations to model parameters that may result from a monomer 
  design change. Overall, we hope that this work can advance the reader's 
  understanding of how one can apply time series analysis in order to understand 
  complex diffusive behavior in any system of interest.
  
%

  \section*{Acknowledgments}
  
  We thank Richard Noble for helping us make the connection between the empirical 
  mean first passage time distribution and the analytical equation to which we fit
  the distributions (Equation~\ref{eqn:passage_times}).
  
  This work was supported in part by the ACS Petroleum Research Fund grant
  \#59814-ND7 and the Graduate Assistance in Areas of National Need (GAANN)
  fellowship which is funded by the U.S. Department of Education.  Molecular
  simulations were performed using the Extreme Science and Engineering
  Discovery Environment (XSEDE), which is supported by National Science
  Foundation grant number ACI-1548562. Specifically, it used the Bridges
  system, which is supported by NSF award number ACI-1445606, at the Pittsburgh
  Supercomputing Center (PSC). This work also utilized the RMACC Summit
  supercomputer, which is supported by the National Science Foundation (awards
  ACI-1532235 and ACI-1532236), the University of Colorado Boulder, and
  Colorado State University. The Summit supercomputer is a joint effort of the
  University of Colorado Boulder and Colorado State University.
  
  \section*{Supporting Information}
  
  \input si_arxiv_v2.tex
 
  \clearpage
  \bibliography{stochastic_transport}

%

\end{document}

%% file: si_arxiv_v2.tex
\begin{appendix}
  
  \section{Setup and analysis scripts}\label{S-section:python_scripts}

  \begin{table*}[!htb]
  \centering
  \newcolumntype{A}{ >{\centering\arraybackslash} m{2.5in} }
  \newcolumntype{B}{ >{\centering\arraybackslash} m{0.5in} }
  \newcolumntype{C}{m{3in}}
  \begin{tabular}{ABC}
  \hline
  \hline
  \textbf{Script Name} & \textbf{Section} & ~~~~~~~~~~~~~~~~~~~~~\textbf{Description} \\
  \hline

  \texttt{/setup/parameterize.py}            & 2.1 & Parameterize liquid crystal monomers and solutes with GAFF \\
  \texttt{/setup/build.py}                   & 2.1 & Build simulation unit cell \\
  \texttt{/setup/place\_solutes\_pores.py}   & 2.1 & Place equispaced solutes in the pore centers of a unit cell \\
  \texttt{/setup/equil.py}                   & 2.1 & Equilibrate unit cell and run production simulation \\
  \texttt{/analysis/solute\_partitioning.py} & 2.1 & Determine time evolution of partition of solutes between pores and tails \\
  \texttt{/timeseries/msd.py}                & 2.2 & Calculate the mean squared displacement of solutes \\
  \texttt{/analysis/sfbm\_parameters.py}     & 2.2 & Get subordinated fractional Brownian motion parameters by fitting to a solute's dwell and hop length distributions and positional autocorrelation function. \\
  \texttt{/timeseries/ctrwsim.py}            & 2.2 & Generate realizations of a continuous time random walk with the user's choice of dwell and hop distributions \\
  \texttt{/timeseries/forecast\_ctrw.py}     & 2.2 & Combines classes from \texttt{sfbm\_parameters.py} and \texttt{ctrwsim.py} to parameterize and predict MSD in one shot. \\
  \texttt{/analysis/ Markov\_state\_dependent\_dynamics.py} & 2.3 & Identify frame-by-frame state of each solute, construct a transition matrix and simulate realizations of the MSDDM model. \\
  \texttt{/timeseries/mfpt\_pore.py}          & 2.4 & Simulate mean first passage time distributions using the AD approach. \\
  \hline
  \hline
  \end{tabular}

  \caption{The first column provides the names of the python scripts available in
  the \texttt{LLC\_Membranes} GitHub repository that were used for system setup and
  post-simulation trajectory analysis. Paths preceding script names are relative to the
  \texttt{LLC\_Membranes/LLC\_Membranes} directory. The third column
  gives a brief description of the purpose of each script.
  }~\label{S-table:python_scripts}
  \end{table*}

  All python and bash scripts used to set up systems and conduct post-simulation trajectory
  analysis are available online at \\
  \texttt{https://github.com/shirtsgroup/LLC\_Membranes}.
  Documentation for the \texttt{LLC\_Membranes} repository is available at \\
  \texttt{https://llc-membranes.readthedocs.io/en/latest/}. Table~\ref{S-table:python_scripts}
  provides more detail about specific scripts used for each type of analysis performed in
  the main text.
  
  \section{Solute Equilibration}\label{S-section:equilibration}
  
\begin{figure*}[hb]
  \centering
  \begin{subfigure}{0.45\textwidth}
  \includegraphics[width=\textwidth]{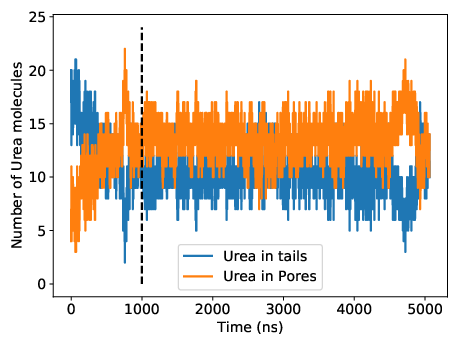}
  \caption{}\label{S-fig:URE_equilibration}
  \end{subfigure}
  \begin{subfigure}{0.45\textwidth}
  \includegraphics[width=\textwidth]{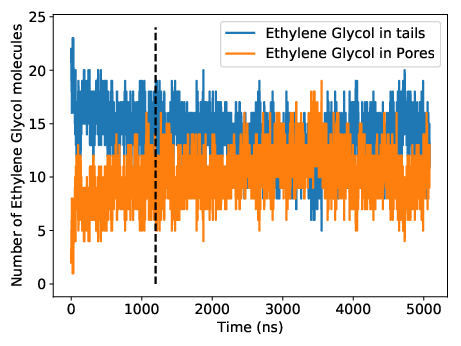}
  \caption{}\label{S-fig:GCL_equilibration}
  \end{subfigure}
  \begin{subfigure}{0.45\textwidth}
  \includegraphics[width=\textwidth]{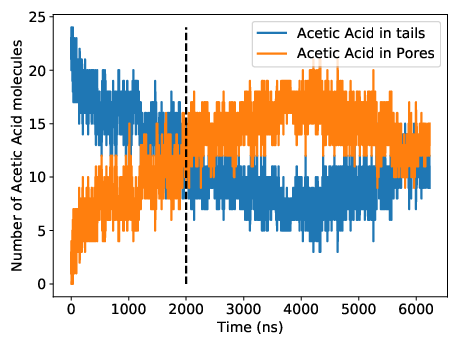}
  \caption{}\label{S-fig:ACH_equilibration}
  \end{subfigure}
  \begin{subfigure}{0.45\textwidth}
  \includegraphics[width=\textwidth]{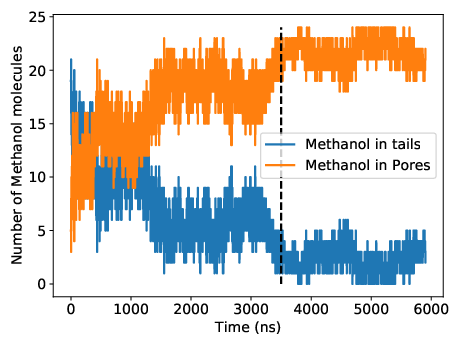}
  \caption{}\label{S-fig:MET_equilibration}
  \end{subfigure}
  \caption{We considered a system to be equilibrated when the partition of solutes
  between the tails and pore plateaued. Our chosen equilibration point for each
  solute is indicated by the vertical black dashed line. (a) Urea equilibrates
  the fastest, after 1000 ns. (b) Ethylene glycol equilibrates after 1200 ns 
  (c) The partition of acetic acid appears oscillate slowly. We considered it to be
  equilibrated after 2000 ns. (d) We considered methanol to be equilibrated after
  3500 ns. Methanol nearly completely partitions into the tails.}\label{S-fig:equilibration}
  \end{figure*}

  We collected all data used for model generation after the solutes were 
  equilibrated. We assumed a solute to be equilibrated when the partition of
  solutes in and out of the pore region stopped changing. The pore region is
  defined as within 0.75 nm of the pore center. We plot the partition
  versus time in Figure~\ref{S-fig:equilibration} and indicated the chosen
  equilibration time points. 
  
  
  \section{Mean Squared Displacement}\label{S-section:msd}
  
  In this study, we primarily use the MSD as a tool for characterizing the average
  dynamic behavior of solute trajectories. Rather than using them to calculate 
  diffusion constants or to relate our simulations to experimental measurements, we
  compare MSDs calculated from MD simulations to those generated from our models 
  in order to validate those models. Therefore, it is only important that we use a
  consistent definition for calculating the MSD between modeled trajectories and
  directly observed MD trajectories.

  One can measure MSD in two ways. The ensemble averaged MSD measures 
  displacements with respect to a particle's initial position:
  \begin{equation}
  \langle z^2(t) \rangle = \langle z(t) - z(0) \rangle^2
  \label{S-eqn:ensemble_msd}
  \end{equation}
  Fits to the ensemble averaged MSD will always reproduce the form of 
  Equation~\ref{eqn:msd_form} of the main text. The time-averaged MSD measures all observed 
  displacements over time lag $\tau$: 
  \begin{equation}
  \overline{z^2(\tau)} = \dfrac{1}{T - \tau}\int_{0}^{T - \tau} (z(t + \tau) - z(t))^2 dt
  \label{S-eqn:tamsd}
  \end{equation}
  where T is the length of the trajectory. 
  
  The time averaged and ensemble averaged MSDs will give identical results 
  unless a system displays non-ergodic behavior. For a pure CTRW, the power 
  law distribution of trapping times leads to weak ergodicity breaking.
  In this case, the time-averaged MSD is linear while the ensemble averaged
  MSD has the form of Equation~\ref{eqn:msd_form} of the main text.~\cite{meroz_toolbox_2015} With
  power law trapping behavior, the time between hops diverges so there is no 
  characteristic measurement time scale of solute motion. In fact, as measurement 
  time increases, the average MSD of a CTRW tends to decrease, a phenomenon called
  aging, because trajectories with trapping times on the order of the measurement 
  time get incorporated into the calculation.~\cite{bel_weak_2005}
  
  We chose to use just the time-averaged MSD to compare MD trajectories with
  modeled trajectories, because, compared to the ensemble average, it is a more
  statistically robust measure of the average distance a solute travels over
  time. The ensemble MSD of only 24 solute trajectories would have much higher
  uncertainties.
  
  \section{Estimating the Hurst Parameter}\label{S-section:H_estimate}
  
  We chose to estimate the Hurst parameter, $H$ by a least squares fit to the analytical
  autocorrelation function for fractional Brownian motion (the variance-normalized version 
  of Equation~\ref{eqn:fbm_autocorrelation} in the main text):
  
  \begin{equation}
    \gamma(k) = \dfrac{1}{2}\bigg[|k-1|^{2H} - 2|k|^{2H} + |k+1|^{2H}\bigg]
  \label{S-eqn:fbm_autocorrelation}
  \end{equation}  
  
  There are many methods for estimating the Hurst parameter for a time series.
  \cite{clegg_practical_2006} It can be a difficult task because 
  Equation~\ref{S-eqn:fbm_autocorrelation} decays slowly to zero, especially when 
  $H > 0.5$, meaning one needs to study large time lags with high frequency.
  
\begin{figure*}
  \centering
  \begin{subfigure}{0.45\textwidth}
  \includegraphics[width=\textwidth]{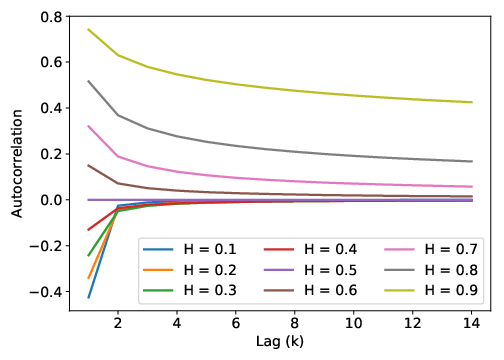}
  \caption{}\label{S-fig:hurst_autocorrelation}
  \end{subfigure}
  \begin{subfigure}{0.45\textwidth}
  \includegraphics[width=\textwidth]{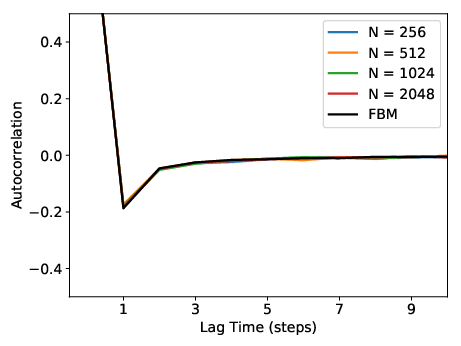}
  \caption{}\label{S-fig:flm_autocorrelation}
  \end{subfigure}  
  \caption{(a) The analytical autocorrelation function of FBM decays to zero faster
  when H $<$ 0.5 compared to when H $>$ 0.5. (b) The autocorrelation function of an 
  FLM process does not change with increasing sequence length (N). It shares the same
  autocorrelation function as fractional Brownian motion (FBM). Note that all lines
  plotted lie on top of each other. All sequences used to make this plot were generated
  using $H$=0.35 and, for FLM, $\alpha$=1.4.}\label{S-fig:hurst_parameters}
  \end{figure*}

  Fortunately, from a mathematical perspective, all of our solutes show anti-correlated
  motion, so most of the information in Equation~\ref{S-eqn:fbm_autocorrelation} is contained
  within the first few time lags. In Figure~\ref{S-fig:hurst_autocorrelation}, we plotted
  Equation~\ref{S-eqn:fbm_autocorrelation} for different values of $H$. When $H > 0.5$, 
  Equation~\ref{S-eqn:fbm_autocorrelation} decays slowly to zero meaning one needs to study
  large time lags with high frequency in order to accurately estimate $H$ from the data. 
  However, when $H < 0.5$, the autocorrelation function quickly decays towards zero.

  The autocovariance function of fractional L\'evy motion is different from fractional
  Brownian motion (see Equations~\ref{eqn:fbm_autocorrelation}
  and~\ref{eqn:flm_autocovariance} of the main text), but their autocorrelation 
  structures are the same. The autocovariance function of FLM is dependent on the 
  expected value of squared draws from the underlying L\'evy distribution, $E\big[L(1)^2\big]$. 
  This is effectively the distribution's variance, which is undefined for most 
  L\'evy stable distributions due to their heavy tails. As a consequence, one should 
  expect $E\big[L(1)^2\big]$ to grow as more samples are drawn from the distribution
  with the autocovariance function responding accordingly. 
  However, we are only interested in the autocorrelation function. In order to predict
  the Hurst parameter from the autocorrelation function, we must show that it has 
  a well-defined structure and is independent of the coefficient in 
  Equation~\ref{eqn:flm_autocovariance} of the main text. In Figure~\ref{S-fig:flm_autocorrelation}, 
  we plot the average autocorrelation function from an FLM process with an increasing 
  number of observations per generated sequence. For all simulations we set $H$=0.35 
  and $\alpha$=1.4. The variance-normalized autocovariance function, i.e. the autocorrelation
  function, does not change with increasing sequence length. Additionally, the
  autocorrelation function of FBM, with the same $H$, is the same. Therefore we are 
  confident that we can use the same Hurst parameter as an input to both FBM and FLM
  simulations.
  
  \section{Simulating Fractional L\'evy Motion}\label{S-section:sFLM}
  
  \subsection{Truncated L\'evy stable hop distributions}\label{S-section:truncation}
  
\begin{figure*}
  \centering
  \includegraphics[width=0.75\textwidth]{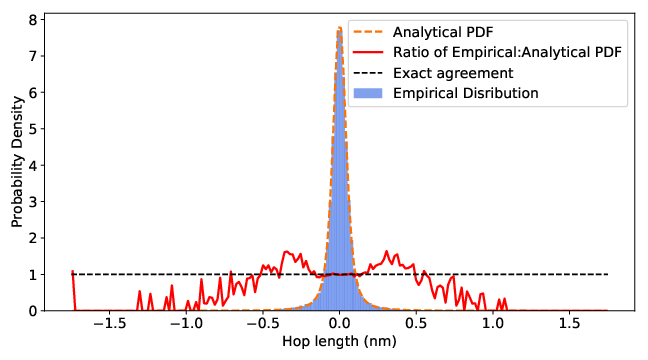}
  \caption{The ratio between the empirical and maximum likelihood theoretical
  distribution quantifies the quality of fit as function of hop length. The fit
  is near-perfect close to the mean. Intermediate hop lengths are over sampled, 
  and the tails are undersampled. We used this type of plot to determine the
  appropriate place to truncate the L\'evy stable distributions.}\label{S-fig:truncation_cutoff}
  \end{figure*}

  \textit{Determining where to truncate the hop distribution:} A pure
  L\'evy stable distribution has heavy tails which can lead to arbitrarily
  long hop lengths. Our distribution of hop lengths fits well to a L\'evy
  distribution near the mean, but under-samples the tails. In 
  Figure~\ref{S-fig:truncation_cutoff} we compare the empirically 
  measured transition emission distribution of the MSDDM for urea to its maximum likelihood fit to
  a L\'evy stable distribution. The ratio between the two distributions 
  at each bin is nearly 1 close to the center, indicating a near-perfect
  fit, larger than 1 slightly further from the center, suggesting that 
  we slightly over sample intermediate hop lengths, and below 1 far from 
  the center, indicating undersampling of extremely long hop lengths.
  Based on the plot, we chose a cut-off of 1 nm in order to compensate for
  over sampled intermediate hop lengths. We chose the same cut-off for all
  solutes.
  
  \textit{Generating FLM realizations from a truncated L\'evy distribution:}
  To generate realizations from an uncorrelated truncated L\'evy process, one would
  randomly sample from the base distribution and replace values that are too large
  with new random samples from the base distribution, repeating the process until
  all samples are under the desired cut-off. 
  
  This procedure is complicated by the correlation structure of FLM. At a high level,
  Stoev and Taqqu use Riemann-sum approximations of the stochastic integrals defining
  FLM in order to generate realizations.~\cite{stoev_simulation_2004} They do this efficiently with the help of 
  Fast Fourier Transforms. In practice, this requires one to Fourier transform a zero-padded
  vector of random samples drawn from the appropriate L\'evy stable distribution, multiply
  the vector in Fourier space by a kernel function and invert back to real space. The end
  result is a correlated vector of fractional L\'evy noise.
  
\begin{figure*}
  \centering
  \begin{subfigure}{0.45\textwidth}
  \includegraphics[width=\textwidth]{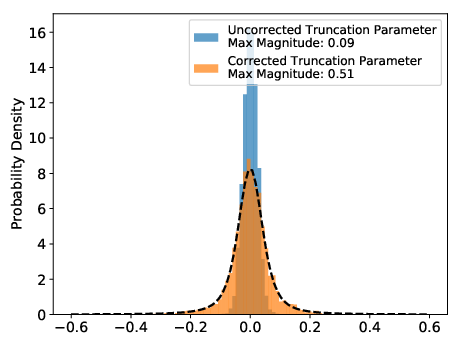}
  \caption{}\label{S-fig:truncation_correction}
  \end{subfigure}
  \begin{subfigure}{0.45\textwidth}
  \includegraphics[width=\textwidth]{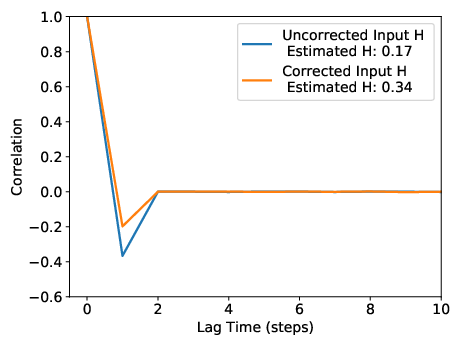}
  \caption{}\label{S-fig:hurst_correction}
  \end{subfigure}
  \caption{(a) We can accurately truncate the marginal distribution of FLM innovations by
  applying a correction to the input truncation parameter. We generated FLM sequences
  and truncated the initial L\'evy stable distribution (before Fourier transforming) at 
  a value of 0.5. After correlation structure is added, the width of the distribution 
  of fractional L\'evy noise decreases significantly. We corrected the input truncation
  parameter with our database resulting in a distribution close to the theoretical
  distribution (black dashed line) with a maximum value close to 0.5.
  (b) Correcting the Hurst parameter input to the algorithm of Stoev and Taqqu
  results in an FLM process with a more accurate correlation structure. We generated
  sequences with an input $H$ of 0.35. We estimated $H$ by fitting the autocorrelation
  function. Without the correction, $H$ is underestimated, meaning realizations are 
  more negatively correlated than they should be. }
  \end{figure*}

  We are unaware of a technique for simulating truncated FLM, therefore we devised our
  own based on the above discussion. If one is to truncate an FLM process, one can apply
  the simple procedure above for drawing uncorrelated values from the marginal L\'evy 
  stable distribution, \textit{but}, after adding correlation, the maximum drawn value is typically lower than the limit set 
  by the user. Additionally, the shape of the distribution itself changes. Therefore, 
  we created a database meant to correct the input truncation parameter (the maximum desired draw). The database
  returns the value of the truncation parameter that will properly truncate the
  output marginal distribution based on $H$, $\alpha$ and $\sigma$ (the width parameter).
  Figure~\ref{S-fig:truncation_correction} shows the result of applying our correction.
  Note that generating this database requires a significant amount of simulation and
  still likely doesn't perfectly correct the truncation parameter. The output leads
  to a somewhat fuzzy, rather than abrupt, cut-off of the output distribution. This 
  is likely beneficial since we observe a small proportion of hops longer the chosen
  truncation cut-off. When the cut-off value is close to the L\'evy stable $\sigma$ parameter,
  as it is in our anomalous diffusion models, we observed that the tails of the 
  truncated distribution tend to be undersampled. In order to maintain the distribution's 
  approximate shape up to the cut-off value we recommend ensuring that the cut-off
  value is at least 2 times $\sigma$. However, this may lead to a slight over-prediction
  of the MSD.
  
  \subsection{Achieving the right correlation structure}\label{S-section:flm_correlation}
  
  We simulated FLM using the algorithm of Stoev and Taqqu~\cite{stoev_simulation_2004}.
  There are no known exact methods for simulating FLM. As a consequence, passing a
  value of $H$ and $\alpha$ to the algorithm does not necessarily result in the correct
  correlation structure, although the marginal L\'evy stable distribution is correct. 
  We applied a database-based empirical correction in order to use the
  algorithm to achieve the correct marginal distribution and correlation structure.
  
  Stoev and Taqqu note that the transition between negatively and positively correlated
  draws occurs when $H = 1/ \alpha$. When $\alpha=2$, the marginal distribution is 
  Gaussian and the transition occurs at $H=0.5$ as expected from FBM. We corrected 
  the input $H$ so that the value of $H$ measured based on the output sequence equaled
  the desired $H$. We first adjusted the value of $H$ by adding ($1 / \alpha - 0.5$),
  effectively recentering the correlation sign transition for any value of $1 \leq \alpha \leq 2$.
  This correction alone does a good job for input $H$ values near 0.5, but is
  insufficient if one desires a low value of $H$. The exact correction to $H$ is 
  not obvious so we created a database of output $H$ values tabulated as a function
  of input $H$ and $\alpha$ values. Figure~\ref{S-fig:hurst_correction} demonstrates the
  results of applying our correction. Without the correction, FLM realizations are
  more negatively correlated. This would result in under-predicted mean squared
  displacements when applying the model.
  
  
  \section{Verifying Markovianity}\label{S-section:markov_validation}
  
\begin{figure*}
  \centering
  \begin{subfigure}{\textwidth}
  \includegraphics[width=\textwidth]{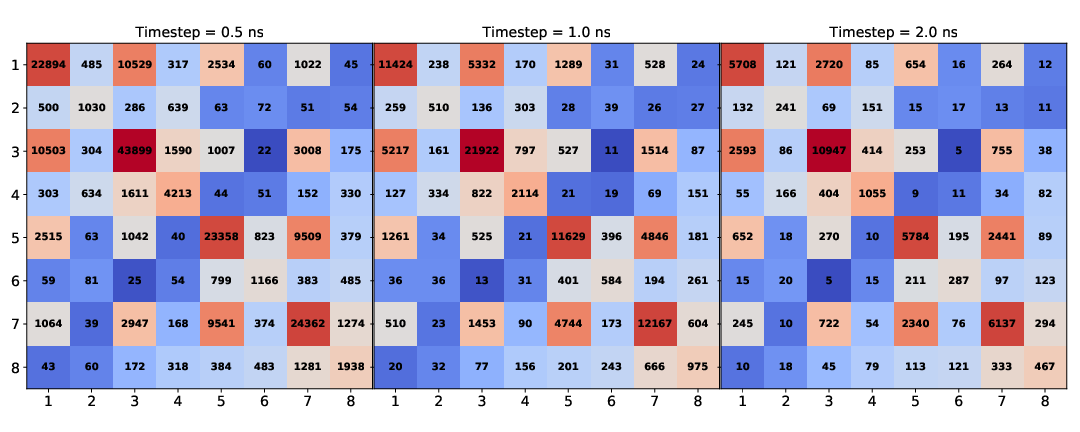}
  \caption{Urea}\label{S-fig:URE_counts}
  \end{subfigure}
  \begin{subfigure}{\textwidth}
  \includegraphics[width=\textwidth]{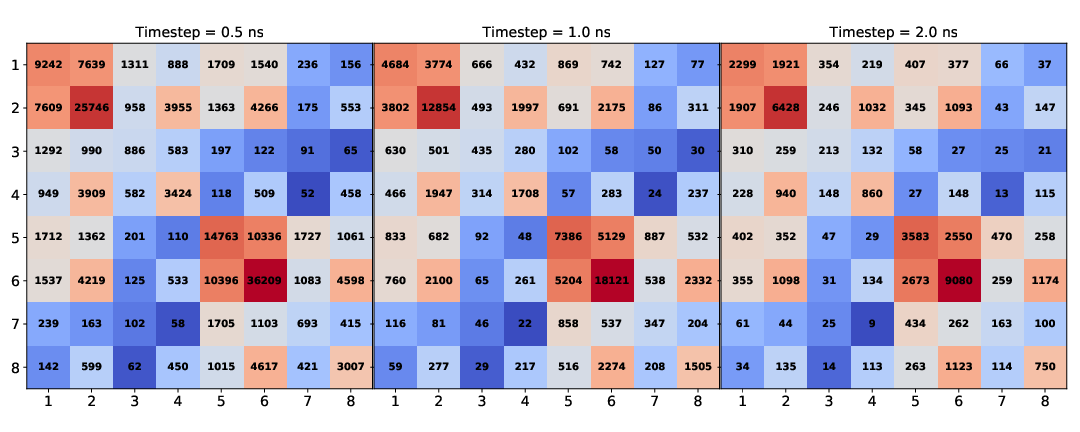}
  \caption{Ethylene Glycol}\label{S-fig:GCL_counts}
  \end{subfigure}
  \caption{The number of transitions from state $i$ to $j$ and $j$ to $i$ are very close
  indicating that our process obeys detailed balance. Detailed balance is conserved for 
  different sized time steps.}\label{S-fig:counts}
  \end{figure*}

  We verified the Markovianity of our transition matrix, $T$, in two ways. First we 
  ensured that the process satisfied detailed balance:
  \begin{equation}
  T_{i,j}P_i(t=\infty) = T_{j,i}P_j(t=\infty)
  \end{equation}
  where $P$ is the equilibrium distribution of states. This implies that the number
  of transitions from state $i$ to $j$ and from state $j$ to $i$ should be equal. Graphical 
  representations of the count matrices show that this is true in Figure~\ref{S-fig:counts}. 
  
\begin{figure*}
  \centering
  \begin{subfigure}{\textwidth}
  \includegraphics[width=\textwidth]{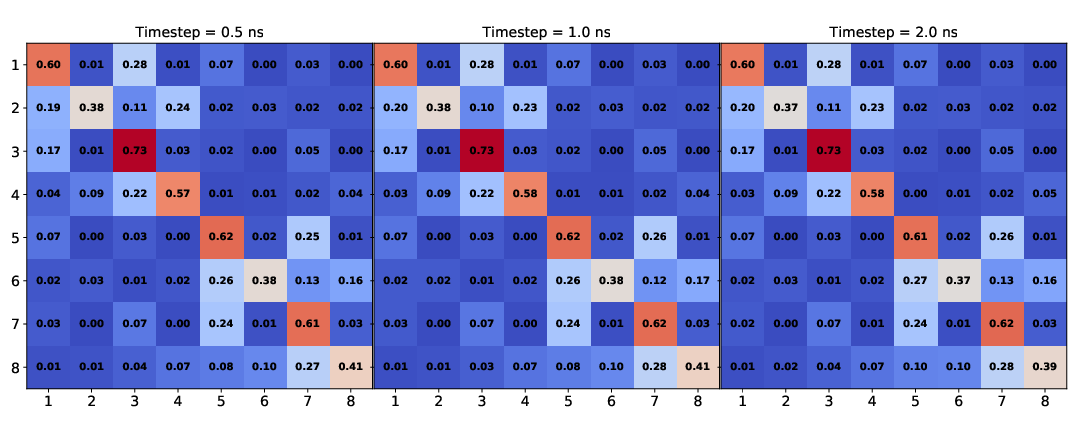}
  \caption{Urea}\label{S-fig:URE_transitions}
  \end{subfigure}
  \begin{subfigure}{\textwidth}
  \includegraphics[width=\textwidth]{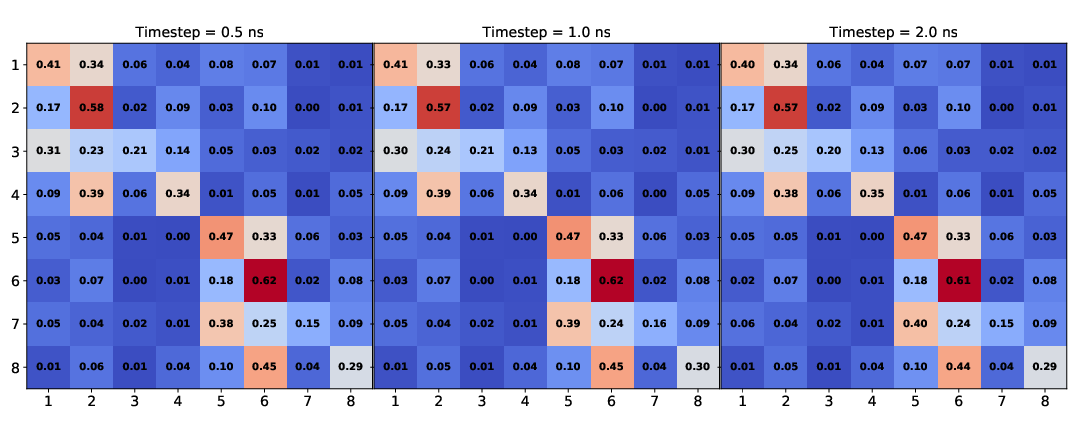}
  \caption{Ethylene Glycol}\label{S-fig:GCL_transitions}
  \end{subfigure}
  \caption{As the timestep between observations increases, the probability
  transition matrix does not change significantly.}\label{S-fig:transitions}
  \end{figure*}

  Second, we ensured that the transition matrix did not change on coarser time scales.
  In Figures~\ref{S-fig:counts} and~\ref{S-fig:transitions}, we show that increasing the 
  length of time between samples does not change the properties of the count or
  probability transition matrices.
  
  \section{Derivation of Passage Time Distributions}\label{S-section:fpt_derivation}
  
  To derive an analytical equation for the mean first passage time 
  (Equation~\ref{eqn:passage_times} of the main text), first consider an 
  initial pulse spreading out over time with a fixed mean. We can solve for the 
  time-dependent probability density of particle positions, $p$, by solving the one
  dimensional diffusion equation:
  \begin{equation}
  \frac{\partial p}{\partial t} = D \frac{\partial^2 p}{\partial z^2}
  \end{equation}
  The appropriate initial and boundary conditions are:
  $$BC1: t > 0, z = \infty, p = 0$$
  $$BC2: t > 0, z = 0, \frac{\partial p}{\partial z} = 0$$
  $$IC: t = 0, c = \delta(z)$$ 
  It has been shown 
  elsewhere that the solution to this equation is:~\cite{cussler_diffusion:_2009}
  \begin{equation}
  p(z, t) = \frac{1}{\sqrt{4 \pi D t}}\exp\bigg(\frac{-z^2}{4Dt}\bigg)
  \end{equation}
  We can make the substitution $z = z - vt$, where $v$ represents a constant average
  velocity, in order to linearly shift the mean as a function of time:
  \begin{equation}
  p(z, t) = \frac{1}{\sqrt{4 \pi D t}}\exp\bigg(\frac{-(z - vt)^2}{4Dt}\bigg)
  \end{equation}
  One can track the fraction of particles, $F$, that have crossed the pore 
  boundary by integrating:
  \begin{equation}
  F(t) = \int_L^\infty p~dz = \mathrm{erfc}\bigg(\frac{L - vt}{2\sqrt{D t}}\bigg)
  \end{equation}
  where $L$ is the pore length. This represents the cumulative first passage 
  time distribution so we take its derivative in order to arrive at the first
  passage time distribution:
  \begin{equation}
  P(t) = -\frac{1}{\sqrt{\pi}}e^{-(L - vt)^2 / (4Dt)}\bigg(-\frac{D(L - vt)}{4(Dt)^{3/2}} - \frac{v}{2\sqrt{Dt}}\bigg)
  \label{S-eqn:passage_times}
  \end{equation} 
  where the only free parameters for fitting are $v$ and $D$. We calculated the
  expected value of Equation~\ref{S-eqn:passage_times} in order to get the MFPT. Specifically,
  we used the python package \texttt{scipy.integrate.quad} to numerically integrate:
  \begin{equation}
  E[t] = \int_0^\infty tP(t) dt
  \end{equation}
  
  \section{Solute hopping and trapping behavior}\label{S-section:sfbm_other_solutes}
  
\begin{figure*}[htb!]
  \centering
  \begin{subfigure}{0.3\textwidth}
  \includegraphics[width=\textwidth]{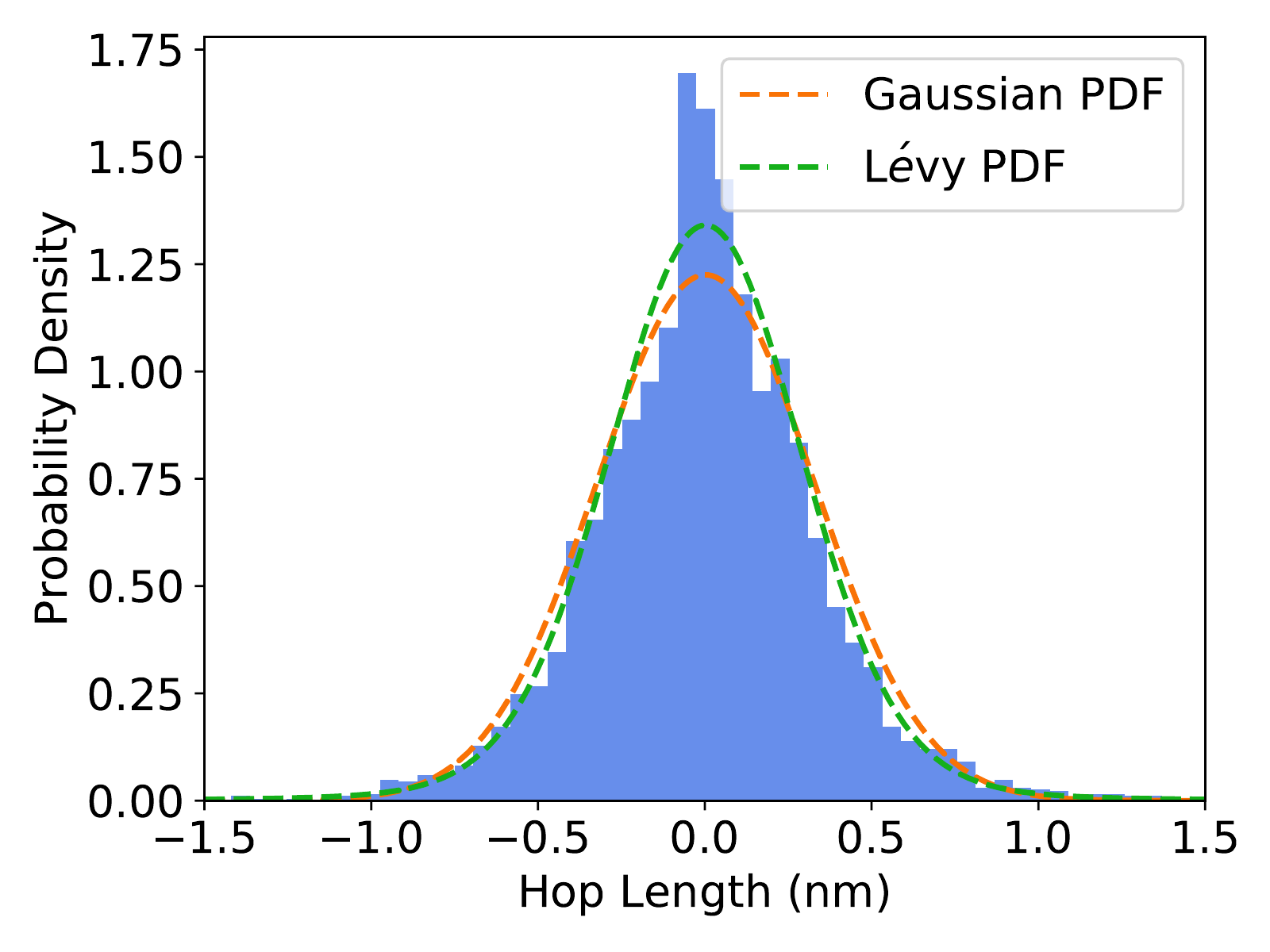}
  \caption{}\label{S-fig:URE_hop_distribution_comparison}
  \end{subfigure}
  \begin{subfigure}{0.3\textwidth}
  \includegraphics[width=\textwidth]{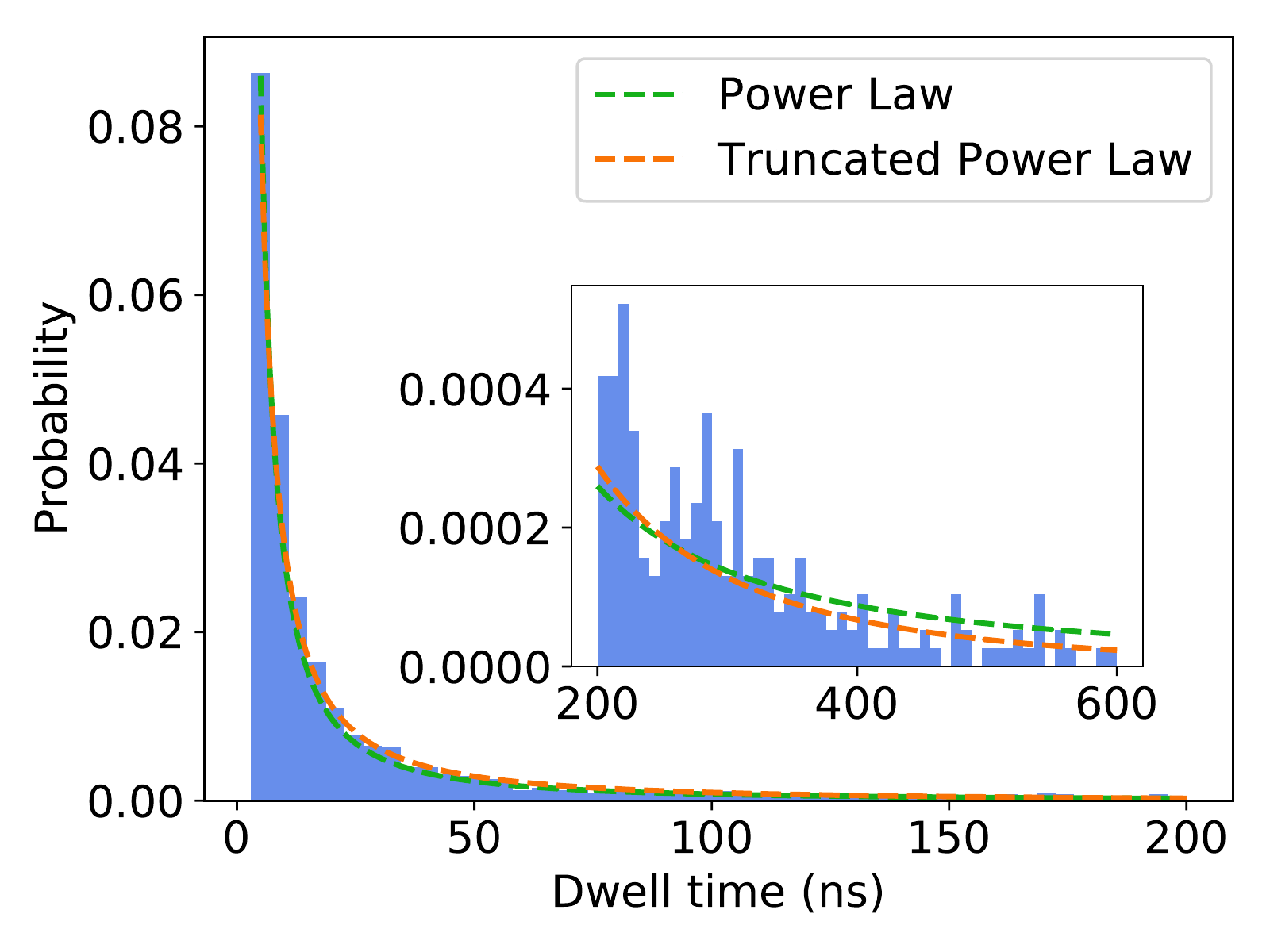}
  \caption{}\label{S-fig:URE_powerlaw}
  \end{subfigure}
  \begin{subfigure}{0.3\textwidth}
  \includegraphics[width=\textwidth]{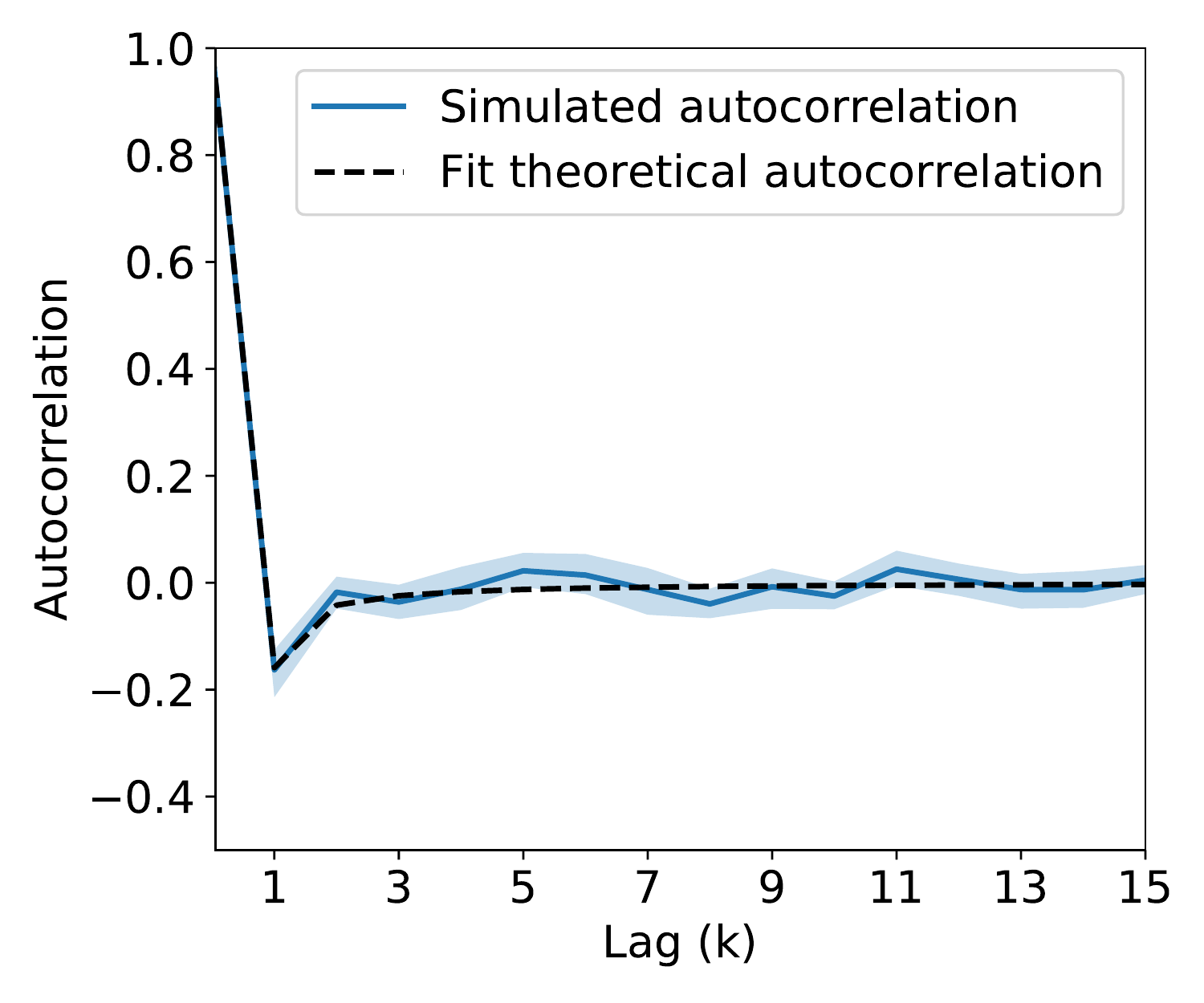}
  \caption{}\label{S-fig:URE_hop_acf}
  \end{subfigure}
  \begin{subfigure}{0.3\textwidth}
  \includegraphics[width=\textwidth]{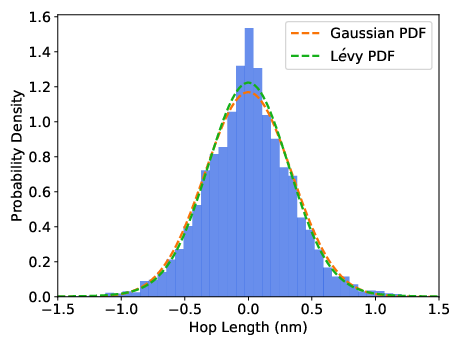}
  \caption{}\label{S-fig:GCL_hop_distribution_comparison}
  \end{subfigure}
  \begin{subfigure}{0.3\textwidth}
  \includegraphics[width=\textwidth]{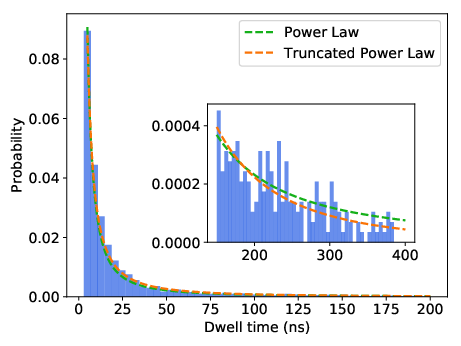}
  \caption{}\label{S-fig:GCL_powerlaw}
  \end{subfigure}
  \begin{subfigure}{0.3\textwidth}
  \includegraphics[width=\textwidth]{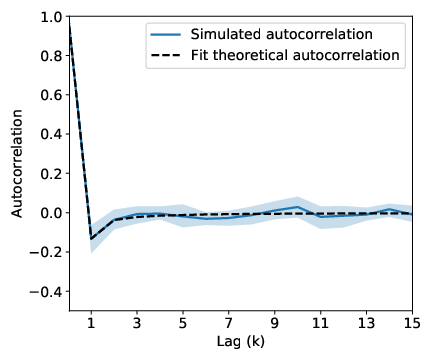}
  \caption{}\label{S-fig:GCL_hop_acf}
  \end{subfigure}
  \begin{subfigure}{0.3\textwidth}
  \includegraphics[width=\textwidth]{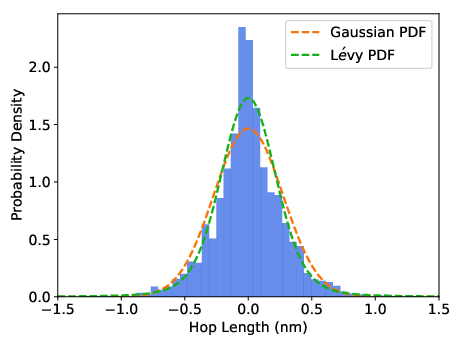}
  \caption{}\label{S-fig:ACH_hop_distribution_comparison}
  \end{subfigure}
  \begin{subfigure}{0.3\textwidth}
  \includegraphics[width=\textwidth]{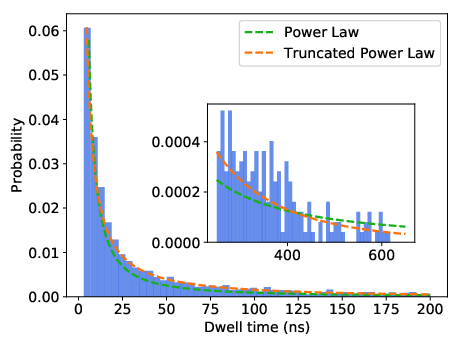}
  \caption{}\label{S-fig:ACH_powerlaw}
  \end{subfigure}
  \begin{subfigure}{0.3\textwidth}
  \includegraphics[width=\textwidth]{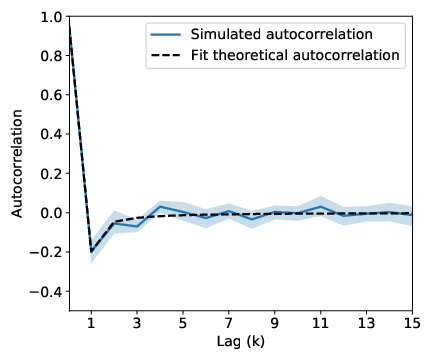}
  \caption{}\label{S-fig:ACH_hop_acf}
  \end{subfigure}
  \begin{subfigure}{0.3\textwidth}
  \includegraphics[width=\textwidth]{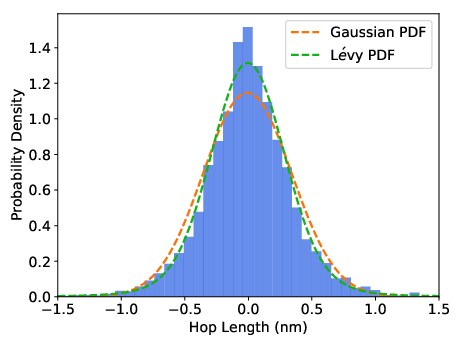}
  \caption{}\label{S-fig:MET_hop_distribution_comparison}
  \end{subfigure}
  \begin{subfigure}{0.3\textwidth}
  \includegraphics[width=\textwidth]{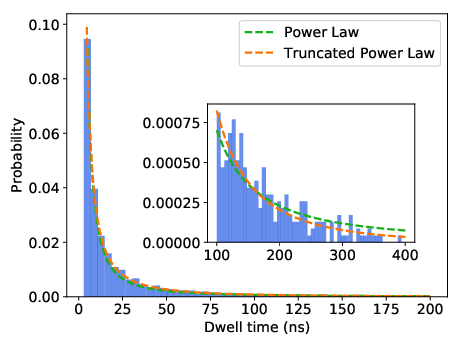}
  \caption{}\label{S-fig:MET_powerlaw}
  \end{subfigure}
  \begin{subfigure}{0.3\textwidth}
  \includegraphics[width=\textwidth]{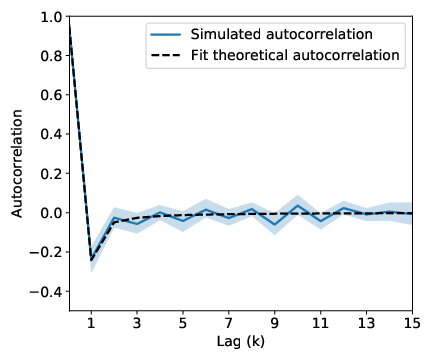}
  \caption{}\label{S-fig:MET_hop_acf}
  \end{subfigure}
  \caption{Hop length distributions, dwell time distributions and hop autocorrelation functions
  respectively for urea (a-c), ethylene glycol (d-f), acetic acid (g-i), and methanol (j-l).
  }\label{S-fig:anticorrelated_hops}
  \end{figure*}

  Figure~\ref{S-fig:anticorrelated_hops} demonstrates that all solutes exhibit the
  same kind of anti-correlated hopping and trapping behavior.
  
  \section{AD model MSD Predictions with Pure Power Law Dwell Times}\label{S-section:pure_power_law}
  
\begin{figure*}[h]
  \centering
  \begin{subfigure}{0.45\textwidth}
  \includegraphics[width=\textwidth]{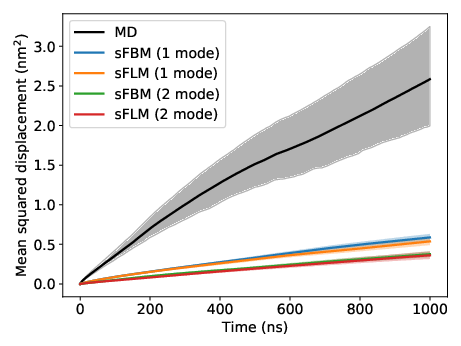}
  \caption{Urea}\label{S-fig:bothmode_msd_comparison_URE}
  \end{subfigure}
  \begin{subfigure}{0.45\textwidth}
  \includegraphics[width=\textwidth]{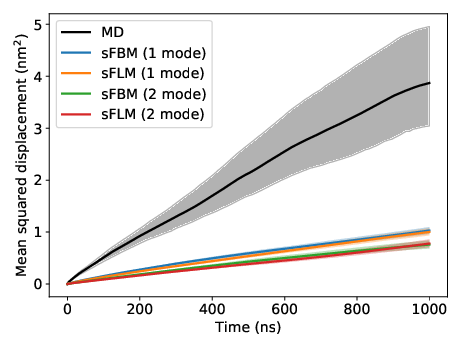}
  \caption{Ethylene Glycol}\label{S-bothmode_msd_comparison_GCL}
  \end{subfigure}
  \begin{subfigure}{0.45\textwidth}
  \includegraphics[width=\textwidth]{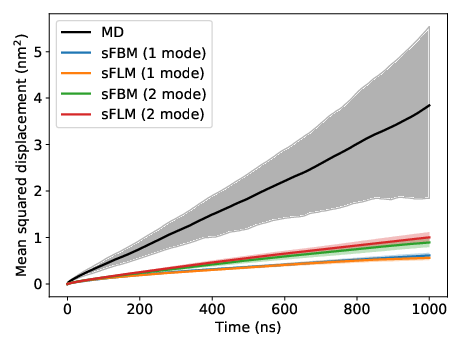}
  \caption{Methanol}\label{S-fig:bothmode_msd_comparison_MET}
  \end{subfigure}
  \begin{subfigure}{0.45\textwidth}
  \includegraphics[width=\textwidth]{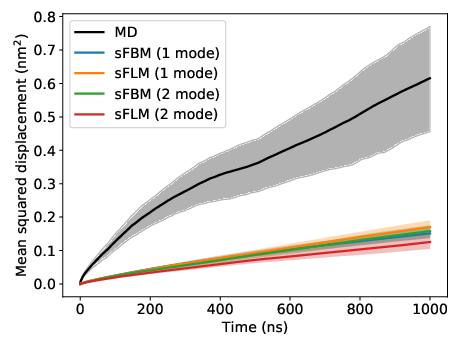}
  \caption{Acetic Acid}\label{S-fig:bothmode_msd_comparison_ACH}
  \end{subfigure}
  \caption{When we do not apply an exponential cut-off to the power law
  distribution of dwell times,  MSDs are consistently under-predicted by the
  AD model.
  }\label{S-fig:anomalous_msds_bothmode}
  \end{figure*}

\begin{figure*}
  \centering
  \begin{subfigure}{0.45\textwidth}
  \includegraphics[width=\textwidth]{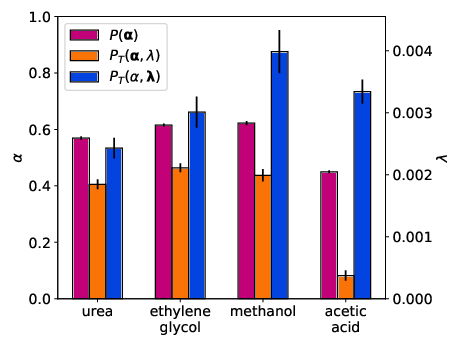}
  \caption{}\label{S-fig:1mode_AD_dwells}
  \end{subfigure}
  \begin{subfigure}{0.45\textwidth}
  \includegraphics[width=\textwidth]{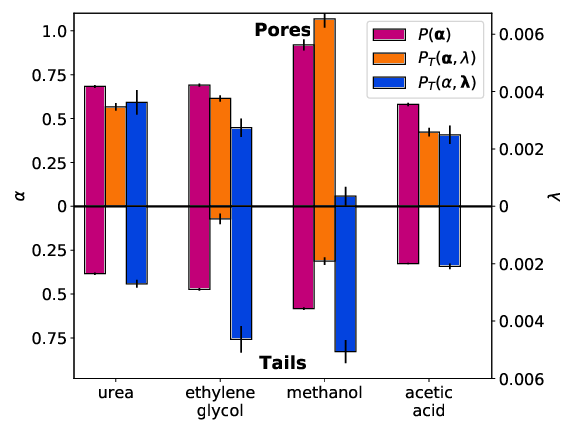}
  \caption{}\label{S-fig:2mode_AD_dwells}
  \end{subfigure}
  \caption{We can parameterize the dwell time distribution in two ways: as a pure
  power law (P($\alpha$)) and as a power law with an exponential cut-off (P($\alpha$, $\lambda$)).
  Pure power laws have an infinite variance which allows extremely long dwell
  times to be sampled (see Figure~\ref{S-fig:anomalous_msds_bothmode}.
  }\label{S-fig:pure_power_law_params}
  \end{figure*}

  When we use a pure power law distribution to parameterize the dwell time 
  distributions of the one and two mode AD models, the MD MSDs are severely 
  under-predicted because we are incorporating dwell times on the order of 
  the simulation length into simulated trajectories 
  (see Figure~\ref{S-fig:anomalous_msds_bothmode}). The parameters of the pure power law 
   distribution are included in Figure~\ref{S-fig:pure_power_law_params}. 
  
  \section{Stationarity of Solute Trajectories}\label{S-section:msd_comparison}

  We observe that in some cases, solute trajectories extracted from our MD
  simulations display non-stationary behavior. We defined the perceived
  equilibration time point for each solute based on the time at which the
  number of solutes inside the pores and tails stabilized
  (Figure~\ref{S-fig:equilibration}). With this definition, we observe
  evidence of non-stationary solute behavior after the perceived equilibration
  point, on the $\mu$s timescale. 

\begin{figure*}[h]
  \centering
  \begin{subfigure}{0.45\textwidth}
  \includegraphics[width=\textwidth]{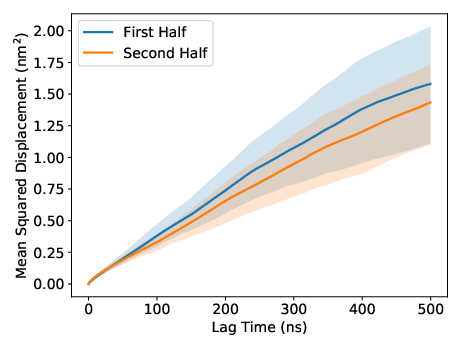}
  \caption{Urea}\label{S-fig:URE_MSD_halves}
  \end{subfigure}
  \begin{subfigure}{0.45\textwidth}
  \includegraphics[width=\textwidth]{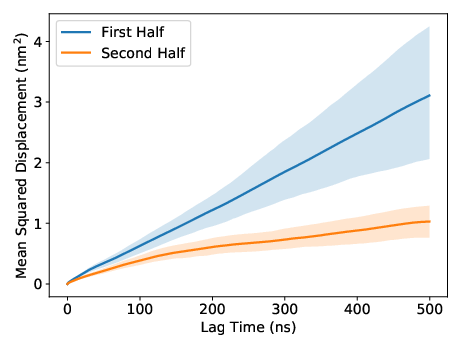}
  \caption{Ethylene Glycol}\label{S-fig:GCL_MSD_halves}
  \end{subfigure}
  \begin{subfigure}{0.45\textwidth}
  \includegraphics[width=\textwidth]{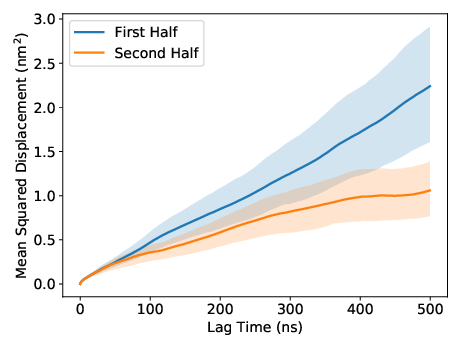}
  \caption{Methanol}\label{S-fig:MET_MSD_halves}
  \end{subfigure}
  \begin{subfigure}{0.45\textwidth}
  \includegraphics[width=\textwidth]{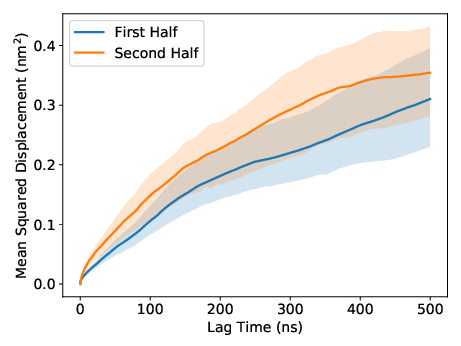}
  \caption{Acetic Acid}\label{S-fig:ACH_MSD_halves}
  \end{subfigure}
  \caption{The ensemble of solute trajectories may be stationary if the MSD
  calculated from different portions of the trajectory are the same. Here we
  plot the MSD calculated up to a 500 ns time lag of the first and second
  halves of the equilibrated solute trajectories. Urea and acetic acid have
  similar MSDs, providing evidence of stationarity, while the MSDs of 
  ethylene glycol and methanol are different suggesting that they are not.}\label{S-fig:msd_comparison}
  \end{figure*}

  We trained
  the model parameters on the first half of the equilibrated MD trajectory data and
  then compared the MSD calculated from AD model realizations to the MSD calculated
  from the second half of the equilibrated MD trajectory data. This metric is
  only meaningful if the ensemble of solute trajectories is stationary. In 
  Figure~\ref{S-fig:msd_comparison}, we show that urea and acetic acid show acceptable
  stationary behavior while methanol and ethylene glycol do not.
  
\begin{figure*}
  \centering
  \begin{subfigure}{0.45\textwidth}
  \includegraphics[width=\textwidth]{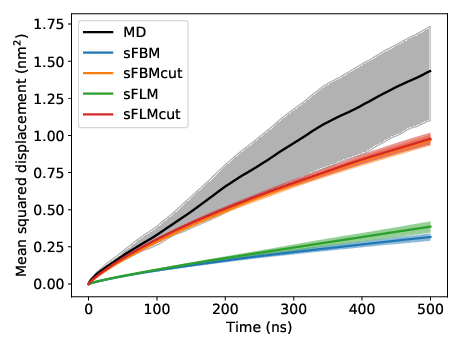}
  \caption{Urea (1 mode)}\label{S-fig:1mode_msd_comparison_URE_train_front}
  \end{subfigure}
  \begin{subfigure}{0.45\textwidth}
  \includegraphics[width=\textwidth]{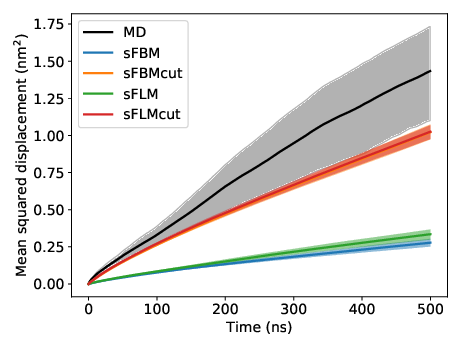}
  \caption{Urea (2 modes)}\label{S-fig:2mode_msd_comparison_URE_train_front}
  \end{subfigure}
  \begin{subfigure}{0.45\textwidth}
  \includegraphics[width=\textwidth]{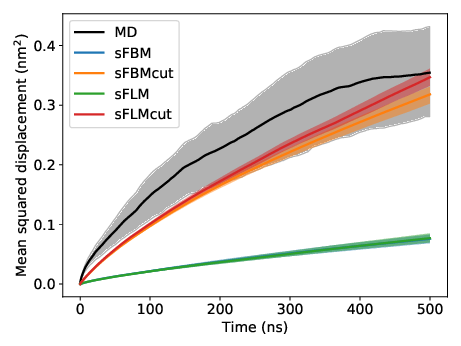}
  \caption{Acetic acid (1 mode)}\label{S-fig:1mode_msd_comparison_ACH_train_front}
  \end{subfigure}
  \begin{subfigure}{0.45\textwidth}
  \includegraphics[width=\textwidth]{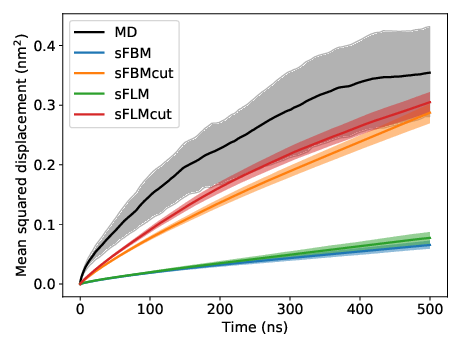}
  \caption{Acetic acid (2 modes)}\label{S-fig:2mode_msd_comparison_ACH_train_front}
  \end{subfigure} 
  \caption{In most cases, when using power laws with exponential cut-offs 
  (sFBMcut and sFLMcut), the MSD curves predicted by the AD model trained on 
  the first half of the equilibrated data lie within the 1$\sigma$ confidence
  intervals of the MD MSD curves generated from the second half of the equilibrated
  solute data. Over- and under-estimated curvature of the of urea and acetic acid's 
  MSD curves respectively causes the magnitude of urea's predicted MSDs to be 
  under-predicted at long timescales and those of acetic acid to be under-predicted
  at short timescales. The models which use pure power laws systematically under-predict
  the MD MSD curves. 
  }\label{S-fig:train_test}
  \end{figure*}

  We validated both the one and two mode AD models with urea and acetic acid, since their
  trajectories appear stationary. The MSDs resulting from 1000 realizations of the AD
  model are shown in Figure~\ref{S-fig:train_test}. We consider the model's prediction 
  to match well if the MSD lies within the 1$\sigma$ confidence intervals of the MD MSDs.
  We also look for qualitative agreement in the shape of the curves.
  
  The models are capable of reasonably predicting the MD MSD values of the second
  half of the solute trajectories based on parameters generated from the first half
  when the dwell time distributions are parameterized by a power law with an 
  exponential cut-off. At long timescales, the MSD of urea is under-predicted 
  for both the one and two mode models with the same true of acetic acid on short 
  timescales. Without truncation of the power law distribution, the MD 
  MSDs are underestimated in all cases because dwell times on the order of the
  MD simulation length are sampled and incorporated into the simulated anomalous
  diffusion trajectories. 
  
  This brief analysis suggests that we may be operating on the border of the
  minimum amount of data required to accurately parameterize AD approach
  models. Working with only half of the data we collected ($\sim$ 2 $\mu$s
  post-equilibration) may not always be sufficient for extracting reliable
  parameter estimates. Therefore, in the main text, we employ parameters fit
  to the entire equilibrated portion of the solute trajectories. Even doubling
  the data might not be good enough for molecules with statistical 
  non-stationarity, meaning the predictive and interpretive power of the time
  series modeling applied to these trajectories will be lower.
  
  \section{Tables of Anomalous Diffusion Parameters}\label{S-section:tabular_AD_params}

  The tables in this section are tabular representations of the parameters depicted in
  Figures~\ref{fig:1mode_parameters} and~\ref{fig:2mode_parameters} of the main text
  and used to generate AD approach realizations whose MSDs are shown in Figure~\ref{fig:anomalous_msds}
  of the main text.
  
  \begin{table*}[h]
  \centering
  \begin{tabular}{|M{2cm}|M{1.65cm}|M{1.95cm}|M{1.95cm}|M{1.95cm}|M{1.95cm}|}
  \hline
  1 Mode Model  & Parameters                & Urea         & Ethylene Glycol &   Methanol   & Acetic Acid  \\\hline
  Dwell         & $P(\alpha_d)$             & 0.57         & 0.62            & 0.62         & 0.45         \\\cline{2-6}
  Distributions & $P_T(\alpha_d, \lambda)$  & 0.40, 0.0024 & 0.47, 0.0030    & 0.44, 0.0040 & 0.08, 0.0033 \\\hline
  Hop           & $\mathcal{N}(\sigma)$     & 0.33         & 0.34            & 0.35         & 0.27         \\\cline{2-6}
  Distributions & $L(\sigma, \alpha_h)$     & 0.21, 1.84   & 0.23, 1.92      & 0.22, 1.80   & 0.16, 1.72   \\\hline
  Correlation   & $\gamma(H)$               & 0.37         & 0.40            & 0.30         & 0.34         \\
  \hline 
  \end{tabular}
  \caption{Parameters of the one mode AD approach models. See the main
  text for further details.}\label{S-table:sfbm_params}
  \end{table*}
  
  \begin{table*}[h]
  \centering
  \begin{tabular}{|M{2cm}|M{1.65cm}|M{.75cm}|M{1.95cm}|M{1.95cm}|M{1.95cm}|M{1.95cm}|M{1.95cm}|M{1.95cm}|}
  \hline
  \multicolumn{7}{|c|}{2 Mode Model} \\\hline
                & Parameters                               & Mode & Urea         & Ethylene Glycol &  Methanol    & Acetic Acid \\
  \hline
                &\multirow{2}{*}{$P(\alpha_d)$}            & 1    & 0.69         &  0.69           & 0.90         & 0.58         \\\cline{3-7}
  Dwell         &                                          & 2    & 0.38         &  0.48           & 0.58         & 0.33         \\\cline{2-7}
  Distributions &\multirow{2}{*}{$P_T(\alpha_d, \lambda)$} & 1    & 0.56, 0.0037 &  0.62, 0.0026   & 1.04, 0.0006 & 0.41, 0.0026 \\\cline{3-7}
                &                                          & 2    & 0.00, 0.0027 &  0.06, 0.0049   & 0.30, 0.0054 & 0.00, 0.0021 \\\hline
                &\multirow{2}{*}{$\mathcal{N}(\sigma)$}    & 1    & 0.35         &  0.38           & 0.45         & 0.32         \\\cline{3-7}
  Hop           &                                          & 2    & 0.24         &  0.23           & 0.32         & 0.17         \\\cline{2-7}
  Distributions &\multirow{2}{*}{$L(\sigma, \alpha_h)$}    & 1    & 0.24, 1.91   &  0.26, 1.99     & 0.31, 1.97   & 0.21, 1.91   \\\cline{3-7}
                &                                          & 2    & 0.12, 1.50   &  0.15, 1.90     & 0.20, 1.85   & 0.09, 1.50   \\\hline
  Correlation   & $\gamma(H)$                              & --   & 0.37         &  0.40           & 0.30         & 0.34         \\\cline{3-7}
  \hline 
  \end{tabular}
  \caption{Parameters of the 2 mode AD approach models. See the main text for further
  details.}\label{S-table:sfbm_params_2mode}
  \end{table*}
  
  \section{MSDDM parameters}\label{S-section:msddm_params}
  
\begin{figure*}
  \centering
  \begin{subfigure}{0.49\textwidth}
  \includegraphics[width=\textwidth]{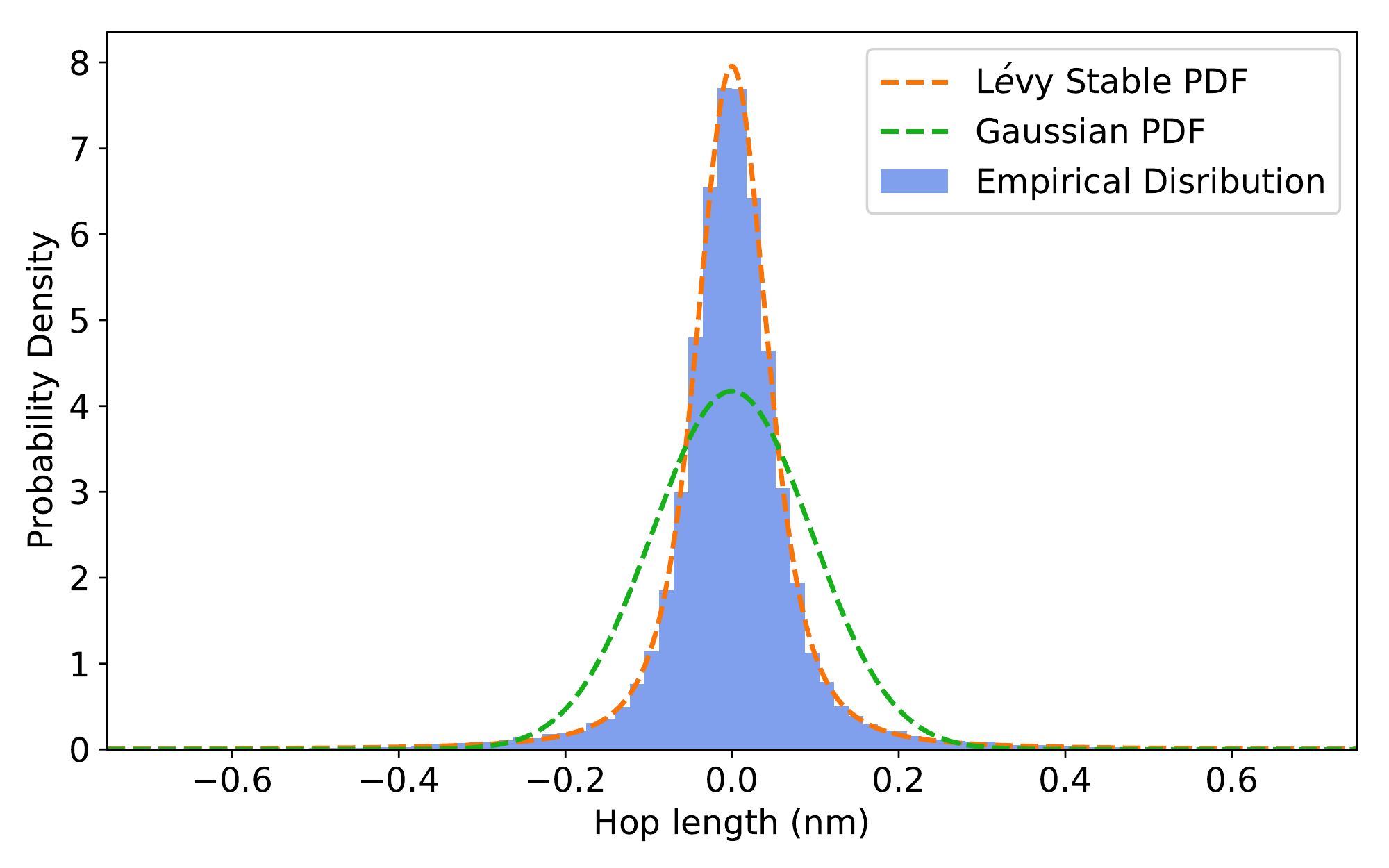}
  \caption{}\label{S-fig:gaussian_levy_comparison}
  \end{subfigure}
  \begin{subfigure}{0.42\textwidth}
  \includegraphics[width=\textwidth]{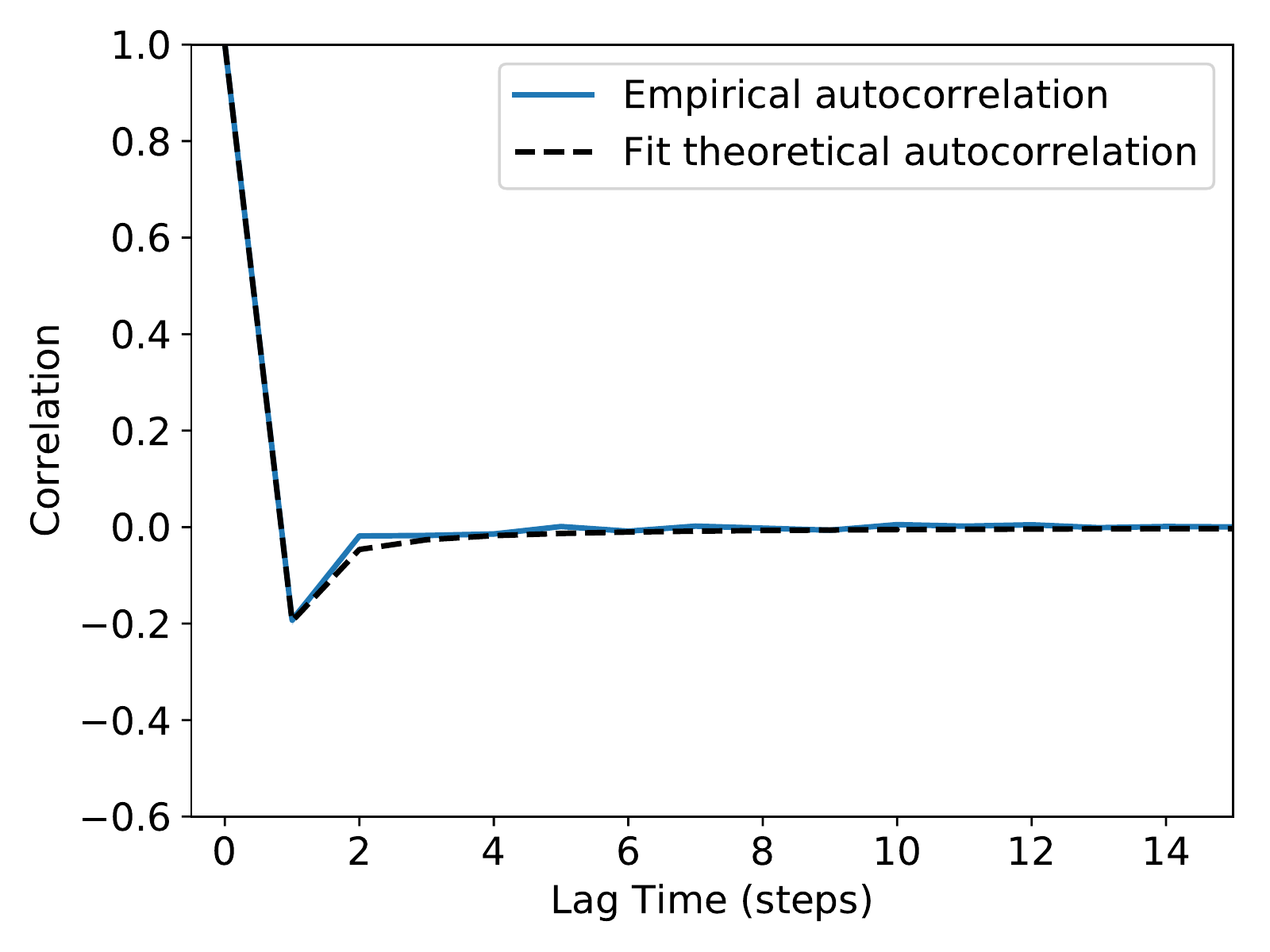}
  \caption{}\label{S-fig:msddm_acf}
  \end{subfigure}
  \caption{(a) The emission distributions of hop lengths are non-Gaussian and
	  heavy-tailed. Shown here is the emission distribution for transitions
	  between states. The maximum likelihood Gaussian fit severely
	  underestimates the empirical density of hops near and far from zero
	  while overestimating the density of hops at intermediate values. (b)
	  Jumps drawn from the transition distribution are negatively
	  correlated to each other. The normalized version of
	  Equation~\ref{eqn:flm_autocovariance} of the main text fits well to the data
	  suggesting FLM is an appropriate way to model jumps.}
	  \label{S-fig:msddm_emissions}
  \end{figure*}

  We observe correlated emissions drawn from L\'evy stable distributions. The
  deviation of the emission distributions from Gaussian behavior is far more
  pronounced than that seen in the hop length distributions of the AD model.
  (see Figure~\ref{S-fig:gaussian_levy_comparison}) We therefore did not consider
  the Gaussian case for the MSDDM. The correlation structure between hops is 
  consistent with that of FLM (Figure~\ref{S-fig:msddm_acf}).  
  
  The following table is a tabular representation of the parameters 
  depicted in Figure~\ref{fig:msddm_parameters} of the main text.
  
  \begin{table*}[h]
  \centering
  \begin{tabular}{|c|c|c|c|c|c|c|c|c|c|c|c|c|}
  \hline
  & \multicolumn{3}{c|}{Urea} & \multicolumn{3}{c|}{Ethylene Glycol} & \multicolumn{3}{c|}{Methanol} & \multicolumn{3}{c|}{Acetic Acid} \\\hline
  State & H     &$\alpha_h$& $\sigma$ & H    &$\alpha_h$& $\sigma$   & H     &$\alpha_h$& $\sigma$ & H    &$\alpha_h$& $\sigma$ \\\hline
  1     & 0.10  & 1.79     & 0.034    & 0.09 & 1.68     & 0.045      & 0.11  & 1.56     & 0.052    & 0.10 & 1.78     & 0.035    \\
  2     & 0.06  & 1.80     & 0.033    & 0.09 & 1.75     & 0.037      & 0.07  & 1.63     & 0.043    & 0.08 & 1.88     & 0.032    \\
  3     & 0.11  & 1.88     & 0.030    & 0.11 & 1.86     & 0.030      & 0.02  & 1.80     & 0.036    & 0.04 & 2.00     & 0.030    \\
  4     & 0.10  & 1.95     & 0.027    & 0.04 & 1.91     & 0.028      & 0.02  & 1.75     & 0.036    & 0.04 & 2.00     & 0.027    \\
  5     & 0.19  & 1.34     & 0.048    & 0.15 & 1.40     & 0.062      & 0.10  & 1.28     & 0.074    & 0.13 & 1.47     & 0.048    \\
  6     & 0.15  & 1.45     & 0.040    & 0.11 & 1.52     & 0.040      & 0.03  & 1.50     & 0.042    & 0.09 & 1.70     & 0.038    \\
  7     & 0.15  & 1.61     & 0.032    & 0.05 & 1.60     & 0.040      & 0.28  & 1.20     & 0.043    & 0.08 & 1.77     & 0.031    \\
  8     & 0.11  & 1.71     & 0.028    & 0.05 & 1.74     & 0.030      & 0.04  & 1.83     & 0.037    & 0.01 & 2.00     & 0.030    \\
  T     & 0.34  & 1.42     & 0.036    & 0.37 & 1.44     & 0.045      & 0.35  & 1.45     & 0.057    & 0.34 & 1.54     & 0.040    \\\hline
  \end{tabular}
  \caption{We calculated values of $H$, $\alpha_h$ and $\sigma$ from MD simulation
  trajectories and used them to generate realizations of our MSDDM model. The states
  are defined in Table~\ref{table:states} of the main text except state T which 
  describes the transition emissions.}\label{S-table:msddm_params}
  \end{table*}  
  
  \section*{Analytical fits to MFPT distributions}\label{S-section:mfpt_fits}
  
\begin{figure*}[h]
  \centering
  \begin{subfigure}{0.45\textwidth}
  \includegraphics[width=\textwidth]{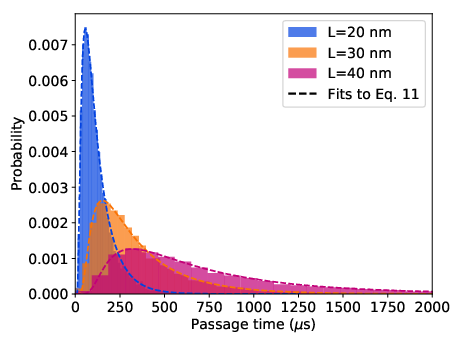}
  \caption{Urea}\label{S-fig:URE_ad_fpt_distributions}
  \end{subfigure}
  \begin{subfigure}{0.45\textwidth}
  \includegraphics[width=\textwidth]{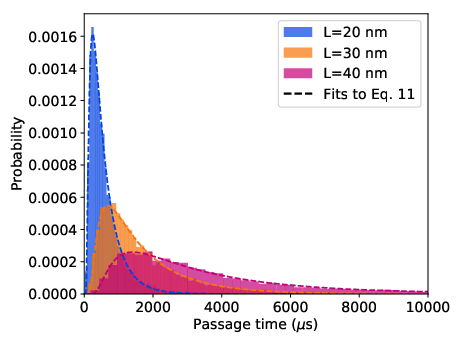}
  \caption{Acetic Acid}\label{S-fig:ACH_ad_fpt_distributions}
  \end{subfigure}
  \begin{subfigure}{0.45\textwidth}
  \includegraphics[width=\textwidth]{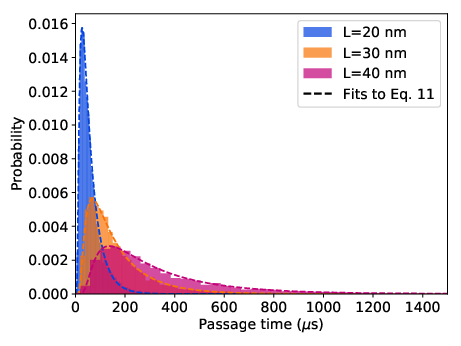}
  \caption{Ethylene Glycol}\label{S-fig:GCL_ad_fpt_distributions}
  \end{subfigure}
  \begin{subfigure}{0.45\textwidth}
  \includegraphics[width=\textwidth]{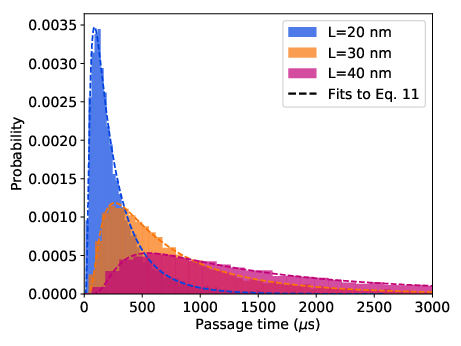}
  \caption{Methanol}\label{S-fig:MET_ad_fpt_distributions}
  \end{subfigure}
  \caption{We fit Equation~\ref{eqn:passage_times} of the main text to the
  first passage time distributions generated by 10,000 realizations of the 
  anomalous diffusion model. }\label{S-fig:ad_fpt_fits}
  \end{figure*}

\begin{figure*}
  \centering
  \begin{subfigure}{0.45\textwidth}
  \includegraphics[width=\textwidth]{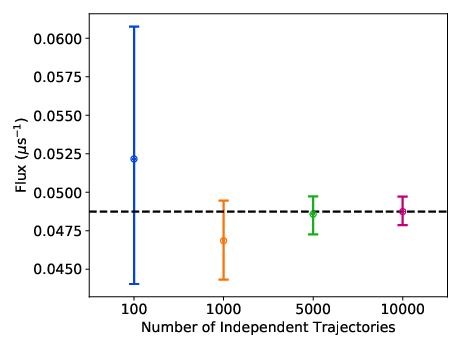}
  \caption{urea}\label{S-fig:Nsensitivity_URE}
  \end{subfigure}
  \begin{subfigure}{0.45\textwidth}
  \includegraphics[width=\textwidth]{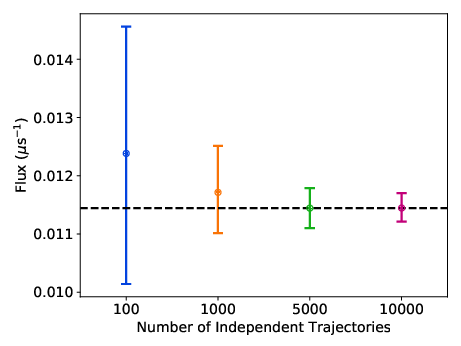}
  \caption{acetic acid}\label{S-fig:Nsensitivity_ACH}
  \end{subfigure}
  \begin{subfigure}{0.45\textwidth}
  \includegraphics[width=\textwidth]{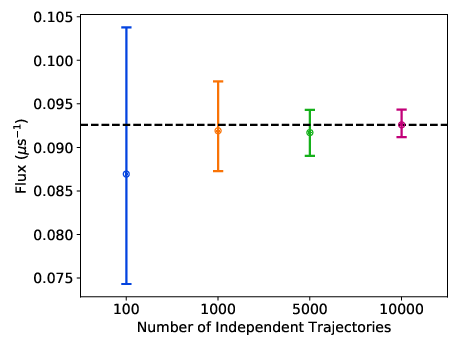}
  \caption{ethylene glycol}\label{S-fig:Nsensitivity_GCL}
  \end{subfigure}
  \begin{subfigure}{0.45\textwidth}
  \includegraphics[width=\textwidth]{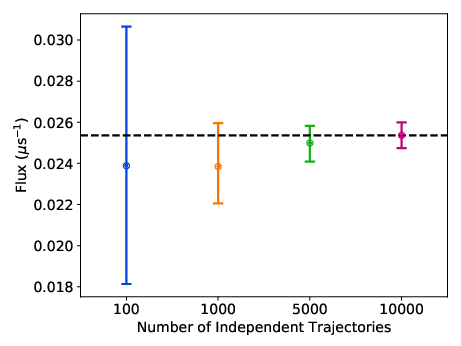}
  \caption{methanol}\label{S-fig:Nsensitivity_MET}
  \end{subfigure}
  \caption{Even using a small number of independent trajectories, one can
  reliably estimate solute flux across a pore 10 nm long. The uncertainty in 
  the flux values decreases as we add more independent trajectories.
  }\label{S-fig:flux_curve_sensitivity}
  \end{figure*}

\begin{figure*}
  \centering
  \begin{subfigure}{0.45\textwidth}
  \includegraphics[width=\textwidth]{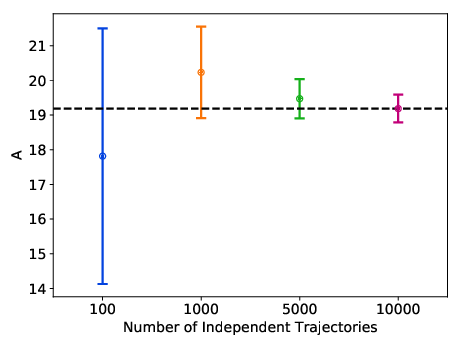}
  \caption{urea}\label{S-fig:Asensitivity_URE}
  \end{subfigure}
  \begin{subfigure}{0.45\textwidth}
  \includegraphics[width=\textwidth]{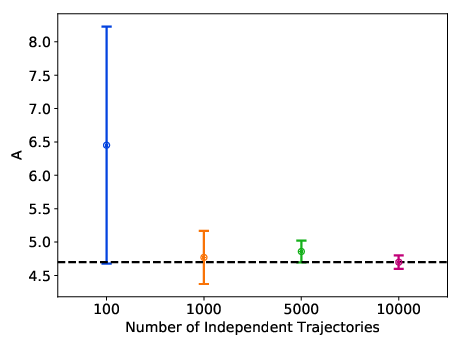}
  \caption{acetic acid}\label{S-fig:Asensitivity_ACH}
  \end{subfigure}
  \begin{subfigure}{0.45\textwidth}
  \includegraphics[width=\textwidth]{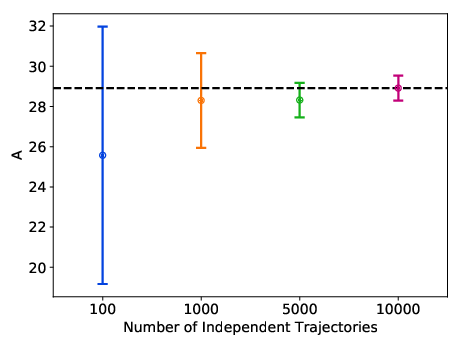}
  \caption{ethylene glycol}\label{S-fig:Asensitivity_GCL}
  \end{subfigure}
  \begin{subfigure}{0.45\textwidth}
  \includegraphics[width=\textwidth]{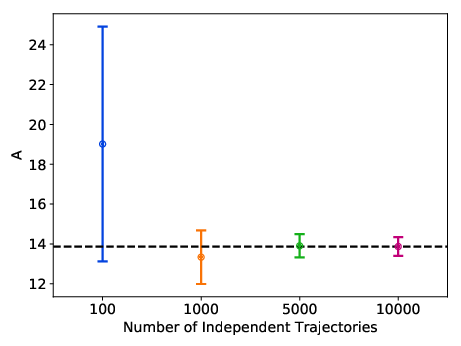}
  \caption{methanol}\label{S-fig:Asensitivity_MET}
  \end{subfigure}
  \caption{The flux scaling parameter ($A$) can be reliably estimated using
  as few as 100 independent realizations of the sFBMcut AD model. To estimate
  $A$, we fit Equation~\ref{eqn:flux_decay} of the main text to a series of
  flux measurements made with 10, 15, 20, 25, 30, 35, 40, 45 and 50 nm pores (see
  Figure~\ref{fig:flux_curves_ad} of the main text).
  }\label{S-fig:A_sensitivity}
  \end{figure*}

\begin{figure*}
  \centering
  \begin{subfigure}{0.45\textwidth}
  \includegraphics[width=\textwidth]{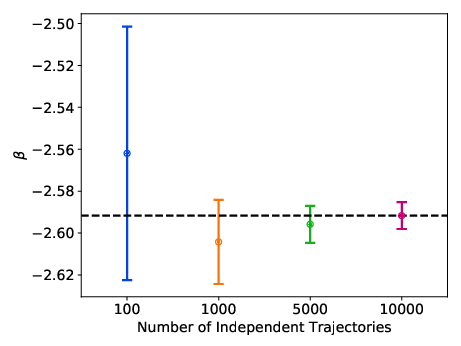}
  \caption{urea}\label{S-fig:betasensitivity_URE}
  \end{subfigure}
  \begin{subfigure}{0.45\textwidth}
  \includegraphics[width=\textwidth]{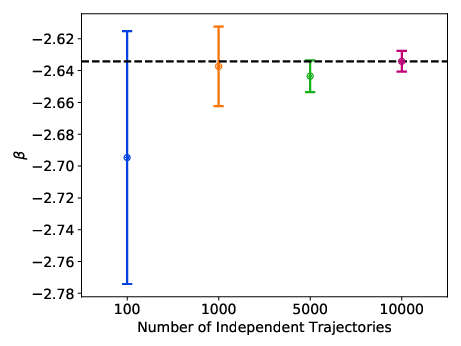}
  \caption{acetic acid}\label{S-fig:betasensitivity_ACH}
  \end{subfigure}
  \begin{subfigure}{0.45\textwidth}
  \includegraphics[width=\textwidth]{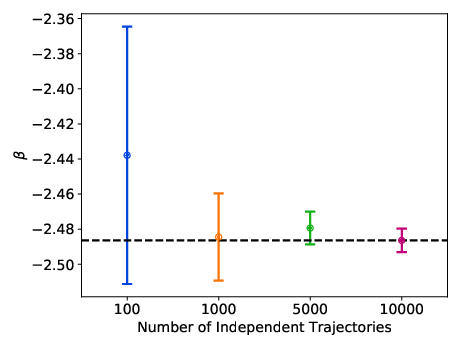}
  \caption{ethylene glycol}\label{S-fig:betasensitivity_GCL}
  \end{subfigure}
  \begin{subfigure}{0.45\textwidth}
  \includegraphics[width=\textwidth]{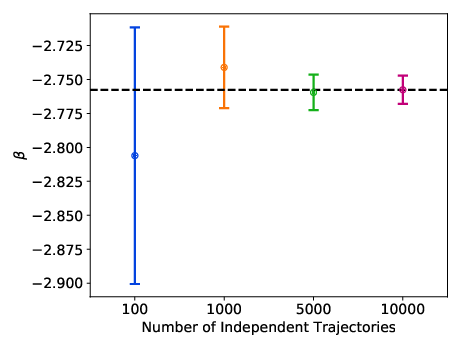}
  \caption{methanol}\label{S-fig:betasensitivity_MET}
  \end{subfigure}
  \caption{Similar to $A$, the parameter which describes the scaling of solute
  flux with pore length, $\beta$, can be reliably estimated using as few as 100
  independent realizations of the sFBMcut AD model. We estimated $\beta$ and
  $A$ simultaneously as described in Figure~\ref{S-fig:A_sensitivity}.
  }\label{S-fig:beta_sensitivity}
  \end{figure*}

  In Figure~\ref{S-fig:ad_fpt_fits} 
  we demonstrate
  the high quality of our analytical fits of Equation~\ref{S-eqn:passage_times} to 
  the distribution of solute first passage times derived from both the AD
  and MSDDM models. 
  In Figures~\ref{S-fig:flux_curve_sensitivity}--~\ref{S-fig:beta_sensitivity}, we
  show that one can reliably fit Equation~\ref{S-eqn:passage_times}
  to the passage time distributions with as few as 100 independent trajectory
  realizations at each pore length. For higher precision, we recommend using at least
  1000 trajectories.
  
\end{appendix}

%% file: ms.bbl
\providecommand{\latin}[1]{#1}
\makeatletter
\providecommand{\doi}
  {\begingroup\let\do\@makeother\dospecials
  \catcode`\{=1 \catcode`\}=2 \doi@aux}
\providecommand{\doi@aux}[1]{\endgroup\texttt{#1}}
\makeatother
\providecommand*\mcitethebibliography{\thebibliography}
\csname @ifundefined\endcsname{endmcitethebibliography}
  {\let\endmcitethebibliography\endthebibliography}{}
\begin{mcitethebibliography}{60}
\providecommand*\natexlab[1]{#1}
\providecommand*\mciteSetBstSublistMode[1]{}
\providecommand*\mciteSetBstMaxWidthForm[2]{}
\providecommand*\mciteBstWouldAddEndPuncttrue
  {\def\EndOfBibitem{\unskip.}}
\providecommand*\mciteBstWouldAddEndPunctfalse
  {\let\EndOfBibitem\relax}
\providecommand*\mciteSetBstMidEndSepPunct[3]{}
\providecommand*\mciteSetBstSublistLabelBeginEnd[3]{}
\providecommand*\EndOfBibitem{}
\mciteSetBstSublistMode{f}
\mciteSetBstMaxWidthForm{subitem}{(\alph{mcitesubitemcount})}
\mciteSetBstSublistLabelBeginEnd
  {\mcitemaxwidthsubitemform\space}
  {\relax}
  {\relax}

\bibitem[Werber \latin{et~al.}(2016)Werber, Osuji, and
  Elimelech]{werber_materials_2016}
Werber,~J.~R.; Osuji,~C.~O.; Elimelech,~M. Materials for {Next}-{Generation}
  {Desalination} and {Water} {Purification} {Membranes}. \emph{Nat. Rev.
  Mater.} \textbf{2016}, \emph{1}, 16018\relax
\mciteBstWouldAddEndPuncttrue
\mciteSetBstMidEndSepPunct{\mcitedefaultmidpunct}
{\mcitedefaultendpunct}{\mcitedefaultseppunct}\relax
\EndOfBibitem
\bibitem[Barbosa \latin{et~al.}(2016)Barbosa, Moreira, Ribeiro, Pereira, and
  Silva]{barbosa_occurrence_2016}
Barbosa,~M.~O.; Moreira,~N. F.~F.; Ribeiro,~A.~R.; Pereira,~M. F.~R.; Silva,~A.
  M.~T. Occurrence and {Removal} of {Organic} {Micropollutants}: {An}
  {Overview} of the {Watch} {List} of {Eu} {Decision} 2015/495. \emph{Water
  Research} \textbf{2016}, \emph{94}, 257--279\relax
\mciteBstWouldAddEndPuncttrue
\mciteSetBstMidEndSepPunct{\mcitedefaultmidpunct}
{\mcitedefaultendpunct}{\mcitedefaultseppunct}\relax
\EndOfBibitem
\bibitem[Gin \latin{et~al.}(2008)Gin, Bara, Noble, and
  Elliot]{gin_polymerized_2008}
Gin,~D.~L.; Bara,~J.~E.; Noble,~R.~D.; Elliot,~B.~J. Polymerized {Lyotropic}
  {Liquid} {Crystal} {Assemblies} for {Membrane} {Applications}.
  \emph{Macromol. Rapid Commun.} \textbf{2008}, \emph{29}, 367--389\relax
\mciteBstWouldAddEndPuncttrue
\mciteSetBstMidEndSepPunct{\mcitedefaultmidpunct}
{\mcitedefaultendpunct}{\mcitedefaultseppunct}\relax
\EndOfBibitem
\bibitem[Dischinger \latin{et~al.}(2017)Dischinger, Rosenblum, Noble, Gin, and
  Linden]{dischinger_application_2017}
Dischinger,~S.~M.; Rosenblum,~J.; Noble,~R.~D.; Gin,~D.~L.; Linden,~K.~G.
  Application of a {Lyotropic} {Liquid} {Crystal} {Nanofiltration} {Membrane}
  for {Hydraulic} {Fracturing} {Flowback} {Water}: {Selectivity} and
  {Implications} for {Treatment}. \emph{J. Membr. Sci.} \textbf{2017},
  \emph{543}, 319--327\relax
\mciteBstWouldAddEndPuncttrue
\mciteSetBstMidEndSepPunct{\mcitedefaultmidpunct}
{\mcitedefaultendpunct}{\mcitedefaultseppunct}\relax
\EndOfBibitem
\bibitem[Dischinger \latin{et~al.}(2017)Dischinger, McGrath, Bourland, Noble,
  and Gin]{dischinger_effect_2017}
Dischinger,~S.~M.; McGrath,~M.~J.; Bourland,~K.~R.; Noble,~R.~D.; Gin,~D.~L.
  Effect of {Post}-{Polymerization} {Anion}-{Exchange} on the {Rejection} of
  {Uncharged} {Aqueous} {Solutes} in {Nanoporous}, {Ionic}, {Lyotropic}
  {Liquid} {Crystal} {Polymer} {Membranes}. \emph{J. Membr. Sci.}
  \textbf{2017}, \emph{529}, 72--79\relax
\mciteBstWouldAddEndPuncttrue
\mciteSetBstMidEndSepPunct{\mcitedefaultmidpunct}
{\mcitedefaultendpunct}{\mcitedefaultseppunct}\relax
\EndOfBibitem
\bibitem[Feng \latin{et~al.}(2014)Feng, Tousley, Cowan, Wiesenauer, Nejati,
  Choo, Noble, Elimelech, Gin, and Osuji]{feng_scalable_2014}
Feng,~X.; Tousley,~M.~E.; Cowan,~M.~G.; Wiesenauer,~B.~R.; Nejati,~S.;
  Choo,~Y.; Noble,~R.~D.; Elimelech,~M.; Gin,~D.~L.; Osuji,~C.~O. Scalable
  {Fabrication} of {Polymer} {Membranes} with {Vertically} {Aligned} 1 nm
  {Pores} by {Magnetic} {Field} {Directed} {Self}-{Assembly}. \emph{ACS Nano}
  \textbf{2014}, \emph{8}, 11977--11986\relax
\mciteBstWouldAddEndPuncttrue
\mciteSetBstMidEndSepPunct{\mcitedefaultmidpunct}
{\mcitedefaultendpunct}{\mcitedefaultseppunct}\relax
\EndOfBibitem
\bibitem[Feng \latin{et~al.}(2016)Feng, Nejati, Cowan, Tousley, Wiesenauer,
  Noble, Elimelech, Gin, and Osuji]{feng_thin_2016}
Feng,~X.; Nejati,~S.; Cowan,~M.~G.; Tousley,~M.~E.; Wiesenauer,~B.~R.;
  Noble,~R.~D.; Elimelech,~M.; Gin,~D.~L.; Osuji,~C.~O. Thin {Polymer} {Films}
  with {Continuous} {Vertically} {Aligned} 1 nm {Pores} {Fabricated} by {Soft}
  {Confinement}. \emph{ACS Nano} \textbf{2016}, \emph{10}, 150--158\relax
\mciteBstWouldAddEndPuncttrue
\mciteSetBstMidEndSepPunct{\mcitedefaultmidpunct}
{\mcitedefaultendpunct}{\mcitedefaultseppunct}\relax
\EndOfBibitem
\bibitem[Carter \latin{et~al.}(2012)Carter, Wiesenauer, Hatakeyama, Barton,
  Noble, and Gin]{carter_glycerol-based_2012}
Carter,~B.~M.; Wiesenauer,~B.~R.; Hatakeyama,~E.~S.; Barton,~J.~L.;
  Noble,~R.~D.; Gin,~D.~L. Glycerol-{Based} {Bicontinuous} {Cubic} {Lyotropic}
  {Liquid} {Crystal} {Monomer} {System} for the {Fabrication} of {Thin}-{Film}
  {Membranes} with {Uniform} {Nanopores}. \emph{Chem. Mater.} \textbf{2012},
  \emph{24}, 4005--4007\relax
\mciteBstWouldAddEndPuncttrue
\mciteSetBstMidEndSepPunct{\mcitedefaultmidpunct}
{\mcitedefaultendpunct}{\mcitedefaultseppunct}\relax
\EndOfBibitem
\bibitem[Hatakeyama \latin{et~al.}(2010)Hatakeyama, Wiesenauer, Gabriel, Noble,
  and Gin]{hatakeyama_nanoporous_2010}
Hatakeyama,~E.~S.; Wiesenauer,~B.~R.; Gabriel,~C.~J.; Noble,~R.~D.; Gin,~D.~L.
  Nanoporous, {Bicontinuous} {Cubic} {Lyotropic} {Liquid} {Crystal} {Networks}
  via {Polymerizable} {Gemini} {Ammonium} {Surfactants}. \emph{Chem. Mater.}
  \textbf{2010}, \emph{22}, 4525--4527\relax
\mciteBstWouldAddEndPuncttrue
\mciteSetBstMidEndSepPunct{\mcitedefaultmidpunct}
{\mcitedefaultendpunct}{\mcitedefaultseppunct}\relax
\EndOfBibitem
\bibitem[Smith \latin{et~al.}(1997)Smith, Fischer, and Gin]{smith_ordered_1997}
Smith,~R.~C.; Fischer,~W.~M.; Gin,~D.~L. Ordered {Poly}(p-phenylenevinylene)
  {Matrix} {Nanocomposites} {Via} {Lyotropic} {Liquid}-{Crystalline}
  {Monomers}. \emph{J. Am. Chem. Soc.} \textbf{1997}, \emph{119},
  4092--4093\relax
\mciteBstWouldAddEndPuncttrue
\mciteSetBstMidEndSepPunct{\mcitedefaultmidpunct}
{\mcitedefaultendpunct}{\mcitedefaultseppunct}\relax
\EndOfBibitem
\bibitem[Zhou \latin{et~al.}(2003)Zhou, Gu, Xu, Pecinovsky, and
  Gin]{zhou_assembly_2003}
Zhou,~W.; Gu,~W.; Xu,~Y.; Pecinovsky,~C.~S.; Gin,~D.~L. Assembly of {Acidic}
  {Amphiphiles} into {Inverted} {Hexagonal} {Phases} {Using} an
  {L}-{Alanine}-{Based} {Surfactant} as a {Structure}-{Directing} {Agent}.
  \emph{Langmuir} \textbf{2003}, \emph{19}, 6346--6348\relax
\mciteBstWouldAddEndPuncttrue
\mciteSetBstMidEndSepPunct{\mcitedefaultmidpunct}
{\mcitedefaultendpunct}{\mcitedefaultseppunct}\relax
\EndOfBibitem
\bibitem[Resel \latin{et~al.}(2000)Resel, Leising, Markart, Kriechbaum, Smith,
  and Gin]{resel_structural_2000}
Resel,~R.; Leising,~G.; Markart,~P.; Kriechbaum,~M.; Smith,~R.; Gin,~D.
  Structural {Properties} of {Polymerised} {Lyotropic} {Liquid} {Crystals}
  {Phases} of 3,4,5-{Tris}(ω-acryloxyalkoxy)benzoate {Salts}. \emph{Macromol.
  Chem. Phys.} \textbf{2000}, \emph{201}, 1128--1133\relax
\mciteBstWouldAddEndPuncttrue
\mciteSetBstMidEndSepPunct{\mcitedefaultmidpunct}
{\mcitedefaultendpunct}{\mcitedefaultseppunct}\relax
\EndOfBibitem
\bibitem[Coscia and Shirts(2019)Coscia, and Shirts]{coscia_chemically_2019}
Coscia,~B.~J.; Shirts,~M.~R. Chemically {Selective} {Transport} in a
  {Cross}-{Linked} {HII} {Phase} {Lyotropic} {Liquid} {Crystal} {Membrane}.
  \emph{J. Phys. Chem. B} \textbf{2019}, \relax
\mciteBstWouldAddEndPunctfalse
\mciteSetBstMidEndSepPunct{\mcitedefaultmidpunct}
{}{\mcitedefaultseppunct}\relax
\EndOfBibitem
\bibitem[Vinh-Thang and Kaliaguine(2013)Vinh-Thang, and
  Kaliaguine]{vinh-thang_predictive_2013}
Vinh-Thang,~H.; Kaliaguine,~S. Predictive {Models} for {Mixed}-{Matrix}
  {Membrane} {Performance}: {A} {Review}. \emph{Chem. Rev.} \textbf{2013},
  \emph{113}, 4980--5028\relax
\mciteBstWouldAddEndPuncttrue
\mciteSetBstMidEndSepPunct{\mcitedefaultmidpunct}
{\mcitedefaultendpunct}{\mcitedefaultseppunct}\relax
\EndOfBibitem
\bibitem[Geens \latin{et~al.}(2006)Geens, Van~der Bruggen, and
  Vandecasteele]{geens_transport_2006}
Geens,~J.; Van~der Bruggen,~B.; Vandecasteele,~C. Transport {Model} for
  {Solvent} {Permeation} {Through} {Nanofiltration} {Membranes}. \emph{Sep.
  Purif. Technol.} \textbf{2006}, \emph{48}, 255--263\relax
\mciteBstWouldAddEndPuncttrue
\mciteSetBstMidEndSepPunct{\mcitedefaultmidpunct}
{\mcitedefaultendpunct}{\mcitedefaultseppunct}\relax
\EndOfBibitem
\bibitem[Darvishmanesh and Van~der Bruggen(2016)Darvishmanesh, and Van~der
  Bruggen]{darvishmanesh_mass_2016}
Darvishmanesh,~S.; Van~der Bruggen,~B. Mass {Transport} through
  {Nanostructured} {Membranes}: {Towards} a {Predictive} {Tool}.
  \emph{Membranes} \textbf{2016}, \emph{6}\relax
\mciteBstWouldAddEndPuncttrue
\mciteSetBstMidEndSepPunct{\mcitedefaultmidpunct}
{\mcitedefaultendpunct}{\mcitedefaultseppunct}\relax
\EndOfBibitem
\bibitem[Wijmans and Baker(1995)Wijmans, and
  Baker]{wijmans_solution-diffusion_1995}
Wijmans,~J.~G.; Baker,~R.~W. The {Solution}-{Diffusion} {Model}: {A} {Review}.
  \emph{J. Membr. Sci.} \textbf{1995}, \emph{107}, 1--21\relax
\mciteBstWouldAddEndPuncttrue
\mciteSetBstMidEndSepPunct{\mcitedefaultmidpunct}
{\mcitedefaultendpunct}{\mcitedefaultseppunct}\relax
\EndOfBibitem
\bibitem[Paul(1974)]{paul_diffusive_1974}
Paul,~D.~R. In \emph{Permeability of {Plastic} {Films} and {Coatings}: {To}
  {Gases}, {Vapors}, and {Liquids}}; Hopfenberg,~H.~B., Ed.; Polymer {Science}
  and {Technology}; Springer US: Boston, MA, 1974; pp 35--48\relax
\mciteBstWouldAddEndPuncttrue
\mciteSetBstMidEndSepPunct{\mcitedefaultmidpunct}
{\mcitedefaultendpunct}{\mcitedefaultseppunct}\relax
\EndOfBibitem
\bibitem[Manzo and Garcia-Parajo(2015)Manzo, and
  Garcia-Parajo]{manzo_review_2015}
Manzo,~C.; Garcia-Parajo,~M.~F. A {Review} of {Progress} in {Single} {Particle}
  {Tracking}: {From} {Methods} to {Biophysical} {Insights}. \emph{Rep. Prog.
  Phys.} \textbf{2015}, \emph{78}, 124601\relax
\mciteBstWouldAddEndPuncttrue
\mciteSetBstMidEndSepPunct{\mcitedefaultmidpunct}
{\mcitedefaultendpunct}{\mcitedefaultseppunct}\relax
\EndOfBibitem
\bibitem[Maginn \latin{et~al.}(2018)Maginn, Messerly, Carlson, Roe, and
  Elliott]{maginn_best_2018}
Maginn,~E.~J.; Messerly,~R.~A.; Carlson,~D.~J.; Roe,~D.~R.; Elliott,~J.~R. Best
  {Practices} for {Computing} {Transport} {Properties} 1. {Self}-{Diffusivity}
  and {Viscosity} from {Equilibrium} {Molecular} {Dynamics} [article {V1}.0].
  \emph{LiveCoMS} \textbf{2018}, \emph{1}, 6324\relax
\mciteBstWouldAddEndPuncttrue
\mciteSetBstMidEndSepPunct{\mcitedefaultmidpunct}
{\mcitedefaultendpunct}{\mcitedefaultseppunct}\relax
\EndOfBibitem
\bibitem[Metzler and Klafter(2000)Metzler, and Klafter]{metzler_random_2000}
Metzler,~R.; Klafter,~J. The {Random} {Walk}'s {Guide} to {Anomalous}
  {Diffusion}: {A} {Fractional} {Dynamics} {Approach}. \emph{Phys. Rep.}
  \textbf{2000}, \emph{339}, 1--77\relax
\mciteBstWouldAddEndPuncttrue
\mciteSetBstMidEndSepPunct{\mcitedefaultmidpunct}
{\mcitedefaultendpunct}{\mcitedefaultseppunct}\relax
\EndOfBibitem
\bibitem[Bouchaud and Georges(1990)Bouchaud, and
  Georges]{bouchaud_anomalous_1990}
Bouchaud,~J.-P.; Georges,~A. Anomalous {Diffusion} in {Disordered} {Media}:
  {Statistical} {Mechanisms}, {Models} and {Physical} {Applications}.
  \emph{Phys. Rep.} \textbf{1990}, \emph{195}, 127--293\relax
\mciteBstWouldAddEndPuncttrue
\mciteSetBstMidEndSepPunct{\mcitedefaultmidpunct}
{\mcitedefaultendpunct}{\mcitedefaultseppunct}\relax
\EndOfBibitem
\bibitem[Gorenflo and Mainardi(1997)Gorenflo, and
  Mainardi]{gorenflo_fractional_1997}
Gorenflo,~R.; Mainardi,~F. In \emph{Fractals and {Fractional} {Calculus} in
  {Continuum} {Mechanics}}; Carpinteri,~A., Mainardi,~F., Eds.; International
  {Centre} for {Mechanical} {Sciences}; Springer Vienna: Vienna, 1997; pp
  223--276\relax
\mciteBstWouldAddEndPuncttrue
\mciteSetBstMidEndSepPunct{\mcitedefaultmidpunct}
{\mcitedefaultendpunct}{\mcitedefaultseppunct}\relax
\EndOfBibitem
\bibitem[Klages \latin{et~al.}(2008)Klages, Radons, and
  Sokolov]{klages_anomalous_2008}
Klages,~R.; Radons,~G.; Sokolov,~I.~M. \emph{Anomalous {Transport}:
  {Foundations} and {Applications}}; John Wiley \& Sons, 2008; Google-Books-ID:
  TDch5DfNgWoC\relax
\mciteBstWouldAddEndPuncttrue
\mciteSetBstMidEndSepPunct{\mcitedefaultmidpunct}
{\mcitedefaultendpunct}{\mcitedefaultseppunct}\relax
\EndOfBibitem
\bibitem[Mandelbrot and Van~Ness(1968)Mandelbrot, and
  Van~Ness]{mandelbrot_fractional_1968}
Mandelbrot,~B.; Van~Ness,~J. Fractional {Brownian} {Motions}, {Fractional}
  {Noises} and {Applications}. \emph{SIAM Rev.} \textbf{1968}, \emph{10},
  422--437\relax
\mciteBstWouldAddEndPuncttrue
\mciteSetBstMidEndSepPunct{\mcitedefaultmidpunct}
{\mcitedefaultendpunct}{\mcitedefaultseppunct}\relax
\EndOfBibitem
\bibitem[Jeon and Metzler(2010)Jeon, and Metzler]{jeon_fractional_2010}
Jeon,~J.-H.; Metzler,~R. Fractional {Brownian} {Motion} and {Motion} {Governed}
  by the {Fractional} {Langevin} {Equation} in {Confined} {Geometries}.
  \emph{Phys. Rev. E} \textbf{2010}, \emph{81}, 021103\relax
\mciteBstWouldAddEndPuncttrue
\mciteSetBstMidEndSepPunct{\mcitedefaultmidpunct}
{\mcitedefaultendpunct}{\mcitedefaultseppunct}\relax
\EndOfBibitem
\bibitem[Banks and Fradin(2005)Banks, and Fradin]{banks_anomalous_2005}
Banks,~D.~S.; Fradin,~C. Anomalous {Diffusion} of {Proteins} {Due} to
  {Molecular} {Crowding}. \emph{Biophys. J.} \textbf{2005}, \emph{89},
  2960--2971\relax
\mciteBstWouldAddEndPuncttrue
\mciteSetBstMidEndSepPunct{\mcitedefaultmidpunct}
{\mcitedefaultendpunct}{\mcitedefaultseppunct}\relax
\EndOfBibitem
\bibitem[Montroll and Weiss(1965)Montroll, and Weiss]{montroll_random_1965}
Montroll,~E.~W.; Weiss,~G.~H. Random {Walks} on {Lattices}. {II}. \emph{J.
  Math. Phys.} \textbf{1965}, \emph{6}, 167--181\relax
\mciteBstWouldAddEndPuncttrue
\mciteSetBstMidEndSepPunct{\mcitedefaultmidpunct}
{\mcitedefaultendpunct}{\mcitedefaultseppunct}\relax
\EndOfBibitem
\bibitem[Snow \latin{et~al.}(2005)Snow, Sorin, Rhee, and Pande]{snow_how_2005}
Snow,~C.~D.; Sorin,~E.~J.; Rhee,~Y.~M.; Pande,~V.~S. How {Well} {Can}
  {Simulation} {Predict} {Protein} {Folding} {Kinetics} and {Thermodynamics}?
  \emph{Annu. Rev. Biophys. Biomol. Struct.} \textbf{2005}, \emph{34},
  43--69\relax
\mciteBstWouldAddEndPuncttrue
\mciteSetBstMidEndSepPunct{\mcitedefaultmidpunct}
{\mcitedefaultendpunct}{\mcitedefaultseppunct}\relax
\EndOfBibitem
\bibitem[Chodera \latin{et~al.}(2007)Chodera, Singhal, Pande, Dill, and
  Swope]{chodera_automatic_2007}
Chodera,~J.~D.; Singhal,~N.; Pande,~V.~S.; Dill,~K.~A.; Swope,~W.~C. Automatic
  {Discovery} of {Metastable} {States} for the {Construction} of {Markov}
  {Models} of {Macromolecular} {Conformational} {Dynamics}. \emph{J. Chem.
  Phys.} \textbf{2007}, \emph{126}, 155101\relax
\mciteBstWouldAddEndPuncttrue
\mciteSetBstMidEndSepPunct{\mcitedefaultmidpunct}
{\mcitedefaultendpunct}{\mcitedefaultseppunct}\relax
\EndOfBibitem
\bibitem[Chodera and Noé(2014)Chodera, and Noé]{chodera_markov_2014}
Chodera,~J.~D.; Noé,~F. Markov {State} {Models} of {Biomolecular}
  {Conformational} {Dynamics}. \emph{Curr. Opin. Struct. Biol.} \textbf{2014},
  \emph{25}, 135--144\relax
\mciteBstWouldAddEndPuncttrue
\mciteSetBstMidEndSepPunct{\mcitedefaultmidpunct}
{\mcitedefaultendpunct}{\mcitedefaultseppunct}\relax
\EndOfBibitem
\bibitem[Bowman \latin{et~al.}(2009)Bowman, Huang, and
  Pande]{bowman_using_2009}
Bowman,~G.~R.; Huang,~X.; Pande,~V.~S. Using {Generalized} {Ensemble}
  {Simulations} and {Markov} {State} {Models} to {Identify} {Conformational}
  {States}. \emph{Methods} \textbf{2009}, \emph{49}, 197--201\relax
\mciteBstWouldAddEndPuncttrue
\mciteSetBstMidEndSepPunct{\mcitedefaultmidpunct}
{\mcitedefaultendpunct}{\mcitedefaultseppunct}\relax
\EndOfBibitem
\bibitem[Bekker \latin{et~al.}(1993)Bekker, Berendsen, Dijkstra, Achterop, van
  Drunen, van~der Spoel, Sijbers, Keegstra, Reitsma, and
  Renardus]{bekker_gromacs:_1993}
Bekker,~H.; Berendsen,~H. J.~C.; Dijkstra,~E.~J.; Achterop,~S.; van Drunen,~R.;
  van~der Spoel,~D.; Sijbers,~A.; Keegstra,~H.; Reitsma,~B.; Renardus,~M. K.~R.
  \emph{{GROMACS}: {A} {Parallel} {Computer} for {Molecular} {Dynamics}
  {Simulations}}; Physics Computing '92: Proceedings of the 4th International
  Conference, Praha, Czechoslovakia, Aug 24-28, 1992; World Scientific:
  Singapore, 1993\relax
\mciteBstWouldAddEndPuncttrue
\mciteSetBstMidEndSepPunct{\mcitedefaultmidpunct}
{\mcitedefaultendpunct}{\mcitedefaultseppunct}\relax
\EndOfBibitem
\bibitem[Berendsen \latin{et~al.}(1995)Berendsen, van~der Spoel, and van
  Drunen]{berendsen_gromacs:_1995}
Berendsen,~H. J.~C.; van~der Spoel,~D.; van Drunen,~R. {GROMACS}: {A}
  {Message}-{Passing} {Parallel} {Molecular} {Dynamics} {Implementation}.
  \emph{Comput. Phys. Commun.} \textbf{1995}, \emph{91}, 43--56\relax
\mciteBstWouldAddEndPuncttrue
\mciteSetBstMidEndSepPunct{\mcitedefaultmidpunct}
{\mcitedefaultendpunct}{\mcitedefaultseppunct}\relax
\EndOfBibitem
\bibitem[Van Der~Spoel \latin{et~al.}(2005)Van Der~Spoel, Lindahl, Hess,
  Groenhof, Mark, and Berendsen]{van_der_spoel_gromacs:_2005}
Van Der~Spoel,~D.; Lindahl,~E.; Hess,~B.; Groenhof,~G.; Mark,~A.~E.;
  Berendsen,~H. J.~C. {GROMACS}: {Fast}, {Flexible}, and {Free}. \emph{J.
  Comput. Chem.} \textbf{2005}, \emph{26}, 1701--1718\relax
\mciteBstWouldAddEndPuncttrue
\mciteSetBstMidEndSepPunct{\mcitedefaultmidpunct}
{\mcitedefaultendpunct}{\mcitedefaultseppunct}\relax
\EndOfBibitem
\bibitem[Hess \latin{et~al.}(2008)Hess, Kutzner, van~der Spoel, and
  Lindahl]{hess_gromacs_2008}
Hess,~B.; Kutzner,~C.; van~der Spoel,~D.; Lindahl,~E. {GROMACS} 4: {Algorithms}
  for {Highly} {Efficient}, {Load}-{Balanced}, and {Scalable} {Molecular}
  {Simulation}. \emph{J. Chem. Theory Comput.} \textbf{2008}, \emph{4},
  435--447\relax
\mciteBstWouldAddEndPuncttrue
\mciteSetBstMidEndSepPunct{\mcitedefaultmidpunct}
{\mcitedefaultendpunct}{\mcitedefaultseppunct}\relax
\EndOfBibitem
\bibitem[Truong \latin{et~al.}(2018)Truong, Oudre, and
  Vayatis]{truong_ruptures:_2018}
Truong,~C.; Oudre,~L.; Vayatis,~N. Ruptures: {Change} {Point} {Detection} in
  {Python}. \emph{arXiv} \textbf{2018}, \emph{arXiv:1801.00826}\relax
\mciteBstWouldAddEndPuncttrue
\mciteSetBstMidEndSepPunct{\mcitedefaultmidpunct}
{\mcitedefaultendpunct}{\mcitedefaultseppunct}\relax
\EndOfBibitem
\bibitem[Meroz and Sokolov(2015)Meroz, and Sokolov]{meroz_toolbox_2015}
Meroz,~Y.; Sokolov,~I.~M. A {Toolbox} for {Determining} {Subdiffusive}
  {Mechanisms}. \emph{Phys. Rep.} \textbf{2015}, \emph{573}, 1--29\relax
\mciteBstWouldAddEndPuncttrue
\mciteSetBstMidEndSepPunct{\mcitedefaultmidpunct}
{\mcitedefaultendpunct}{\mcitedefaultseppunct}\relax
\EndOfBibitem
\bibitem[Clauset \latin{et~al.}(2009)Clauset, Shalizi, and
  Newman]{clauset_power-law_2009}
Clauset,~A.; Shalizi,~C.; Newman,~M. Power-{Law} {Distributions} in {Empirical}
  {Data}. \emph{SIAM Rev.} \textbf{2009}, \emph{51}, 661--703\relax
\mciteBstWouldAddEndPuncttrue
\mciteSetBstMidEndSepPunct{\mcitedefaultmidpunct}
{\mcitedefaultendpunct}{\mcitedefaultseppunct}\relax
\EndOfBibitem
\bibitem[Newman(2005)]{newman_power_2005}
Newman,~M. E.~J. Power {Laws}, {Pareto} {Distributions} and {Zipf}'s {Law}.
  \emph{Contemp. Phys.} \textbf{2005}, \emph{46}, 323--351\relax
\mciteBstWouldAddEndPuncttrue
\mciteSetBstMidEndSepPunct{\mcitedefaultmidpunct}
{\mcitedefaultendpunct}{\mcitedefaultseppunct}\relax
\EndOfBibitem
\bibitem[Metzler \latin{et~al.}(2014)Metzler, Jeon, Cherstvy, and
  Barkai]{metzler_anomalous_2014}
Metzler,~R.; Jeon,~J.-H.; Cherstvy,~A.~G.; Barkai,~E. Anomalous {Diffusion}
  {Models} and {Their} {Properties}: {Non}-{Stationarity}, {Non}-{Ergodicity},
  and {Ageing} at the {Centenary} of {Single} {Particle} {Tracking}.
  \emph{Phys. Chem. Chem. Phys.} \textbf{2014}, \emph{16}, 24128--24164\relax
\mciteBstWouldAddEndPuncttrue
\mciteSetBstMidEndSepPunct{\mcitedefaultmidpunct}
{\mcitedefaultendpunct}{\mcitedefaultseppunct}\relax
\EndOfBibitem
\bibitem[Neusius \latin{et~al.}(2009)Neusius, Sokolov, and
  Smith]{neusius_subdiffusion_2009}
Neusius,~T.; Sokolov,~I.~M.; Smith,~J.~C. Subdiffusion in {Time}-{Averaged},
  {Confined} {Random} {Walks}. \emph{Phys. Rev. E} \textbf{2009}, \emph{80},
  011109\relax
\mciteBstWouldAddEndPuncttrue
\mciteSetBstMidEndSepPunct{\mcitedefaultmidpunct}
{\mcitedefaultendpunct}{\mcitedefaultseppunct}\relax
\EndOfBibitem
\bibitem[Tikanmäki and Mishura(2010)Tikanmäki, and
  Mishura]{tikanmaki_fractional_2010}
Tikanmäki,~H.; Mishura,~Y. Fractional {Lévy} {Processes} as a {Result} of
  {Compact} {Interval} {Integral} {Transformation}. \emph{arXiv:1002.0780
  [math]} \textbf{2010}, arXiv: 1002.0780\relax
\mciteBstWouldAddEndPuncttrue
\mciteSetBstMidEndSepPunct{\mcitedefaultmidpunct}
{\mcitedefaultendpunct}{\mcitedefaultseppunct}\relax
\EndOfBibitem
\bibitem[Bishwal(2011)]{bishwal_maximum_2011}
Bishwal,~J. P.~N. Maximum {Quasi}-likelihood {Estimation} in {Fractional}
  {Levy} {Stochastic} {Volatility} {Model}. \emph{JMF} \textbf{2011},
  \emph{01}, 58--62\relax
\mciteBstWouldAddEndPuncttrue
\mciteSetBstMidEndSepPunct{\mcitedefaultmidpunct}
{\mcitedefaultendpunct}{\mcitedefaultseppunct}\relax
\EndOfBibitem
\bibitem[Mantegna and Stanley(1994)Mantegna, and
  Stanley]{mantegna_stochastic_1994}
Mantegna,~R.~N.; Stanley,~H.~E. Stochastic {Process} with {Ultraslow}
  {Convergence} to a {Gaussian}: {The} {Truncated} {Lévy} {Flight}.
  \emph{Phys. Rev. Lett.} \textbf{1994}, \emph{73}, 2946--2949\relax
\mciteBstWouldAddEndPuncttrue
\mciteSetBstMidEndSepPunct{\mcitedefaultmidpunct}
{\mcitedefaultendpunct}{\mcitedefaultseppunct}\relax
\EndOfBibitem
\bibitem[Bacallado \latin{et~al.}(2009)Bacallado, Chodera, and
  Pande]{bacallado_bayesian_2009}
Bacallado,~S.; Chodera,~J.~D.; Pande,~V. Bayesian {Comparison} of {Markov}
  {Models} of {Molecular} {Dynamics} with {Detailed} {Balance} {Constraint}.
  \emph{J. Chem. Phys.} \textbf{2009}, \emph{131}, 045106\relax
\mciteBstWouldAddEndPuncttrue
\mciteSetBstMidEndSepPunct{\mcitedefaultmidpunct}
{\mcitedefaultendpunct}{\mcitedefaultseppunct}\relax
\EndOfBibitem
\bibitem[Flynn(2019)]{flynn_exact_2019}
Flynn,~C.~R. Exact {Methods} for {Simulating} {Fractional} {Brownian} {Motion}
  and {Fractional} {Gaussian} {Noise} in {Python}. 2019;
  \url{https://github.com/crflynn/fbm}\relax
\mciteBstWouldAddEndPuncttrue
\mciteSetBstMidEndSepPunct{\mcitedefaultmidpunct}
{\mcitedefaultendpunct}{\mcitedefaultseppunct}\relax
\EndOfBibitem
\bibitem[Stoev and Taqqu(2004)Stoev, and Taqqu]{stoev_simulation_2004}
Stoev,~S.; Taqqu,~M.~S. Simulation {Methods} for {Linear} {Fractional} {Stable}
  {Motion} and {Farima} {Using} the {Fast} {Fourier} {Transform}.
  \emph{Fractals} \textbf{2004}, \emph{12}, 95--121\relax
\mciteBstWouldAddEndPuncttrue
\mciteSetBstMidEndSepPunct{\mcitedefaultmidpunct}
{\mcitedefaultendpunct}{\mcitedefaultseppunct}\relax
\EndOfBibitem
\bibitem[Pande \latin{et~al.}(2010)Pande, Beauchamp, and
  Bowman]{pande_everything_2010}
Pande,~V.~S.; Beauchamp,~K.; Bowman,~G.~R. Everything {You} {Wanted} to {Know}
  {About} {Markov} {State} {Models} but {Were} {Afraid} to {Ask}.
  \emph{Methods} \textbf{2010}, \emph{52}, 99--105\relax
\mciteBstWouldAddEndPuncttrue
\mciteSetBstMidEndSepPunct{\mcitedefaultmidpunct}
{\mcitedefaultendpunct}{\mcitedefaultseppunct}\relax
\EndOfBibitem
\bibitem[Wehmeyer \latin{et~al.}(2018)Wehmeyer, Scherer, Hempel, Husic, Olsson,
  and Noé]{wehmeyer_introduction_2018}
Wehmeyer,~C.; Scherer,~M.~K.; Hempel,~T.; Husic,~B.~E.; Olsson,~S.; Noé,~F.
  Introduction to {Markov} {State} {Modeling} with the {Pyemma} {Software}
  [article {V1}.0]. \emph{LiveCoMS} \textbf{2018}, \emph{1}, 5965--\relax
\mciteBstWouldAddEndPuncttrue
\mciteSetBstMidEndSepPunct{\mcitedefaultmidpunct}
{\mcitedefaultendpunct}{\mcitedefaultseppunct}\relax
\EndOfBibitem
\bibitem[Luzar and Chandler(1996)Luzar, and Chandler]{luzar_effect_1996}
Luzar,~A.; Chandler,~D. Effect of {Environment} on {Hydrogen} {Bond} {Dynamics}
  in {Liquid} {Water}. \emph{Phys. Rev. Lett.} \textbf{1996}, \emph{76},
  928--931\relax
\mciteBstWouldAddEndPuncttrue
\mciteSetBstMidEndSepPunct{\mcitedefaultmidpunct}
{\mcitedefaultendpunct}{\mcitedefaultseppunct}\relax
\EndOfBibitem
\bibitem[Hill(1989)]{hill_free_1989}
Hill,~T.~L. \emph{Free {Energy} {Transduction} and {Biochemical} {Cycle}
  {Kinetics}}; Springer-Verlag, 1989\relax
\mciteBstWouldAddEndPuncttrue
\mciteSetBstMidEndSepPunct{\mcitedefaultmidpunct}
{\mcitedefaultendpunct}{\mcitedefaultseppunct}\relax
\EndOfBibitem
\bibitem[Cussler(2009)]{cussler_diffusion:_2009}
Cussler,~E.~L. \emph{Diffusion: {Mass} {Transfer} in {Fluid} {Systems}}, 3rd
  ed.; Cambridge University Press, 2009\relax
\mciteBstWouldAddEndPuncttrue
\mciteSetBstMidEndSepPunct{\mcitedefaultmidpunct}
{\mcitedefaultendpunct}{\mcitedefaultseppunct}\relax
\EndOfBibitem
\bibitem[Guo \latin{et~al.}(2004)Guo, Chung, and
  Matsuura]{guo_pervaporation_2004}
Guo,~W.~F.; Chung,~T.-S.; Matsuura,~T. Pervaporation {Study} on the
  {Dehydration} of {Aqueous} {Butanol} {Solutions}: {A} {Comparison} of {Flux}
  {Vs}. {Permeance}, {Separation} {Factor} {Vs}. {Selectivity}. \emph{J. Membr.
  Sci.} \textbf{2004}, \emph{245}, 199--210\relax
\mciteBstWouldAddEndPuncttrue
\mciteSetBstMidEndSepPunct{\mcitedefaultmidpunct}
{\mcitedefaultendpunct}{\mcitedefaultseppunct}\relax
\EndOfBibitem
\bibitem[Kedem and Katchalsky(1963)Kedem, and
  Katchalsky]{kedem_permeability_1963}
Kedem,~O.; Katchalsky,~A. Permeability of {Composite} {Membranes}. {Part}
  1.—{Electric} {Current}, {Volume} {Flow} and {Flow} of {Solute} {Through}
  {Membranes}. \emph{Trans. Faraday Soc.} \textbf{1963}, \emph{59},
  1918--1930\relax
\mciteBstWouldAddEndPuncttrue
\mciteSetBstMidEndSepPunct{\mcitedefaultmidpunct}
{\mcitedefaultendpunct}{\mcitedefaultseppunct}\relax
\EndOfBibitem
\bibitem[Al-Zoubi \latin{et~al.}(2007)Al-Zoubi, Hilal, Darwish, and
  Mohammad]{al-zoubi_rejection_2007}
Al-Zoubi,~H.; Hilal,~N.; Darwish,~N.~A.; Mohammad,~A.~W. Rejection and
  {Modelling} of {Sulphate} and {Potassium} {Salts} by {Nanofiltration}
  {Membranes}: {Neural} {Network} and {Spiegler}–{Kedem} {Model}.
  \emph{Desalination} \textbf{2007}, \emph{206}, 42--60\relax
\mciteBstWouldAddEndPuncttrue
\mciteSetBstMidEndSepPunct{\mcitedefaultmidpunct}
{\mcitedefaultendpunct}{\mcitedefaultseppunct}\relax
\EndOfBibitem
\bibitem[Ernst \latin{et~al.}(2012)Ernst, Hellmann, Köhler, and
  Weiss]{ernst_fractional_2012}
Ernst,~D.; Hellmann,~M.; Köhler,~J.; Weiss,~M. Fractional {Brownian} {Motion}
  in {Crowded} {Fluids}. \emph{Soft Matter} \textbf{2012}, \emph{8},
  4886--4889\relax
\mciteBstWouldAddEndPuncttrue
\mciteSetBstMidEndSepPunct{\mcitedefaultmidpunct}
{\mcitedefaultendpunct}{\mcitedefaultseppunct}\relax
\EndOfBibitem
\bibitem[Bel and Barkai(2005)Bel, and Barkai]{bel_weak_2005}
Bel,~G.; Barkai,~E. Weak {Ergodicity} {Breaking} in the {Continuous}-{Time}
  {Random} {Walk}. \emph{Phys. Rev. Lett.} \textbf{2005}, \emph{94},
  240602\relax
\mciteBstWouldAddEndPuncttrue
\mciteSetBstMidEndSepPunct{\mcitedefaultmidpunct}
{\mcitedefaultendpunct}{\mcitedefaultseppunct}\relax
\EndOfBibitem
\bibitem[Clegg(2006)]{clegg_practical_2006}
Clegg,~R.~G. A {Practical} {Guide} to {Measuring} the {Hurst} {Parameter}.
  \emph{arXiv:math/0610756} \textbf{2006}, arXiv: math/0610756\relax
\mciteBstWouldAddEndPuncttrue
\mciteSetBstMidEndSepPunct{\mcitedefaultmidpunct}
{\mcitedefaultendpunct}{\mcitedefaultseppunct}\relax
\EndOfBibitem
\end{mcitethebibliography}
